\documentclass[11pt,a4paper]{article}

\usepackage[
            top=1.0in,
            bottom=1.0in,
            left=1.0in,
            right=1.0in
            ]{geometry}

\usepackage[natbib=true,
            style=nature,
            sorting=none,
            citestyle=numeric-comp,
            ]{biblatex}

\usepackage[
            automake,
            ]{glossaries-extra}    

\usepackage[
            numbib,
            ]{tocbibind}

\usepackage[
            T1,
            ]{fontenc}

\usepackage[normalem]{ulem}
\usepackage[table]{xcolor}

\usepackage{pythonhighlight}
\usepackage{subcaption}
\usepackage{tcolorbox}
\usepackage{graphicx}
\usepackage{pdfpages}
\usepackage{tabularx}
\usepackage{booktabs}
\usepackage{authblk}
\usepackage{amsmath}
\usepackage{lineno}
\usepackage{lscape}
\usepackage{array}
\usepackage{float}
\usepackage{url}
\usepackage{xr}

\tcbuselibrary{breakable}

\makeglossaries

\newacronym{ml}{ML}{Machine Learning}
\newacronym{ai}{AI}{Artificial Intelligence}

\newacronym[sort=id1]{id}{$ID$}{Inside the Domain}
\newacronym[sort=od1]{od}{$OD$}{Outside the Domain}
\newacronym[sort=idod1]{idod}{\gls{id}/\gls{od}}{\gls{id} or \gls{od}}

\newacronym[sort=id2]{hatid}{$\widehat{ID}$}{Predicted \gls{id}}
\newacronym[sort=od2]{hatod}{$\widehat{OD}$}{Predicted \gls{od}}
\newacronym[sort=idod2]{hatidod}{\gls{hatid}/\gls{hatod}}{Predicted \gls{idod}}

\newacronym{tp}{TP}{True Positive}
\newacronym{tn}{TN}{True Negative}
\newacronym{fp}{FP}{False Positive}
\newacronym{fn}{FN}{False Negative}

\newacronym{kde}{KDE}{Kernel Density Estimate}
\newacronym{gpr}{GPR}{Gaussian Process Regression}

\newacronym{rmse}{RMSE}{Root Mean Squared Error}
\newacronym{auc}{AUC}{Area Under the Curve}

\newacronym{cv}{CV}{Cross Validation}
\newacronym{oob}{OOB}{Out-Of-Bag}
\newacronym{itb}{ITB}{In-The-Bag}
\newacronym{loo}{LOO}{Leave-One-Out}
\newacronym{bloco}{BLOCO}{Bootstrapped Leave-One-Cluster Out}

\newacronym{nn}{NN}{Neural Network}
\newacronym{rf}{RF}{Random Forest}
\newacronym{bsvr}{BSVR}{Bagged Support Vector Regressor}
\newacronym{bnn}{BNN}{Bagged Neural Network}
\newacronym{bols}{BOLS}{Bagged Ordinary Least Squares}

\newacronym{umap}{UMAP}{Uniform Manifold Approximation and Projection}
\newacronym{mastml}{MAST-ML}{Materials Simulation Toolkit for Machine Learning}

\newacronym{dft}{DFT}{Density Functional Theory}

\newacronym{rpv}{RPV}{Reactor Pressure Vessel}
\newacronym{fwodc}{FWODC}{\gls{friedman} WithOut Distinct Clusters}

\newglossaryentry{mprop}
{
    name=\ensuremath{M^{prop}},
    sort={m1},
    description={The model for property prediction},
}
\newglossaryentry{munc}
{
    name=\ensuremath{M^{unc}},
    sort={m2},
    description={The model for uncertainty prediction},
}
\newglossaryentry{mdis}
{
    name=\ensuremath{M^{dis}},
    sort={m3},
    description={The model for measuring dissimilarity through \gls{kde}},
}
\newglossaryentry{mdom}
{
    name=\ensuremath{M^{dom}},
    sort={m4},
    description={The model classifying \gls{hatidod}},
}
\newglossaryentry{mcdom}
{
    name=\ensuremath{M^{cdom}},
    sort={m5},
    description={The composite function \gls{mdom}(\gls{mdis}(\gls{x}))},
}

\newglossaryentry{eres}
{
    name=\ensuremath{E^{|y-\hat{y}|/MAD_{y}}},
    sort={e1},
    description={The absolute residuals normalized by \gls{mady}},
}
\newglossaryentry{ermse}
{
    name=\ensuremath{E^{RMSE/\sigma_{y}}},
    sort={e2},
    description={The \gls{rmse} normalized by \gls{sigmay}},
}
\newglossaryentry{earea}
{
    name=\ensuremath{E^{area}},
    sort={e3},
    description={The miscalibration area},
}
\newglossaryentry{echem}
{
    name=\ensuremath{E^{chem}},
    sort={e4},
    description={The chemical mismatch as defined},
}

\newglossaryentry{eresc}
{
    name=\ensuremath{E^{|y-\hat{y}|/MAD_{y}}_{c}},
    sort={ec1},
    description={The ground truth label cutoff for \gls{eres}},
}
\newglossaryentry{ermsec}
{
    name=\ensuremath{E^{RMSE/\sigma_{y}}_{c}},
    sort={ec2},
    description={The ground truth label cutoff for \gls{ermse}},
}
\newglossaryentry{eareac}
{
    name=\ensuremath{E^{area}_{c}},
    sort={ec3},
    description={The ground truth label cutoff for \gls{earea}},
}
\newglossaryentry{echemc}
{
    name=\ensuremath{E^{chem}_{c}},
    sort={ec4},
    description={The ground truth label cutoff  for \gls{echem}},
}

\newglossaryentry{ares}
{
    name=\ensuremath{A^{|y-\hat{y}|/MAD_{y}}},
    sort={a1},
    description={The assessment of \gls{d} with respect to labels produced with \gls{eresc}},
}
\newglossaryentry{armse}
{
    name=\ensuremath{A^{RMSE/\sigma_{y}}},
    sort={a2},
    description={The assessment of \gls{d} with respect to labels produced with \gls{ermsec}},
}
\newglossaryentry{aarea}
{
    name=\ensuremath{A^{area}},
    sort={a3},
    description={The assessment of \gls{d} with respect to labels produced with \gls{eareac}},
}
\newglossaryentry{achem}
{
    name=\ensuremath{A^{chem}},
    sort={a4},
    description={The assessment of \gls{d} with respect to labels produced with \gls{echemc}},
}

\newglossaryentry{X}
{
    name=\ensuremath{X},
    sort={X},
    description={The features for a set of \gls{x}},
}
\newglossaryentry{x}
{
    name=\ensuremath{\vec{x}},
    sort={x},
    description={The feature values for a single case},
}
\newglossaryentry{Xitb}
{
    name=\ensuremath{X_{\gls{itb}}},
    sort={Xitb},
    description={The \gls{X} for \gls{itb} data},
}
\newglossaryentry{y}
{
    name=\ensuremath{y},
    sort={y1},
    description={The regression target},
}
\newglossaryentry{z}
{
    name=\ensuremath{z},
    sort={z},
    description={Scaled uncertainties defined by (\gls{y}-\gls{haty})/\gls{sigmac}},
}
\newglossaryentry{Z}
{
    name=\ensuremath{Z},
    sort={Z},
    description={A set of \gls{z} values},
}
\newglossaryentry{d}
{
    name=\ensuremath{d},
    sort={d1},
    description={The \gls{kde} dissimilarity measure},
}
\newglossaryentry{dt}
{
    name=\ensuremath{d^{t}},
    sort={d2},
    description={A cutoff value in the interval $[0,1]$ used by \gls{mdom} for class prediction},
}
\newglossaryentry{dc}
{
    name=\ensuremath{d^{t}_{c}},
    sort={d3},
    description={A selected value of \gls{dc} which gives \gls{f1max}},
}
\newglossaryentry{haty}
{
    name=\ensuremath{\hat{y}},
    sort={y2},
    description={The regression target},
}
\newglossaryentry{sigmau}
{
    name=\ensuremath{\sigma_{u}},
    sort={sigmau},
    description={Uncalibrated uncertainty estimates},
}
\newglossaryentry{sigmac}
{
    name=\ensuremath{\sigma_{c}},
    sort={sigmac},
    description={Calibrated uncertainty estimates},
}
\newglossaryentry{sigmay}
{
    name=\ensuremath{\sigma_{y}},
    sort={sigmay},
    description={The standard deviation of a set of \gls{y} values},
}
\newglossaryentry{mady}
{
    name=\ensuremath{MAD_y},
    sort={mady},
    description={The mean absolute deviation of \gls{y} values},
}
\newglossaryentry{bary}
{
    name=\ensuremath{\bar{y}},
    sort={y3},
    description={The mean of a set of \gls{y} values},
}
\newglossaryentry{r}
{
    name=\ensuremath{r},
    sort={r},
    description={The Pearson's correlation coefficient},
}

\newglossaryentry{oz}
{
    name=\ensuremath{O(z)},
    sort={oz},
    description={The observed cumulative distribution of \gls{z} values},
}
\newglossaryentry{phi}
{
    name=\ensuremath{\Phi(0,1)},
    sort={phi},
    description={The cumulative distribution of a standard normal distribution where the means is 0 and the standard deviation is 1 by definition},
}

\newglossaryentry{f1}
{
    name=\ensuremath{F1},
    sort={f11},
    description={The harmonic mean between precision and recall scores},
}
\newglossaryentry{f1max}
{
    name=\ensuremath{F1_{max}},
    sort={f12},
    description={The maximum \gls{f1}},
}
\newglossaryentry{deltaf1}
{
    name=\ensuremath{\Delta F1},
    sort={f13},
    description={The difference between the \gls{f1max} mentioned in the main text (\gls{d} from \gls{kde}) and another measure of \gls{f1max} attained from alternate means},
}

\newglossaryentry{diffusion}
{
    name={Diffusion},
    sort={Diffusion},
    description={The Diffusion data set},
}
\newglossaryentry{fluence}
{
    name={Fluence},
    sort={Fluence},
    description={The Fluence data set},
}
\newglossaryentry{strength}
{
    name={Steel Strength},
    sort={Steel Strength},
    description={The Steel Strength data set},
}
\newglossaryentry{supercond}
{
    name={Superconductor},
    sort={Superconductor},
    description={The Superconductor data set},
}
\newglossaryentry{iron}
{
    name={Iron-Based},
    sort={Iron-Based},
    description={The Iron-Based subset of the \gls{supercond} data set},
}
\newglossaryentry{cuprates}
{
    name={Cuprates},
    sort={Cuprates},
    description={The Cuprate subset of the \gls{supercond} data set},
}
\newglossaryentry{lowtc}
{
    name={Low-\ensuremath{T_{c}}},
    sort={low-tc},
    description={The Low-$T_{c}$ subset of the \gls{supercond} data set},
}
\newglossaryentry{friedman}
{
    name={Friedman},
    sort={Friedman},
    description={The Friedman data set},
}

\title{A General Approach for Determining Applicability Domain of Machine Learning Models}

\author[a*]{Lane E. Schultz}
\author[b]{Yiqi Wang}
\author[a]{Ryan Jacobs}
\author[a]{Dane Morgan}

\affil[a]{University of Wisconsin-Madison, 1509 University Avenue, Madison, WI, 53706, USA}
\affil[b]{Carnegie Mellon University, 5000 Forbes Ave, Pittsburgh, PA 15213, USA}
\affil[*]{Corresponding author. Email: lsschultz@wisc.edu (L.E.S.)}

\date{\today}

\begin{document}

\maketitle 

\begin{abstract}
Knowledge of the domain of applicability of a machine learning model is essential to ensuring accurate and reliable model predictions. In this work, we develop a new and general approach of assessing model domain and demonstrate that our approach provides accurate and meaningful domain designation across multiple model types and material property data sets. Our approach assesses the distance between data in feature space using kernel density estimation, where this distance provides an effective tool for domain determination. We show that chemical groups considered unrelated based on chemical knowledge exhibit significant dissimilarities by our measure. We also show that high measures of dissimilarity are associated with poor model performance (i.e., high residual magnitudes) and poor estimates of model uncertainty (i.e., unreliable uncertainty estimation). Automated tools are provided to enable researchers to establish acceptable dissimilarity thresholds to identify whether new predictions of their own machine learning models are in-domain versus out-of-domain.
\end{abstract}


\section*{Introduction}
\label{introduction}

Machine learning (ML), as one component of the larger umbrella of artificial intelligence (AI), is one of the fastest evolving technologies in the world today. In the context of materials science, thousands of papers using ML are now published each year, and the number of publications using ML has been growing exponentially since around 2015 \supercite{Morgan2020,Morgan2023}. The applications of ML in materials science takes many forms, including materials property prediction \supercite{Long2023}, computer vision-based defect detection and microstructure segmentation \supercite{Gao2022, Jacobs2022}, assimilation of data and knowledge from publications using natural language processing and large language models \supercite{polak2023extracting,ChemDataExtractor}, and fitting of ML-based interatomic potentials representing nearly all elements in the periodic table to enable fast and accurate atomistic simulations \supercite{Chen2022}.

Useful ML models generally require some form of prediction quality quantification because models can experience significant performance degradation when predicting on data that falls outside the model's domain of applicability. This performance degradation can manifest as high errors, unreliable uncertainty estimates, or both. Without some estimation of model domain, one does not know, a priori, whether the results are reliable when making predictions on new test data. More precisely, useful ML models ideally have at least the following three characteristics regarding their prediction quality: (i) accurate prediction, meaning the model has low residual magnitudes, (ii) accurate uncertainty in prediction, meaning the model produces some useful quantification of uncertainty on new predictions (note this requirement does not stipulate that the uncertainties should be small), and (iii) domain classification, meaning the model can reliably determine when predictions are inside a domain ($ID$) versus outside a domain ($OD$) of feature space where the model is trustworthy.

Separate from determining whether data are $ID$ or $OD$ are the techniques used in domain adaptation. In certain situations, domain adaptation techniques enable the fine-tuning of a model or data to transform originally $OD$ data into $ID$ data. The objective of domain adaptation is to adapt a model for prediction on a property (denoted as $M^{prop}$ here) to new data whose distribution may be shifted from training data \supercite{Stahlbock,de2021adapt}. There would be no need to identify $OD$ data if domain adaptation were always effective, but adapting models to initially $OD$ data can be a challenging and intricate process. First, many techniques require re-training models, involving a substantial effort in tuning parameters and validating models. Second, once a model is adapted to one target domain, it may still fail on other unknown domains. It is therefore useful to have a method to identify when a model is applied to problematic domains without having to adapt the model. In this work, we develop a domain classification technique that identifies when predictions are likely $ID$ or $OD$ (equivalently $ID$/$OD$).

The domain classification problem, at least for materials property prediction and many other similar problems, can be formulated as follows: given a trained model $M^{prop}$ and the features of an arbitrary test data point, how can we develop a model to predict if the test data point is $ID$/$OD$ for $M^{prop}$? In this work, we frame this challenge as a classic supervised ML problem for categorization. To develop such a model, we need training data with input features and labels, which labels are $ID$/$OD$, as well as some ML modeling approach for making the label prediction. We will denote this ML model for domain as $M^{dom}$ to distinguish it from $M^{prop}$. Note that the labeled training data for $M^{dom}$ does not necessarily have to match the original $M^{prop}$ model training data.

There is no unique, universal definition for the domain of an $M^{prop}$ model, and therefore no unambiguously defined labels for the $M^{dom}$ training data. In other words, we do not have an absolute ground truth labeling on which to train the $M^{dom}$ model~\supercite{pseudo_labels,Yang2021}. This problem can be solved by imposing some reasonable definition of ground truth for $ID$/$OD$ based on model reliability, as quantified by, e.g., small residual magnitudes and stable predictions under changes in data. In many cases $ID$/$OD$ data points are described in terms of a region of feature space, in which case $M^{dom}$ becomes a trivial check if a data point is in the region or not \supercite{Jaworska2005}. Predictions of $ID$/$OD$ are often checked against chemical intuition of whether data are somehow “similar” to training data, and therefore $ID$, or not, and therefore $OD$. Such checks are an effective way of using field-specific knowledge of similarity to provide a ground truth for $ID$/$OD$ classification.

We are aware of three approaches, developed in Refs.~\cite{panapitiya2023outlierbased}, \cite{Sutton2020}, and \cite{convex_hull}, that effectively find a region in feature space where an $M^{prop}$ model shows performance above some cutoff. Ref.~\cite{panapitiya2023outlierbased} employs threshold values for descriptors to ascertain whether a prediction is $ID/OD$ based on collected prediction errors. While their method for domain classification appears to be effective, several issues regarding complexity and implementation made it difficult for us to implement. Features are prioritized based on their error reduction capacity, and thresholds are assigned to each feature to distinguish between $ID$ and $OD$ points. This requires determining a threshold for each feature based on a cross validation scheme with its own threshold. Also, their approach does not provide a definitive maximum number of features to consider, resulting in $ID/OD$ classifications that vary depending on the number of selected features. In Ref.~\cite{Sutton2020}, the purpose of the developed technique is to establish a continuous feature space where models yield low errors (i.e., points that are $ID$) from model predictions via a tradeoff of feature region coverage and error. They are limited by a single continuous region when multiple areas in feature space could provide low error predictions from models. In the case of Ref.~\cite{convex_hull}, they gather prediction errors from cross validation, reduce the feature dimension to five using principal component analysis, establish an error threshold of the fifth percentile of errors to define the boundary of a convex hull, and finally measure the distance of predicted points from the convex hull. A convex hull defines a boundary that encompasses a set of points, creating a region that may contain areas where training data are sparse or underrepresented in a model's feature space. These approaches are potentially very powerful but have limitations. For methods that create a single connected region to denote as $ID$, multiple disjointed regions in space that could yield perfectly reasonable predictions from $M^{prop}$ will be excluded. A method by which non-connected $ID$ regions are established without a single, pre-defined shape would be advantageous (e.g., with kernel density estimation as done in this work). In addition, the sophisticated approaches in these models introduce significant complexity, which makes them challenging to implement. It is therefore useful to revisit simpler approaches that build directly on the intuition that $ID$ regions of feature space are likely to be those regions close to significant amounts of training data \supercite{Caldeira2020,Janet2019,convex_hull}.

There are many techniques for quantifying closeness in feature spaces, including convex hulls, distance measures, and (probability) density estimates \supercite{Jaworska2005}. While identifying as $ID$ all data within a convex hull of the training data methods is reasonable, such approaches have the major limitation of potentially including large regions with no training data. For example, the convex hull of points on a circle in a two-dimensional feature space includes the entirely empty middle of the circle as $ID$. Distance measures are also a reasonable approach to measure closeness. In Ref.~\cite{Askanazi2024}, distance measures from a number of nearest neighbors was used as a dissimilarity score between a point and a model's training data. They showed that target property prediction errors generally increase with increasing distance, which followed intuition. Nevertheless, distance measures have the limitation of there being no unique measure of distance between two points and no unique single distance of a new point from a set of training data. This creates a vast space of possible ways of measuring two-point distances (e.g., Euclidean, Mahalanobis, etc.) and one to N-point distances (closest point distance, closest k-points distance, weighted average of distances, etc.), making it difficult to find a robust method. In general, approaches based on convex hulls or standard distances between two points do not account naturally for data sparsity, and may consider a point near one outlier training data point or many training data points as almost identically likely to be $ID$.

Kernel density estimation (KDE), and density-based methods in general, offer several advantages vs. other approaches, including (i) a density value that can act as a distance or dissimilarity measure, (ii) a natural accounting for data sparsity, and (iii) trivial treatment of arbitrarily complex geometries of data and $ID$ regions. Techniques utilizing Gaussian process have comparable advantages. But unlike Gaussian process, KDE is relatively fast to fit and evaluate, at least of modest size data sets that are common in materials (see the Supplemental Materials for comparisons between KDE and Gaussian process regression). The work in Ref.~\cite{li2024probing} employed KDEs to demonstrate, using a projection of features, that many assessments of machine learning models were in regions where models had a significant number of training data. Their research demonstrated that numerous assessments previously categorized as extrapolation were, upon closer examination, actually instances of interpolation. The authors also showed that model residuals generally increased in regions of the feature space with little to no training data. While the authors employed KDE in their study, they do not utilize it as a means to categorize new predictions as $ID$/$OD$. In contrast, our research establishes a definition for $ID$/$OD$ classification based on prediction errors. Later, we demonstrate how KDE can effectively differentiate data points that fall within the $ID$ category and those that are considered $OD$. We therefore focus on KDE, as it provides a natural solution to all the issues around topology, distance measures, data sparsity, and has been effectively shown to have a relationship with prediction errors in the past.

Given the absence of any unique ground truth, as noted above, we explore four different approaches for defining $ID$/$OD$. Specifically, we define four domain types, each based on a corresponding ground truth, which are: (i) a chemical domain where test data materials with similar chemical characteristics to the training data are $ID$, (ii) a residual domain where test data with residuals below a chosen threshold are $ID$, (iii) another residual domain where groups (i.e., not single cases) of test data with residuals below a chosen threshold are $ID$, and (iv) an uncertainty domain where groups of test data with differences between predicted and expected uncertainties below a chosen threshold are $ID$. For each of our four domain types, we assessed our models on sets of test data that were increasingly distinct from the training data. Generally, test cases that had low KDE likelihoods were chemically dissimilar, had large residuals, and had inaccurate uncertainties, just as one would hope for an effective method of domain determination.

Here we summarize the structure of the paper. Results starts with some basic details on ML models and our definitions of domain. Four materials property data sets and one commonly used synthetic data set were studied. Various model types including random forest, bagged support vector regressor, bagged neural network, and bagged ordinary least squares models were trained with the aforementioned data sets. We then show the assessment of our domain predictions. We finish with a discussion, methods that show the methodology behind assessing domains and data curation, data resources, code resources, acknowledgements, author contributions, and competing interests. Our findings indicate that KDE likelihoods can provide valuable insights of model applicability domain, enabling effective classification of $ID$/$OD$ for most data sets and models examined here. Importantly, our approach is expected to be generally applicable to regression-based prediction problems involving tabular data and can be easily applied to other regression tasks.

\section*{Results}

Our work presents an approach for determining machine learning model applicability domains. The beginning of this section establishes critical terminology and mathematical foundations that are referenced extensively in the remaining results. Having readers understand these concepts before encountering numerical results significantly enhances the impact of the remaining work. Subsequently, we show how these conceptual tools can build reliable models for domain determination based on several criteria including chemistry, residuals of predictions, and quality of uncertainty estimate measures.

\color{black}

\subsection*{Definition of Model Types}
\label{all_models}

We outline here the types of models, their applications in this work, and define the nomenclature for each. First, $M^{prop}$ is a regression model that uses $X$ to predict a property, $y$. Second, $M^{dom}$ is a classification model that predicts domain labels $ID$/$OD$ given the features of a single data point, $\vec{x}$. The predicted labels of domain will henceforth be called $\widehat{ID}$ and $\widehat{OD}$. Construction of $M^{dom}$ requires other additional models. We define a model called $M^{dis}$($X$,$\vec{x}$) (or just $M^{dis}$ for short) which returns a dissimilarity score between $\vec{x}$ and the data $X$ used in training $M^{prop}$. Finally, we also build a model for predicting uncertainties in $M^{prop}$ predictions, $M^{unc}$, which uses $M^{prop}$. The data used to build or train the models of $M^{prop}$, $M^{unc}$, and $M^{dis}$ will be referred to as In-The-Bag (ITB). Conversely, the data excluded from the training of these models will be termed Out-Of-Bag (OOB). The following subsections provide explanations for how $M^{prop}$, $M^{unc}$, and $M^{dis}$ combine to generate $M^{dom}$. Because of the interdependence of many kinds of models to produce $M^{dom}$, a summary of models, their inputs, and their outputs are provided in Table~\ref{models}.

\begin{table}[H]
    \centering
    \caption{The summary of variables used in model training, variables used in model prediction, and the outputs from each model are covered here. $\{\cdot\}$ represents a set.}
    \label{models}
    \begin{tabular}{lllll}
    \toprule
    \textbf{Model} & \textbf{Description}    & \textbf{Training Inputs}  & \textbf{Deployment Inputs} & \textbf{Outputs} \\
    \midrule
    $M^{prop}$    & Property prediction     & $X,\{y\}$                  & $\vec{x}$    & $\hat{y},\sigma_{u}$           \\
    $M^{unc}$     & Uncertainty estimation  & $\{y,\hat{y},\sigma_{u}\}$ & $\sigma_{u}$ & $\sigma_{c}$                   \\
    $M^{dis}$     & Measuring dissimilarity & $X$                        & $\vec{x}$    & $d$                            \\
    $M^{dom}$     & Classifying domain      & $\{d,ID/OD\}$              & $d$          & $\widehat{ID}/\widehat{OD}$    \\
    \bottomrule
    \end{tabular}
\end{table}

In some cases, these models are very simple, e.g., just a simple function or a check if a value is above or below a cutoff. We make the choice of defining them as models and denoting them with a variable for two reasons. First, it gives a well-defined symbol for each item, which makes the discussion more precise and compact, although at the cost of more variables. Second, defining each of these relationships as models stresses the fact that the specific models we use in this work could easily be replaced by other models, and these could be much more complex. For example, our domain model $M^{dom}$ for predicting whether a test data point with features $\vec{x}$ belongs to the $ID$ domain is a simple check of whether the kernel density value exceeds a cutoff. However, this could be replaced by a more complex ML model based on the KDE or other features. We hope that these definitions will help make the paper clearer and suggest natural ways to improve our approach in the future.

\subsubsection*{Model for Property Regression ($M^{prop}$)}
\label{prop_model}

We investigated a range of model types for $M^{prop}$ including Random Forest (RF), Bagged Support Vector Regressor (BSVR), Bagged Neural Network (BNN), and Bagged Ordinary Least Squares (BOLS). We used bagged versions of all but the RF models (RF is already an ensemble model) as these ensembles were used to generate uncertainty estimates and define the model $M^{unc}$. Note that $M^{dis}$ depends only on the features of ITB data, $X_{ITB}$, and does not depend on the form of $M^{prop}$ nor $M^{unc}$. Two types of errors on property predictions were considered. First, absolute residuals ($|y-\hat{y}|$), normalized by the Mean Absolute Deviation of $y$ ($MAD_y$) was measured with Eq.~\ref{absres_mad} and was named $E^{|y-\hat{y}|/MAD_{y}}$.

\begin{equation}\label{absres_mad}
    E^{|y-\hat{y}|/MAD_{y}} = \frac{|y-\hat{y}|}{MAD_y}
\end{equation}

Second, we denote the Root Mean Squared Error ($RMSE$) of predictions from $M^{prop}$ normalized by the standard deviation of $y$ ($\sigma_{y}$) by the symbol $E^{RMSE/\sigma_{y}}$ (see Eq.~\ref{rmse_sigma}). In both Eqs.~\ref{absres_mad} and \ref{rmse_sigma}, ``$E$'' denotes that it is a type of error, $\hat{y}$ is a prediction from $M^{prop}$, and $\bar{y}$ is the mean of $y$. $E^{|y-\hat{y}|/MAD_{y}}$ can be measured for any individual data point considered, but can be randomly low for a data point known to be $OD$ (i.e., residuals are stochastic). $E^{RMSE/\sigma_{y}}$ considers residuals for groups of data, so data points with low and high values of $E^{|y-\hat{y}|/MAD_{y}}$ are included in a statistical measure. Both $E^{|y-\hat{y}|/MAD_{y}}$ and $E^{RMSE/\sigma_{y}}$ were used in producing $ID$/$OD$ labels to train $M^{dom}$ (see the Defining Ground Truths section).

\begin{equation}\label{rmse_sigma}
    E^{RMSE/\sigma_{y}} = \frac{RMSE}{\sigma_{y}} = \sqrt{\frac{\sum^{N}_{i=1} (y_{i}-\hat{y}_{i})^{2}}{\sum^{N}_{i=1} (y_{i}-\bar{y})^{2}}}
\end{equation}

\subsubsection*{Model for Uncertainty Estimates ($M^{unc}$)}
\label{std_method}

Uncertainty calibration has been extensively studied in both classification and regression settings \supercite{Abdar2021,Scalia2020,Tran2019,Caruana2005,Kull2017}. Our study utilized uncertainty estimates for regression and used the calibration implementation from Palmer et al. \supercite{Palmer2022}. In this approach, we started from an ensemble model to acquire individual predictions from each sub-model comprising an ensemble. The mean of predictions from the sub-models is $\hat{y}$. The standard deviation of predictions, $\sigma_{u}$, was calculated and then calibrated to yield the calibrated uncertainty estimates, denoted as $\sigma_{c}$. Repeated 5-fold Cross Validation (CV) was used throughout the study for producing residual and $\sigma_{u}$ values used for calibration.

We borrowed arguments from Ref.~\cite{Pernot_long} for assessing the quality of calibrated uncertainty estimates. The mean and variance of a set of $z$-scores, $Z$, should be 0 (unbiased) and 1 (unit-scaled), respectively. Each $z$ $\in$ $Z$ was calculated by $z=(y-\hat{y})/\sigma_{c}$. We made the assumption that the distribution of $Z$ is standard normal (i.e., had a mean of 0 and a standard deviation of 1) for reliable uncertainty estimates as done in Refs.~\cite{ling2017high} and \cite{Palmer2022}, which is not done in Ref.~\cite{Pernot_long}. We called the cumulative distribution of both $Z$ and the standard normal distribution $O(z)$ and $\Phi(0,1)$, respectively. We calculated how far any $O(z)$ was from $\Phi(0,1)$ via Eq.~\ref{misscalibration_area}, where $E^{area}$ is the miscalibration area or error in uncertainties. Other measures of the quality of uncertainty measures exist like sharpness, dispersion, etc. (see Ref.~\cite{Gruich2023}) that could be used instead. However, we used Eq.~\ref{misscalibration_area} because it includes information of $\sigma_{c}$, $y$, and $\hat{y}$. $E^{area}$ was important for determining the quality of uncertainties in future sections and was used to produce $ID$/$OD$ labels for training $M^{dom}$ (see the Defining Ground Truths section).

\begin{equation}\label{misscalibration_area}
    E^{area}=\int_{-\infty}^{\infty}|O(z)-\Phi(0,1)|dz
\end{equation}

$E^{area}$ is a comparison of two statistical quantities and is inaccurate for small sample sizes. However, the comparison is still possible. If the sample size of $O(z)$=$\Phi(0,1)$ is 1, then the cumulative distribution will be 0 until the single observation and then become 1. $E^{area}$ will be non-zero in the aforementioned case, but does not represent the population of $O(z)$ well. In other words, even a single value sampled from a standard normal distribution will give a non-zero $E^{area}$. To prevent issues with small samples sizes, we performed multiple splits and binned data as outlined in Model Assessments section to get many points for statistical comparisons.

\subsubsection*{Model for Dissimilarity Measures ($M^{dis}$)}
\label{kde_method}

KDE is a method to approximate a density of points given finite sampling of points from that density. If KDE is fit to a density distribution normalized to an area of one over a region, then KDE can be used to estimate the probability density (often called a likelihood) of observing points from the distribution at a point in that region. KDE works by first placing a local distribution (kernel) with a thickness (bandwidth) on each observed data point. Overlapping regions of kernels are superimposed, providing an estimate of an overall density and can be generalized to any number of dimensions \supercite{multivariate_kde}. In our approach, we developed a KDE of $X_{ITB}$ whose values approximately give the likelihood of finding a data point at any coordinate in feature space.

To implement KDE, we took the following steps. First, $X_{ITB}$ was standardized using scikit-learn's StandardScaler fit (i.e., data were rescaled to have a mean of zero and a standard deviation of one for each feature) \supercite{scikit-learn}. This was done because the KDE implementation in scikit-learn uses a single bandwidth parameter, so all dimensions of $X_{ITB}$ should be set to a similar length scale. Then, that single bandwidth parameter was used with the Epanechnikov kernel to construct the KDE of $X_{ITB}$ data of $M^{prop}$. The bandwidth was estimated automatically using a nearest-neighbors algorithm as implemented in scikit-learn through the sklearn.cluster.estimate bandwidth method \supercite{scikit-learn}. One could potentially obtain better results by more carefully optimizing the bandwidth for each data set and model combination studied. However, we opted for a straightforward automated approach to avoid any possible data leakage into our assessment and to keep the method simple to understand and implement. We used the Epanechnikov kernel because it is widely  used and is a bounded kernel with a value of zero for any observation outside the bandwidth. This means that the likelihood of far away data will be exactly zero. Other kernels like the Gaussian and exponential kernels are unbounded and have non-zero values for any point. More details for these hyperparameter choices are covered in the Supplemental Materials.

Scikit-learn uses the natural logarithm of likelihoods as the inference outputs of KDE. We converted any logarithmic output to likelihoods by exponentiating the values. Eq.~\ref{kde_trans} was then employed to transform the likelihood of any inference data point $\vec{x}$ with respect to the maximum likelihood observed from $X_{ITB}$. The result obtained from Eq.~\ref{kde_trans} served as our dissimilarity measure, $d$. It is important to note that $\vec{x}$ was transformed using the same scaler utilized to build the KDE previously mentioned prior to attaining its likelihood with $KDE(\vec{x})$.

\begin{equation}\label{kde_trans}
    d=1-\dfrac{KDE(\vec{x})}{\max\limits_{\forall \vec{a} \in X_{ITB}}({KDE(\vec{a})})}
\end{equation}

This transformation did not alter the information provided by the KDE and was performed only to produce an easy to interpret number, which ranges from 0 to 1. A $d$ value of 0 corresponds to $\vec{x}$ being in the region of feature space most densely sampled by $X_{ITB}$ (i.e., the peak of the KDE) and a $d$ value of 1 corresponds to $\vec{x}$ being far from $X_{ITB}$, where the density is zero for our choice of kernel.

\subsubsection*{Model for Domain ($M^{dom}$)}
\label{domain_method}

$M^{dom}$ is a classifier which relies only on $d$ as an input and produces $\widehat{ID}$/$\widehat{OD}$ given a cutoff, $d^{t}$. Values $d$ $<$ $d^{t}$ result in the label $\widehat{ID}$ and values $d$ $\geq$ $d^{t}$ result in the label $\widehat{OD}$ (Eq.~\ref{eq_mdom}). $d^{t}$ is a value chosen from interval $[0,1]$. Each $d^{t}$ is associated with a specific set of $\widehat{ID}$/$\widehat{OD}$ predictions. The effectiveness for a given $d^{t}$ is assessed by precision and recall given a ground truth, which we define based on chemistry, $E^{|y-\hat{y}|/MAD_{y}}$, $E^{RMSE/\sigma_{y}}$, or $E^{area}$ (see the Defining Ground Truths section for ground truth labeling).

\begin{equation}\label{eq_mdom}
    \text{Predicted Label} = 
    \begin{cases}
    \widehat{ID}, & \text{if }d < d^{t} \\
    \widehat{OD}, & \text{otherwise}
    \end{cases}
\end{equation}

Here, we explain the procedure for training $M^{dom}$, which amounts to determining a single value called $d^{t}_{c}$. The superscript in $d^{t}_{c}$ does not represent exponentiation, instead it denotes a specific value of $d$ used in predictions by $M^{dom}$. Note that $M^{dom}$ can be trained with data from $E^{|y-\hat{y}|/MAD_{y}}$, $E^{RMSE/\sigma_{y}}$, $E^{area}$, or chemical information, and each will produce a different $M^{dom}$. Assume we have a data set split into ITB and OOB sets. We generated labels on OOB data for training $M^{dom}$ by following the procedures outlined in the Defining Ground Truths section, which produced the class labels of $ID$ and $OD$. For each $d^{t}$, precision and recall were measured by comparing $ID$/$OD$ and $\widehat{ID}$/$\widehat{OD}$ (Eq.~\ref{pr_equation}). The number of True Positives (TP), True Negatives (TN), False Positives (FP), and False Negatives (FN) were acquired, corresponding to when the (ground truth, prediction) are ($ID$, $\widehat{ID}$), ($OD$, $\widehat{OD}$), ($OD$, $\widehat{ID}$), and ($ID$, $\widehat{OD}$), respectively. Note that TN is not used in Eq.~\ref{pr_equation}, but it is included for completeness.

\begin{equation}\label{pr_equation}
\begin{aligned}
    precision = \dfrac{TP}{TP+FP}, \quad recall = \dfrac{TP}{TP+FN}
\end{aligned}
\end{equation}

$M^{dom}$ learns by selecting $d^{t}_{c}$ such that desirable properties are acquired. $d^{t}_{c}$ can be selected to maximize the harmonic mean between precision and recall ($F1_{max}$) or to maximize recall while retaining a precision above a desired value. We used $F1_{max}$ for selecting $d^{t}_{c}$ in this study unless explicitly stated otherwise. We performed many splits of data into ITB and OOB sets using the approaches described in the Model Assessments section to obtain a large set of OOB data. We then find a single $d^{t}_{c}$ to optimize $F1$ for all OOB data together. If all the OOB is $OD$, then $d^{t}_{c}$ is set to less than zero (we effectively use $-\infty$) so that nothing is predicted $\widehat{ID}$. Conversely, $d^{t}_{c}$ is set to greater than 1 (we effectively use $+\infty$) if all OOB data is $ID$ so that all data are predicted $\widehat{ID}$. Each value of $d$ was acquired from the KDE built from each $X_{ITB}$ for each split.

Deployment of $M^{dom}$ only relies on $\vec{x}$ of a single data point and can produce $\widehat{ID}$/$\widehat{OD}$ for that point by checking the $d$ produced by $M^{dis}$. If $d$ $<$ $d^{t}_{c}$, then the prediction is $\widehat{ID}$ and $\widehat{OD}$ otherwise (Eq.~\ref{eq_mdom} when $d^{t}$=$d^{t}_{c}$). This describes how someone can use $M^{dom}$ for domain prediction after training. Note that this prediction does not involve $E^{|y-\hat{y}|/MAD_{y}}$, $E^{RMSE/\sigma_{y}}$, $E^{area}$, nor chemical intuition after the learning process is complete. The overall precision and recall on splits used to find $d^{t}_{c}$ can be stored and provided at inference along with $\widehat{ID}$/$\widehat{OD}$, giving guidance to the user on the confidence they should put in the domain determination.

\subsubsection*{Why not More Complex $M^{dis}$ and $M^{dom}$?}

In this section, we discuss the question ``Why use this particular approach to get domain when there are clear opportunities for something potentially more accurate?'' The composite function $M^{cdom}=M^{dom}(M^{dis}(\vec{x}))$ takes a feature vector for a given data point and returns a class value of $\widehat{ID}$/$\widehat{OD}$. In our approach, we used $M^{dis}$($\vec{x}$) and a simple cutoff as the learned parameter for $M^{dom}$. One could easily replace this approach with a full ML model for $M^{cdom}$, presumably with much more ability to predict $ID$/$OD$. We did not do this for two reasons. First, we were concerned that it would be difficult to avoid overfitting a complex ML model for $M^{cdom}$ with limited data. The present choice for $M^{cdom}$ has almost no adjustable parameters and a strong foundation in our understanding that ML models learn best where there is more training data, which protects from overfitting. Second, we wanted to have a simple approach that was easy to use and reproduce by other researchers. We expect that more complex versions of $M^{cdom}$ will be explored in the future, and we hope our work can provide a baseline for such studies, as our work has yielded a simple but promising method, as shown by our results.

\subsection*{Defining Ground Truths}
\label{ground_truth}

We highlighted the absence of an absolute, unique definition for data being $ID$ in our Introduction section. As we are proposing $M^{dom}$ to predict if a data point will be $ID$ and wish to assess its effectiveness, we must provide some precise quantitative ground truth definitions for our OOB data points being $ID$/$OD$. We will generally define a ground truth for an observation to be either $ID$ or $OD$ of a model based on privileged information, by which we mean information never seen during the creation of $M^{prop}$, $M^{unc}$, nor $M^{dis}$. Chemical differences, residuals, $RMSE$s, and accuracy of uncertainty estimates are examples of privileged information and are represented in our study by errors denoted as $E^{chem}$, $E^{|y-\hat{y}|/MAD_{y}}$, $E^{RMSE/\sigma_{y}}$, and $E^{area}$. Each of these errors has an associated cutoff to provide a ground truth $ID$/$OD$ labeling. The cutoffs for $E^{chem}$, $E^{|y-\hat{y}|/MAD_{y}}$, $E^{RMSE/\sigma_{y}}$, and $E^{area}$ are denoted $E^{chem}_{c}$, $E^{|y-\hat{y}|/MAD_{y}}_{c}$, $E^{RMSE/\sigma_{y}}_{c}$, and $E^{area}_{c}$, respectively. When $E^{i}<E^{i}_{c}$ $(i \in \{chem, |y-\hat{y}|/MAD_{y}, RMSE/\sigma_{y}, area\})$, the ground truth label is $ID$ and $OD$ otherwise (Eq.~\ref{eq_gt}).

\begin{equation}\label{eq_gt}
    \text{Ground Truth} = 
    \begin{cases}
    \text{$ID$}, & \text{if } E^{i} < E^{i}_{c} \\
    \text{$OD$}, & \text{otherwise}
    \end{cases}
\end{equation}

We outline our rationale behind these sets of privileged information and how we used them to produce our ground truth labels in subsequent sections. Note that there are multiple ways one could use different privileged information and define these ground truths, but we feel these form a logical and broad set of ground truths. The key result of this paper is to show that our dissimilarity measure $d$ (see Eq.~\ref{kde_trans}), determined only from the data features, can predict these ground truth $ID$/$OD$ categorizations and therefore clearly contain essential aspects of privileged information. We demonstrate this ability of $d$ through the assessments outlined in the Model Assessments section.

\subsubsection*{Chemical Intuition}
\label{gt_chem}

Here we define $E^{chem}$ and associated cutoffs. We define $E^{chem}$ as the mismatch between the chemistries used to build $M^{prop}$ and those seen during deployment of $M^{prop}$. $E^{chem}_{c}$ implicitly denotes when chemical mismatch ($E^{chem}$) becomes sufficiently large such that the physics governing materials defined as $ID$ are vastly different from those that are $OD$. Both $E^{chem}$ and $E^{chem}_{c}$ depend on the set of studied materials, their governing physics, and empirical observations. By choosing very similar (e.g., materials selected from ITB data) and very different (e.g., materials with totally different composition, phases, controlling physics, etc.) materials, it is easy to produce data that are mostly $ID$/$OD$. For example, an $M^{prop}$ trained on steel alloys (labeled $ID$) should not be applicable to polymers (labeled $OD$). Note that this approach does not require the value for the property being predicted, $y$. Therefore, we can assess $M^{dom}$ on any data point for which we can write down the feature vector ($\vec{x}$) and have a useful intuition about its chemical similarity to ITB data. We view $M^{dom}$ being successful at delineating $ID$/$OD$ based on simple chemical intuition as a necessary, but not sufficient, condition to use KDE for domain determination. In other words, if $M^{dom}$ struggles to delineate basic intuitive chemical domains, it is likely that the approach is not particularly useful for domain determination. Specifics on chemical groups, which implicitly define $E^{chem}_{c}$, are covered in the Data Curation Section.

\subsubsection*{Normalized Absolute Residuals and $RMSE$}
\label{gt_rmse}

Here we define the cutoffs $E^{|y-\hat{y}|/MAD_{y}}_{c}$ and $E^{RMSE/\sigma_{y}}_{c}$ as the associated error metrics were already defined (see Eqs.~\ref{absres_mad} and \ref{rmse_sigma}). For an $M^{prop}$ whose predictions represent the mean of all $y$ (denoted as $\bar{y}$), both $E^{|y-\hat{y}|/MAD_{y}}$ and $E^{RMSE/\sigma_{y}}$ are 1.0 as calculated from Eqs.~\ref{absres_mad} and \ref{rmse_sigma}. We consider this ``predicting the mean'' to be a baseline na{\"i}ve model with respect to which any reasonable $M^{prop}$ model should perform better. If a fit $M^{prop}$ yields better predictions than a baseline, we considered those data to be $ID$, otherwise, those data were considered $OD$ (Eq.~\ref{eq_gt}). This defines $E^{|y-\hat{y}|/MAD_{y}}_{c}$=$E^{RMSE/\sigma_{y}}_{c}$=1. Using either $E^{|y-\hat{y}|/MAD_{y}}_{c}$ or $E^{RMSE/\sigma_{y}}_{c}$ as the ground truth cutoff for $ID$/$OD$ labeling captures the widely invoked idea that $ID$ cases should be predicted better compared to the $OD$ cases. These definitions require residuals and can therefore be assessed on data with $X$ and $y$ for $M^{prop}$.

\subsubsection*{Errors in Predicted Uncertainties}
\label{gt_area}

Here we define the cutoff $E^{area}_{c}$ as the associated error metric was already defined (see Eq.~\ref{misscalibration_area}). Similar to $E^{|y-\hat{y}|/MAD_{y}}_{c}$ and $E^{RMSE/\sigma_{y}}_{c}$, we set $E^{area}_{c}$ such that $M^{unc}$ will show better performance than a na{\"i}ve baseline case. For this case, we assume a baseline $M^{prop}$ model that provides $\bar{y}$ as the prediction for all cases and a baseline $M^{unc}$ that provides the standard deviation of the training target values, $\sigma_{y}$, as the uncertainty for all cases. These baseline predictions can be used in Eq.~\ref{misscalibration_area} to measure $E^{area}$ for any given set of data, and we take this predicted baseline $E^{area}$ to be $E^{area}_{c}$. We defined $ID$ cases as those for which the $E^{area}$ $<$ $E^{area}_{c}$ ($OD$ otherwise). Unlike both $E^{|y-\hat{y}|/MAD_{y}}_{c}$ and $E^{RMSE/\sigma_{y}}_{c}$, $E^{area}_{c}$ changes depending on the evaluated set of $y$ and ranged from approximately 0.2 to 0.3 in our study. In other words, Eq.~\ref{misscalibration_area} does not yield a single number for naïve models.

It is worth relating that this criterion of $E^{area}$ < $E^{area}_{c}$ for a set of data points tells us that $M^{unc}$ is more accurate than the na{\"i}ve baseline, but not that $M^{prop}$ has small residuals. If $M^{prop}$ has high $E^{RMSE/\sigma_{y}}$ on a set of data points but $M^{unc}$ has accurate estimates of those residuals, then $M^{unc}$ clearly knows significant and useful information about the data, and the data is therefore in some sense $ID$. This situation can be contrasted with the case where $M^{prop}$ has high residuals and $M^{unc}$ has very inaccurate estimates of those residuals, in which case $M^{prop}$ and $M^{unc}$ appear to know nothing about the data, and the data is therefore reasonably considered $OD$. The extent a quality $ID$/$OD$ ground truth definition can be defined by just $E^{area}$, whether $E^{area}$ needs to be combined with the other errors considered, or even whether $E^{area}$ should be excluded entirely from consideration is not established at this point. Regardless, we show that $M^{dom}$ provides an excellent ability to predict an $ID$/$OD$ ground truth label based on $E^{area}$.

\subsubsection*{Summary of Ground Truth Definitions}

The above detailed definitions of $ID$/$OD$ can be confusing, so we summarize the idea again here. We considered four intuitive ways of thinking about being $ID$/$OD$ of a trained model based on intuitive chemical similarity to ITB data, model residuals, model $RMSE$s, and model uncertainty estimates. For each way of thinking, we defined a score of the closeness of the data to being $ID$, which we denoted $E^{chem}$, $E^{|y-\hat{y}|/MAD_{y}}$, $E^{RMSE/\sigma_{y}}$, and $E^{area}$, respectively. We then used chemical intuition (for chemical dissimilarity) or na{\"i}ve baseline model behavior (for model residuals and model uncertainty estimates) to define a cutoff (denoted $E^{i}_{c}$) for each corresponding $E^{i}$ $(i \in \{chem, |y-\hat{y}|/MAD_y, RMSE/\sigma_{y}, area\})$ such that it was reasonable to assume that $E^{i} < E^{i}_{c}$ meant data were $ID$ (Eq.~\ref{eq_gt}). This allowed us to assign the labels of $ID$/$OD$ for data so that we could evaluate the utility of $d$. It is important to realize that these ground truth definitions make extensive use of privileged information (i.e., information not expected to be available during application of the model). However, these ground truth definitions make no direct use of $d$ and $d$ makes no use of the privileged information. The value of our results is that we gain access to $ID$/$OD$ measures that require privileged information by using $d$.

We note that while $E^{chem}$ relies on a researcher's intuition regarding the chemical field and is specific to given fields, definitions relying on $E^{|y-\hat{y}|/MAD_{y}}$, $E^{RMSE/\sigma_{y}}$, and $E^{area}$ are automated numerical approaches and quite general. Therefore, the success of $M^{dom}$ on methods relying on $E^{|y-\hat{y}|/MAD_{y}}$, $E^{RMSE/\sigma_{y}}$, and $E^{area}$ suggests our approach is not limited to materials and chemistry problems, but that in general $d$ gives valuable access to the domain implications of knowledge of residuals. Note that the way we define $E^{RMSE/\sigma_{y}}$ and $E^{area}$ for our automated numerical ground truths require averaging over groupings based on $d$ and is covered in the Methods section. Thus, there is a very modest and indirect path by which the nature of $d$ can influence the ground truth categorization based on $E^{RMSE/\sigma_{y}}$ and $E^{area}$. Although we think this effect is modest, it represents a path of data leakage that could potentially lead to bias in the assessments of the approach. To counteract this concern, we have included $E^{|y-\hat{y}|/MAD_{y}}$, which does not require any binning and is immune to data leakage. This quantity has other limitations as it is quite stochastic, but our success on assessments using the ground truth categorization from $E^{chem}_{c}$ and $E^{|y-\hat{y}|/MAD_{y}}_{c}$ demonstrates that the other successes were not dominantly influenced by data leakage from binning on $d$. Each error and ground truth definition have a corresponding assessment and the details are covered in the Methods section. Assessments on each type of error follow the convention of $A^{i}$ $(i \in \{chem, |y-\hat{y}|/MAD_y, RMSE/\sigma_{y}, area\})$.

\subsection*{Aggregate of Assessment Results}

Now that the conceptual tools and methodology used in this study are established, we can apply those concepts to various data sets and models. We start with $A^{chem}$ and show that $d$ mostly separated $OD$ materials from $ID$ materials. Although the separation is not perfect in many cases, it provided a clear way to flag $OD$ materials. We then cover more automated numerical methods which do not require a priori knowledge of domains from chemical intuition. We show that $d$ can be used to separate cases with high $E^{|y-\hat{y}|/MAD_{y}}$ (i.e., data that are $OD$) and cases with lower $E^{|y-\hat{y}|/MAD_{y}}$ (i.e., data that are $ID$). A similar observation was made with the separation of high and lower $E^{RMSE/\sigma_{y}}$ and $E^{area}$ groups of data, respectively. We end by providing time scaling and notes of caution for use cases where developed methods may not yield desired behavior. All results can be seen in Table~\ref{table_results}.

If the precision is 1 and the recall is low in Table~\ref{table_results}, only some of the data are predicted as $ID$ but many are mislabeled as $OD$. This means that we discard many reliable predictions, but the remaining $ID$ predictions are reliable. If the recall is 1 and the precision is low, then nearly all data are predicted as $ID$. This is indeed a problem and the chosen $d_{c}^{t}$ should be changed such that a desired precision is reached as discussed in the Model for Domain ($M^{dom}$) section. Although $F1_{max}$ is a standard metric in classification, it has its issues. In addition to $F1_{max}$, we ask readers to note AUC-Baseline (defined in the Model Assessments section) because it considers multiple values of $d$, its corresponding precision and recall, and how much above a baseline the measure of $d$ provides. AUC-Baseline is covered in the Model Assessments section.

{
\setlength\LTleft{-0.5in}
\setlength\LTright{-0.5in}
\begin{longtable}{llllllllll}
    \caption{Classification metrics are tabulated for $A^{chem}$, $A^{|y-\hat{y}|/MAD_{y}}$, $A^{RMSE/\sigma_{y}}$, and $A^{area}$. The $d^{t}_{c}$, precision, recall, and $F1$ are for $F1_{max}$. $A^{chem}$ entries do not require $M^{prop}$ and are left empty.} \label{table_results} \\
    \hline
    Data & $M^{prop}$ & Assessment &  Baseline &  AUC &  AUC-Baseline &  $d^{t}_{c}$ &  Precision &  Recall &  $F1$ \\
    \hline
         Diffusion &            &                $A^{chem}$ &      0.50 & 0.68 &          0.18 &      1.00 &       0.64 &    0.80 &  0.71 \\
    Steel Strength &            &                $A^{chem}$ &      0.50 & 0.99 &          0.49 &      1.00 &       1.00 &    0.98 &  0.99 \\
           Fluence &            &                $A^{chem}$ &      0.50 & 1.00 &          0.50 &      0.99 &       1.00 &    1.00 &  1.00 \\
          Cuprates &            &                $A^{chem}$ &      0.50 & 0.97 &          0.47 &      0.99 &       0.94 &    0.98 &  0.96 \\
        Iron-Based &            &                $A^{chem}$ &      0.50 & 0.84 &          0.34 &      0.97 &       0.58 &    0.98 &  0.73 \\
       Low-$T_{c}$ &            &                $A^{chem}$ &      0.50 & 0.96 &          0.46 &      0.80 &       0.98 &    0.80 &  0.88 \\
         Diffusion &        BNN & $A^{|y-\hat{y}|/MAD_{y}}$ &      0.53 & 0.90 &          0.37 &      1.00 &       0.85 &    0.86 &  0.86 \\
         Diffusion &       BOLS & $A^{|y-\hat{y}|/MAD_{y}}$ &      0.59 & 0.93 &          0.34 &      0.99 &       0.95 &    0.83 &  0.88 \\
         Diffusion &       BSVR & $A^{|y-\hat{y}|/MAD_{y}}$ &      0.48 & 0.96 &          0.48 &      1.00 &       0.88 &    0.98 &  0.93 \\
         Diffusion &         RF & $A^{|y-\hat{y}|/MAD_{y}}$ &      0.57 & 0.95 &          0.38 &      1.00 &       0.96 &    0.89 &  0.92 \\
           Fluence &        BNN & $A^{|y-\hat{y}|/MAD_{y}}$ &      0.65 & 0.92 &          0.26 &      1.00 &       0.88 &    0.91 &  0.89 \\
           Fluence &       BOLS & $A^{|y-\hat{y}|/MAD_{y}}$ &      0.71 & 0.86 &          0.15 &      1.00 &       0.71 &    1.00 &  0.83 \\
           Fluence &       BSVR & $A^{|y-\hat{y}|/MAD_{y}}$ &      0.63 & 0.86 &          0.23 &      1.00 &       0.80 &    0.82 &  0.81 \\
           Fluence &         RF & $A^{|y-\hat{y}|/MAD_{y}}$ &      0.74 & 0.90 &          0.16 &      1.00 &       0.74 &    1.00 &  0.85 \\
          Friedman &        BNN & $A^{|y-\hat{y}|/MAD_{y}}$ &      0.71 & 0.97 &          0.27 &      1.00 &       0.98 &    0.89 &  0.93 \\
          Friedman &       BOLS & $A^{|y-\hat{y}|/MAD_{y}}$ &      0.75 & 0.96 &          0.21 &      1.00 &       0.94 &    0.89 &  0.91 \\
          Friedman &       BSVR & $A^{|y-\hat{y}|/MAD_{y}}$ &      0.64 & 0.98 &          0.34 &      1.00 &       0.99 &    0.91 &  0.95 \\
          Friedman &         RF & $A^{|y-\hat{y}|/MAD_{y}}$ &      0.73 & 0.95 &          0.22 &      1.00 &       0.93 &    0.86 &  0.89 \\
    Steel Strength &        BNN & $A^{|y-\hat{y}|/MAD_{y}}$ &      0.49 & 0.68 &          0.19 &      1.00 &       0.64 &    0.80 &  0.71 \\
    Steel Strength &       BOLS & $A^{|y-\hat{y}|/MAD_{y}}$ &      0.45 & 0.69 &          0.25 &      1.00 &       0.62 &    0.87 &  0.72 \\
    Steel Strength &       BSVR & $A^{|y-\hat{y}|/MAD_{y}}$ &      0.49 & 0.62 &          0.13 &      1.00 &       0.49 &    1.00 &  0.66 \\
    Steel Strength &         RF & $A^{|y-\hat{y}|/MAD_{y}}$ &      0.64 & 0.83 &          0.19 &      1.00 &       0.64 &    1.00 &  0.78 \\
    Superconductor &        BNN & $A^{|y-\hat{y}|/MAD_{y}}$ &      0.57 & 0.74 &          0.18 &      1.00 &       0.57 &    1.00 &  0.72 \\
    Superconductor &       BOLS & $A^{|y-\hat{y}|/MAD_{y}}$ &      0.43 & 0.61 &          0.19 &      0.99 &       0.61 &    0.68 &  0.65 \\
    Superconductor &       BSVR & $A^{|y-\hat{y}|/MAD_{y}}$ &      0.62 & 0.70 &          0.08 &      1.00 &       0.62 &    1.00 &  0.77 \\
    Superconductor &         RF & $A^{|y-\hat{y}|/MAD_{y}}$ &      0.62 & 0.79 &          0.17 &      1.00 &       0.62 &    1.00 &  0.77 \\
         Diffusion &        BNN &     $A^{RMSE/\sigma_{y}}$ &      0.47 & 1.00 &          0.53 &      0.96 &       1.00 &    1.00 &  1.00 \\
         Diffusion &       BOLS &     $A^{RMSE/\sigma_{y}}$ &      0.53 & 1.00 &          0.47 &      1.00 &       1.00 &    1.00 &  1.00 \\
         Diffusion &       BSVR &     $A^{RMSE/\sigma_{y}}$ &      0.53 & 1.00 &          0.47 &      1.00 &       1.00 &    1.00 &  1.00 \\
         Diffusion &         RF &     $A^{RMSE/\sigma_{y}}$ &      0.53 & 1.00 &          0.47 &      1.00 &       1.00 &    1.00 &  1.00 \\
           Fluence &        BNN &     $A^{RMSE/\sigma_{y}}$ &      0.64 & 1.00 &          0.36 &      1.00 &       1.00 &    1.00 &  1.00 \\
           Fluence &       BOLS &     $A^{RMSE/\sigma_{y}}$ &      0.57 & 0.99 &          0.41 &      1.00 &       0.89 &    1.00 &  0.94 \\
           Fluence &       BSVR &     $A^{RMSE/\sigma_{y}}$ &      0.43 & 1.00 &          0.57 &      0.91 &       1.00 &    1.00 &  1.00 \\
           Fluence &         RF &     $A^{RMSE/\sigma_{y}}$ &      0.64 & 1.00 &          0.36 &      1.00 &       1.00 &    1.00 &  1.00 \\
          Friedman &        BNN &     $A^{RMSE/\sigma_{y}}$ &      0.65 & 1.00 &          0.35 &      1.00 &       1.00 &    1.00 &  1.00 \\
          Friedman &       BOLS &     $A^{RMSE/\sigma_{y}}$ &      0.71 & 1.00 &          0.29 &      1.00 &       1.00 &    1.00 &  1.00 \\
          Friedman &       BSVR &     $A^{RMSE/\sigma_{y}}$ &      0.62 & 1.00 &          0.38 &      1.00 &       1.00 &    1.00 &  1.00 \\
          Friedman &         RF &     $A^{RMSE/\sigma_{y}}$ &      0.67 & 1.00 &          0.33 &      1.00 &       1.00 &    1.00 &  1.00 \\
    Steel Strength &        BNN &     $A^{RMSE/\sigma_{y}}$ &      0.47 & 1.00 &          0.53 &      0.88 &       1.00 &    1.00 &  1.00 \\
    Steel Strength &       BOLS &     $A^{RMSE/\sigma_{y}}$ &      0.36 & 0.96 &          0.61 &      0.78 &       0.83 &    1.00 &  0.91 \\
    Steel Strength &       BSVR &     $A^{RMSE/\sigma_{y}}$ &      0.20 & 1.00 &          0.80 &      0.43 &       1.00 &    1.00 &  1.00 \\
    Steel Strength &         RF &     $A^{RMSE/\sigma_{y}}$ &      0.47 & 1.00 &          0.53 &      0.85 &       1.00 &    1.00 &  1.00 \\
    Superconductor &        BNN &     $A^{RMSE/\sigma_{y}}$ &      0.36 & 1.00 &          0.64 &      0.97 &       1.00 &    1.00 &  1.00 \\
    Superconductor &       BOLS &     $A^{RMSE/\sigma_{y}}$ &      0.36 & 1.00 &          0.64 &      0.97 &       1.00 &    1.00 &  1.00 \\
    Superconductor &       BSVR &     $A^{RMSE/\sigma_{y}}$ &      0.54 & 0.85 &          0.31 &      0.97 &       1.00 &    0.67 &  0.80 \\
    Superconductor &         RF &     $A^{RMSE/\sigma_{y}}$ &      0.45 & 0.78 &          0.34 &      0.93 &       1.00 &    0.61 &  0.76 \\
         Diffusion &        BNN &                $A^{area}$ &      0.06 & 0.31 &          0.25 &      0.31 &       0.50 &    1.00 &  0.67 \\
         Diffusion &       BOLS &                $A^{area}$ &      0.53 & 0.50 &         -0.03 &      1.00 &       0.53 &    1.00 &  0.69 \\
         Diffusion &       BSVR &                $A^{area}$ &      0.18 & 1.00 &          0.82 &      0.46 &       1.00 &    1.00 &  1.00 \\
         Diffusion &         RF &                $A^{area}$ &      0.18 & 1.00 &          0.82 &      0.46 &       1.00 &    1.00 &  1.00 \\
           Fluence &        BNN &                $A^{area}$ &      0.43 & 1.00 &          0.57 &      0.92 &       1.00 &    1.00 &  1.00 \\
           Fluence &       BOLS &                $A^{area}$ &      0.29 & 0.43 &          0.14 &      0.97 &       0.57 &    1.00 &  0.73 \\
           Fluence &       BSVR &                $A^{area}$ &      0.43 & 1.00 &          0.57 &      0.91 &       1.00 &    1.00 &  1.00 \\
           Fluence &         RF &                $A^{area}$ &      0.43 & 1.00 &          0.57 &      0.92 &       1.00 &    1.00 &  1.00 \\
          Friedman &        BNN &                $A^{area}$ &      0.21 & 0.30 &          0.09 &      0.98 &       0.43 &    1.00 &  0.60 \\
          Friedman &       BOLS &                $A^{area}$ &      0.00 & 0.00 &          0.00 & $-\infty$ &       0.00 &    0.00 &  0.00 \\
          Friedman &       BSVR &                $A^{area}$ &      0.23 & 1.00 &          0.77 &      0.34 &       1.00 &    1.00 &  1.00 \\
          Friedman &         RF &                $A^{area}$ &      0.29 & 1.00 &          0.71 &      0.65 &       1.00 &    1.00 &  1.00 \\
    Steel Strength &        BNN &                $A^{area}$ &      0.13 & 0.13 &         -0.01 &      0.99 &       0.22 &    1.00 &  0.36 \\
    Steel Strength &       BOLS &                $A^{area}$ &      0.14 & 0.29 &          0.15 &      0.78 &       0.33 &    1.00 &  0.50 \\
    Steel Strength &       BSVR &                $A^{area}$ &      0.07 & 0.31 &          0.24 &      0.30 &       0.50 &    1.00 &  0.67 \\
    Steel Strength &         RF &                $A^{area}$ &      0.33 & 1.00 &          0.67 &      0.65 &       1.00 &    1.00 &  1.00 \\
    Superconductor &        BNN &                $A^{area}$ &      0.18 & 0.32 &          0.14 &      0.99 &       0.40 &    1.00 &  0.57 \\
    Superconductor &       BOLS &                $A^{area}$ &      0.27 & 0.90 &          0.63 &      0.97 &       0.75 &    1.00 &  0.86 \\
    Superconductor &       BSVR &                $A^{area}$ &      0.09 & 0.31 &          0.22 &      0.56 &       0.50 &    1.00 &  0.67 \\
    Superconductor &         RF &                $A^{area}$ &      0.27 & 1.00 &          0.73 &      0.93 &       1.00 &    1.00 &  1.00 \\
    \hline
\end{longtable}
}

\begin{figure}[H]
    \centering

        \begin{subfigure}{0.5\textwidth}
            \includegraphics[width=\linewidth]{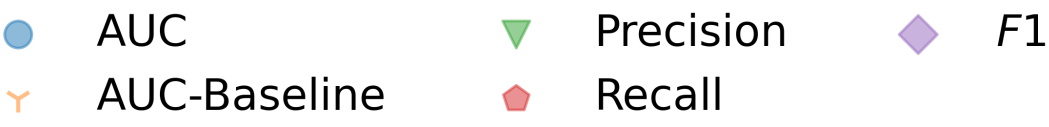}
            \caption*{}
        \end{subfigure}

        \vspace{-0.5cm}

        \begin{subfigure}{0.48\textwidth}
            \includegraphics[width=\linewidth]{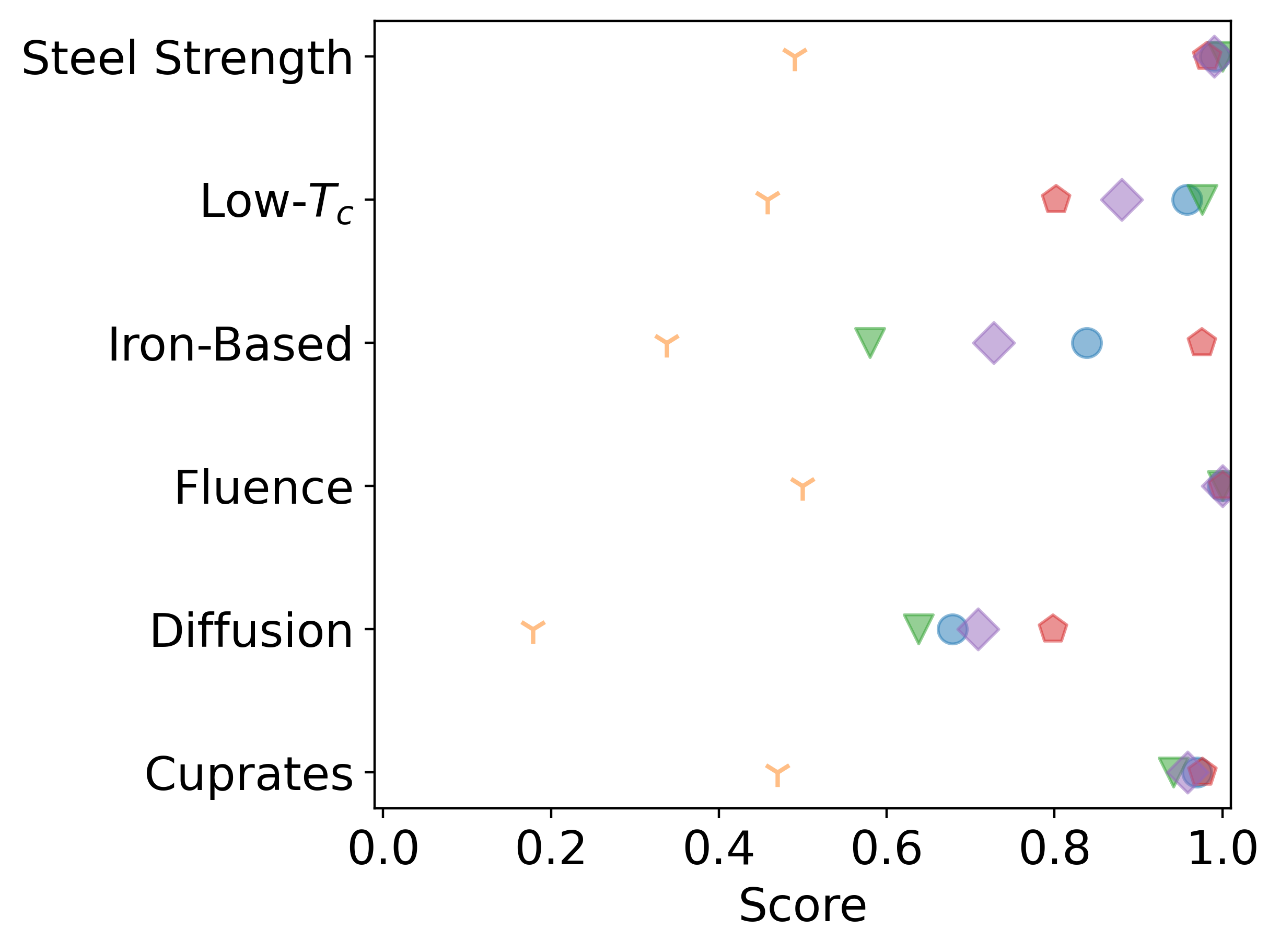}
            \caption{$A^{chem}$}
            \label{Achem}
        \end{subfigure}
        \hfill
        \begin{subfigure}{0.48\textwidth}
            \includegraphics[width=\linewidth]{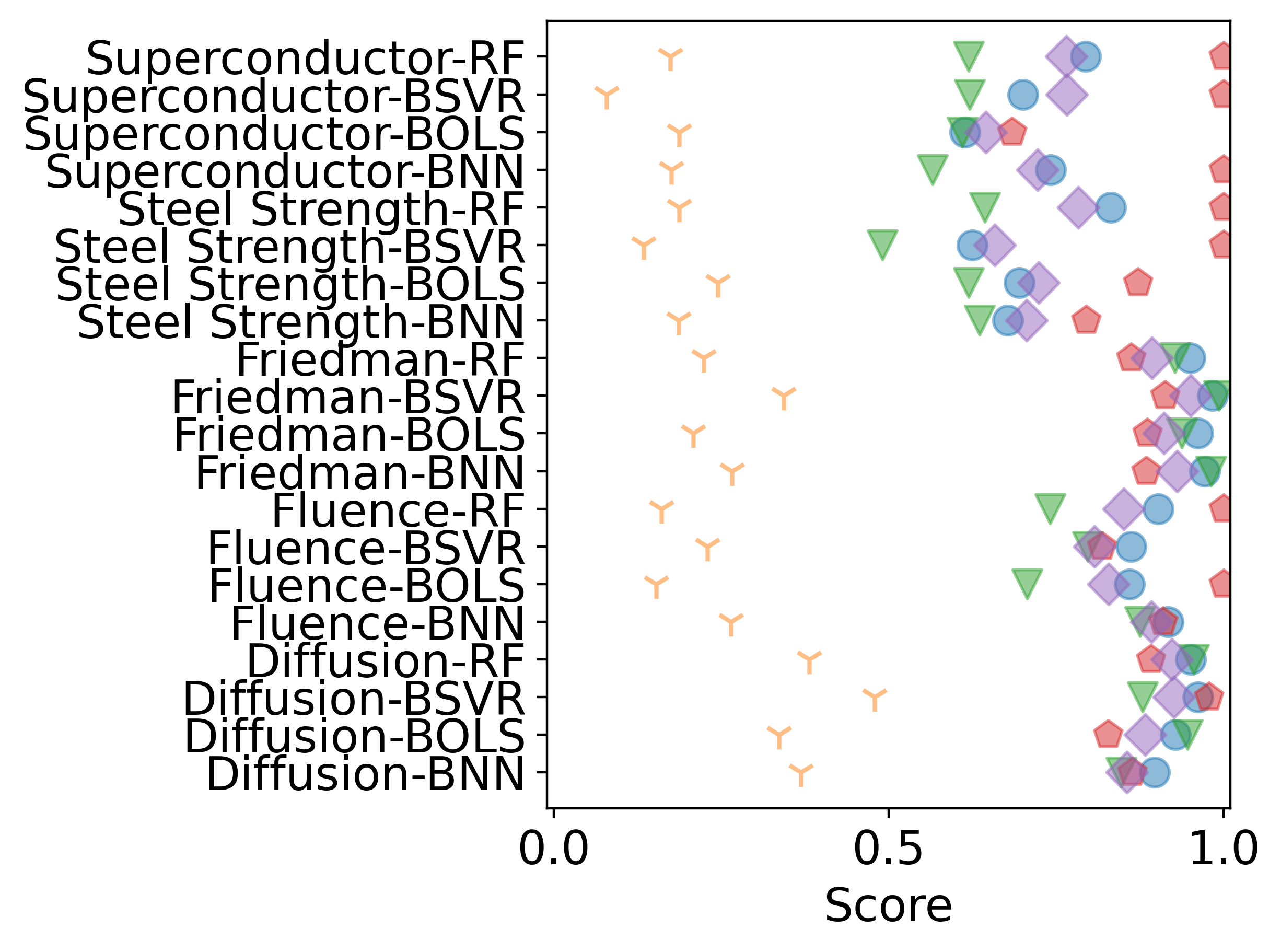}
            \caption{$A^{|y-\hat{y}|/MAD_{y}}$}
            \label{Aabsres}
        \end{subfigure}
    
        \vspace{1cm}
    
        \begin{subfigure}{0.48\textwidth}
            \includegraphics[width=\linewidth]{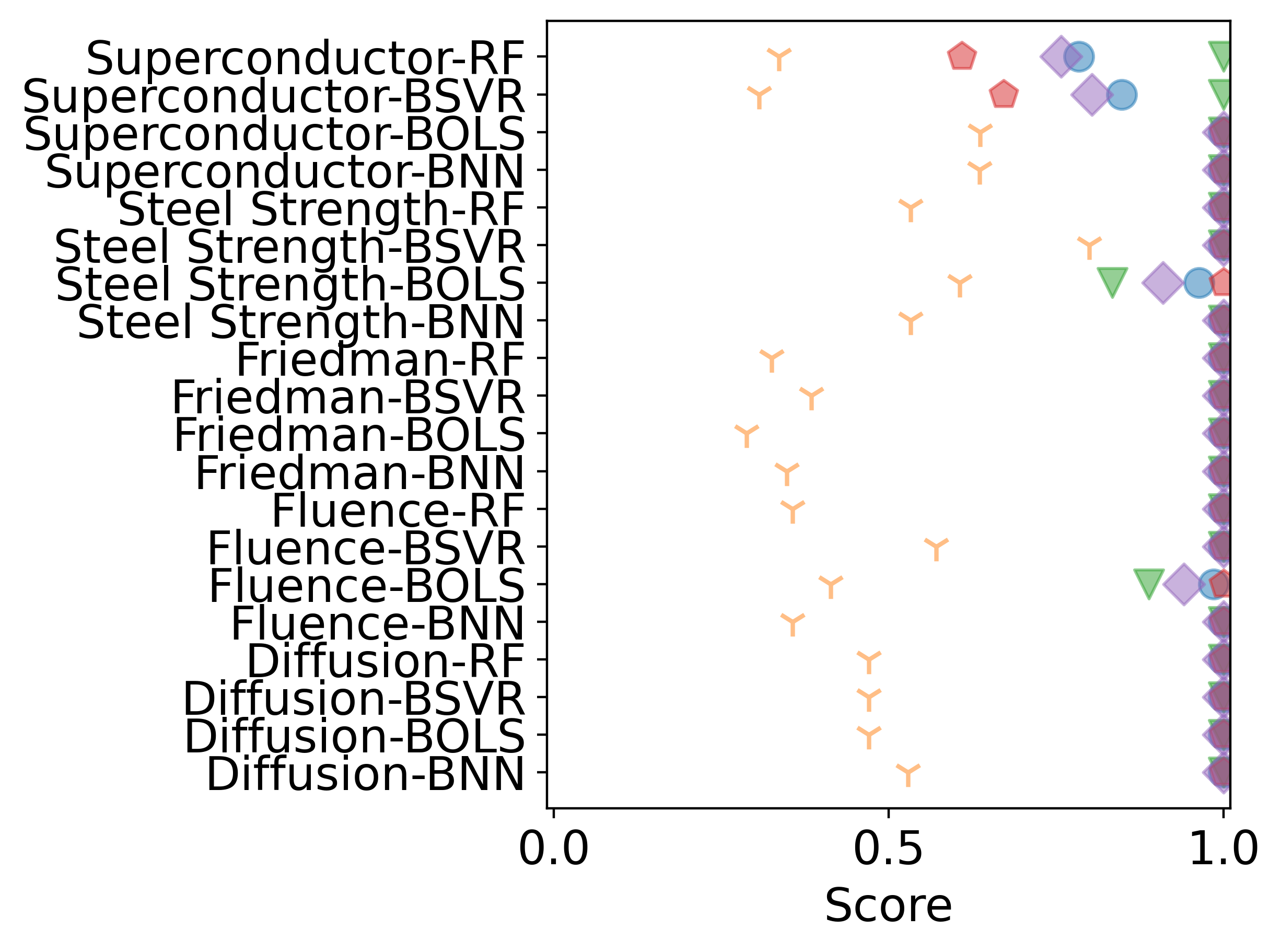}
            \caption{$A^{RMSE/\sigma_{y}}$}
            \label{Armse}
        \end{subfigure}
        \hfill
        \begin{subfigure}{0.48\textwidth}
            \includegraphics[width=\linewidth]{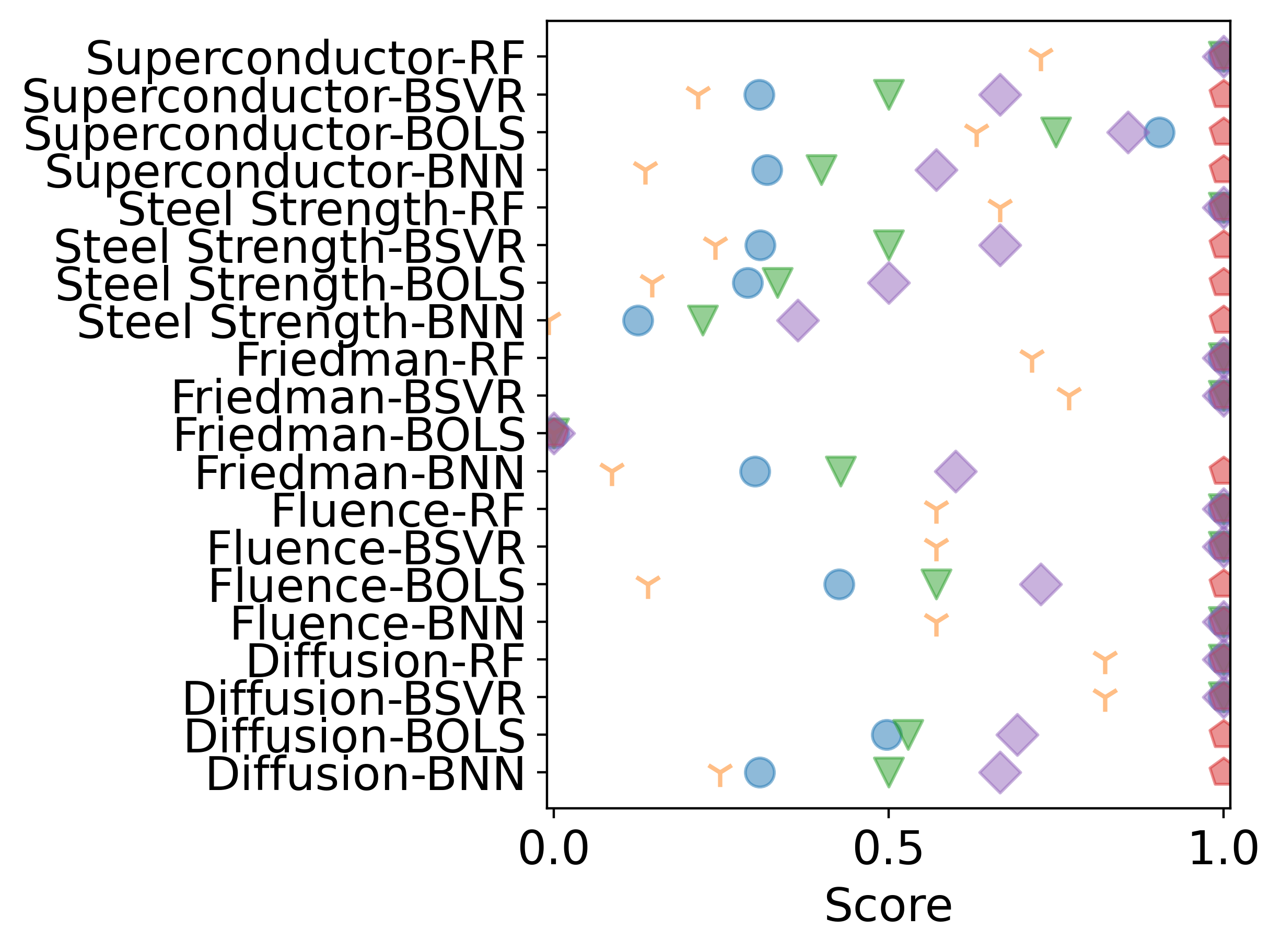}
            \caption{$A^{area}$}
            \label{Aarea}
        \end{subfigure}
    
        \caption{\textbf{Visulaization of scores from Table~\ref{table_results}.} For all points shown in these figures, a score closer to 1 is better, with decreasing values being worse. AUC, AUC-Baseline, precision, recall, and $F1_{max}$ are shown by the blue, orange, green, red, and purple markers, respectively. The type of assessment is noted by the subfigure caption.}
        \label{big_table_visuals}
\end{figure}

Because interpreting the data in Table~\ref{table_results} can be difficult due its size, we have included visualizations of the data in Fig.~\ref{big_table_visuals}. Figs.~\ref{Achem} through \ref{Aarea} plot the AUC, AUC-Baseline, precision, recall, and $F1_{max}$ scores from the table. Values closer to 1 are better with decreasing values signaling worsening scores.

\subsection*{Relationship Between Chemical Dissimilarity ($E^{chem}$) and Distance ($d$) from the Chemical Assessment ($A^{chem}$)}
\label{results_chem}

\begin{figure}[H]
\centering

    \begin{subfigure}{0.42\textwidth}
        \includegraphics[width=\linewidth]{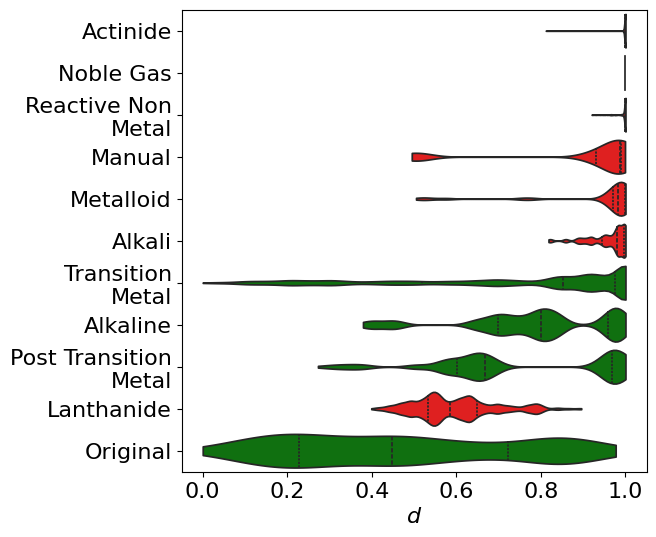}
        \caption{Diffusion}
        \label{diffusion_chemical}
    \end{subfigure}
    \hfill
    \begin{subfigure}{0.42\textwidth}
        \includegraphics[width=\linewidth]{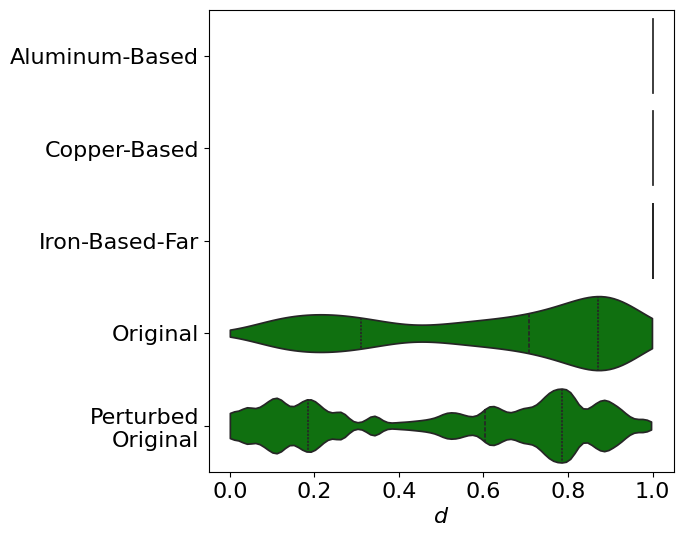}
        \caption{Fluence}
        \label{fluence_chemical}
    \end{subfigure}

    \vspace{1cm}

    \begin{subfigure}{0.42\textwidth}
        \includegraphics[width=\linewidth]{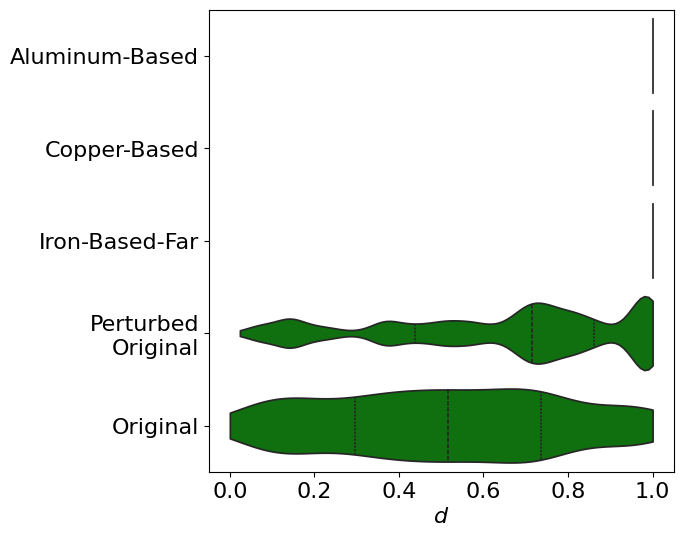}
        \caption{Steel Strength}
        \label{steel_chemical}
    \end{subfigure}
    \hfill
    \begin{subfigure}{0.42\textwidth}
        \includegraphics[width=\linewidth]{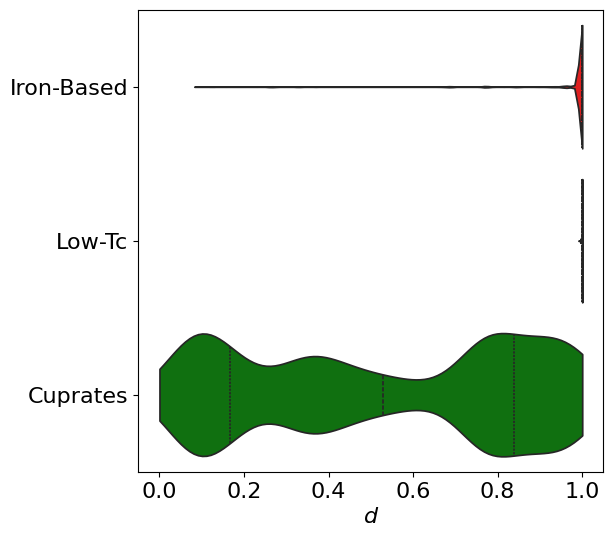}
        \caption{Cuprates}
        \label{cond_cuprates}
    \end{subfigure}

    \vspace{1cm}

    \begin{subfigure}{0.42\textwidth}
        \includegraphics[width=\linewidth]{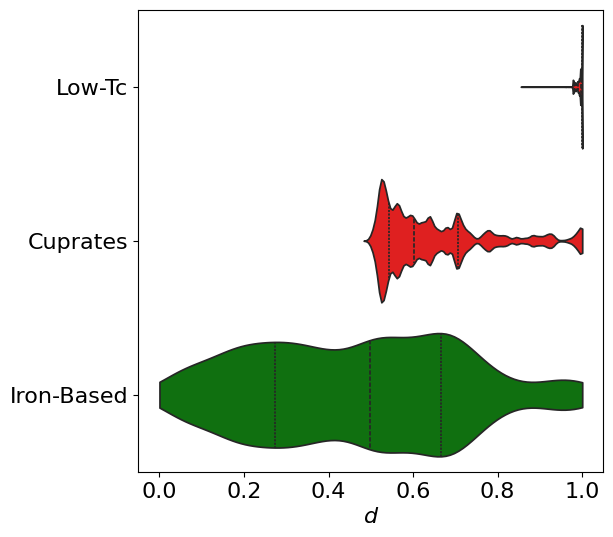}
        \caption{Iron-Based}
        \label{cond_fe}
    \end{subfigure}
    \hfill
    \begin{subfigure}{0.42\textwidth}
        \includegraphics[width=\linewidth]{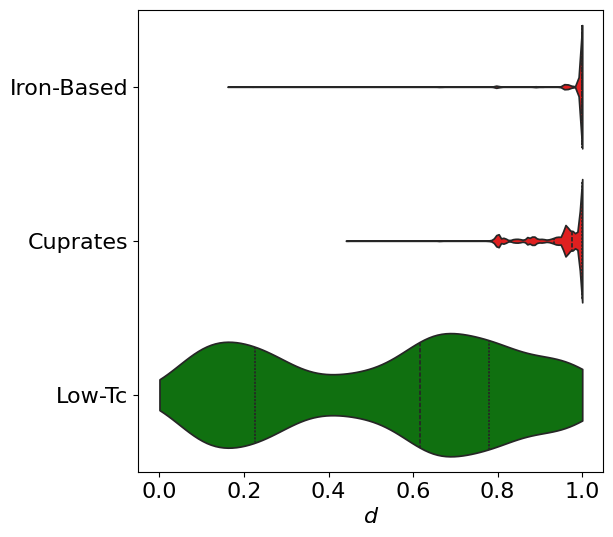}
        \caption{Low-${T_{c}}$}
        \label{cond_low_tc}
    \end{subfigure}

    \caption{\textbf{KDE separates distinct materials.} We show the violin plot for all the $d$ scores separated by chemical groups. The first, second, and third vertical lines within each violin denote the separations between the first, second, third, and fourth quartiles. Values to the left are more likely to be observed compared to values to the right. All violins were forced to have the same width for visual purposes (i.e., the actual number of observations are not reflected by the visual). Green and red violins denote $ID$ and $OD$ groups, respectively. The data set is denoted by the captions.}
    \label{chemical_test_results}
\end{figure}

For all chemical data sets analyzed through $A^{chem}$, violin plots for $d$ with respect to chemical groups were generated and shown in Fig.~\ref{chemical_test_results}. A positive result shows that $ID$ materials have lower values of $d$ compared to $OD$ materials. Values to the left are more likely to be observed (i.e., be $ID$), and values to the right become less likely to be observed (i.e., be $OD$). All sets of $A^{chem}$ generally show the aforementioned trend. Data are organized based on their median $d$ values, with lower values positioned at the bottom and higher values at the top in Fig.~\ref{chemical_test_results}. The classification metrics are tabulated in Table~\ref{table_results}. Our $F1_{max}$ scores range from 0.71 to 1.00, which are rather high. We show how $d$ is useful for flagging $OD$ materials from predictions in subsequent text. The figures for all precision and recall curves are provided in the Supplemental Materials. Details on chemical groupings and their naming are in the Data Curation section.

\pagebreak
\pagebreak

The violin plot for the Diffusion data is shown in Fig.~\ref{diffusion_chemical}. Intuitively dissimilar materials to the Original data subset, such as noble gases, are to the very right of the figure. Materials that were mixed across chemical groups (Manual set) also show high dissimilarities compared to the Original set. Conversely, more similar materials like transition metals, alkaline earths, and post transition metals have their median $d$ value closer to the Original subset compared to the aforementioned groups. Lanthanides, however, appear closer to our Original set than expected. If we want 95\% of data to come from a trustworthy chemical domain while maximizing recall for the Diffusion data, then the corresponding threshold of $d^{t}_{c}$=0.45 can act as a decision boundary. Using this threshold for $M^{dom}$, chemical groups such as noble gases, actinides, reactive non-metals, etc. are mostly excluded from $\widehat{ID}$, but our recall is only 0.20. At a value slightly below 1.00 for $d^{t}_{c}$, our $F1$ score is maximized and can still exclude most noble gases, reactive non-metals, and actinides. At $F1_{max}$, our recall and precision are 0.80 and 0.64 respectively. One can increase the precision of predictions by choosing a lower value of $d^{t}_{c}$ at the cost of recall and vice versa according to the needs of $M^{dom}$ application.

We have a similar observation for our Fluence data (see Fig.~\ref{fluence_chemical}). Note that the Perturbed Original set have their quartiles shifted to the left of the Original data, meaning that values from the Perturbed Original data are predicted by the KDE to be more likely to be observed than test cases from the Original data, which is surprising. This is likely just an anomaly due to the modest size of the data and is not a concern given that both sets of data are $ID$. The important consideration is that all $OD$ data have $d$ values at or near 1.00. All $OD$ data are far away from $ID$ data. Any prediction with $d^{t}_{c}$$\approx$0.99 is deemed to be untrustworthy and separates $ID$/$OD$ essentially perfectly.

As for our Steel Strength data in Fig.~\ref{steel_chemical}, the Perturbed Original data have a median $d$ value further to the right than the Original set, as expected. Most Perturbed Original data from the Steel Strength data are to the left of other $OD$ groups. In other words, $OD$ data are further away than $ID$ data. The Iron-Based-Far data have cases that are closer to the Original data than the Copper-Based and Aluminum-Based sets, although it is not evident from the figure. The observation is reasonable considering most of the Original data are iron based, but generally outside the range of $Fe$ weight percentages and with more constituent elements than in the Iron-Based-Far data. Like our Fluence data, any prediction at or near $d^{t}_{c}$$\approx$1.00 was deemed to be untrustworthy. But unlike Fluence, $ID$/$OD$ labels were not separated perfectly. Our recall is 0.98 because of a few FN predictions and the precision is 1.00.

Regarding the Superconductor data, we divided data into Cuprates, Iron-Based, and Low-${T_{c}}$ groups and assessed the behavior where each set was taken as the Original data (see the Model Assessments section) with the results shown in Figs.~\ref{cond_cuprates}, \ref{cond_fe}, and \ref{cond_low_tc}, respectively. In all cases, the $ID$ materials in the Original data generally have $d$ values to the left of $OD$ data. $F1_{max}$ scores are not lower than 0.7, which is generally good. Note that we can tune $d^{t}_{c}$ (i.e., non-$F1_{max}$ $d^{t}_{c}$) such that the corresponding precision is near or at 1.00 at the cost of a lower recall for these data.

While our method does not flawlessly differentiate all chemical cases, it serves as a robust tool to eliminate a substantial portion of unreliable predictions. Furthermore, our chemical domains, like all the domain definitions in this paper, are approximate, and it is not expected that any method will provide perfect $ID$/$OD$ predictions. Predicting domain with $d^{t}_{c}$ derived from data analysis like those shown in Figs.~\ref{diffusion_chemical} through \ref{cond_low_tc} required being able to label a large set of data as chemically similar ($ID$) or dissimilar ($OD$), which requires extensive domain expertise and is not always feasible. Our other methods for domain determination rely purely on statistics of the trained model for assigning labels for domain, which makes them more practical for deployment. $A^{chem}$ serves as an initial, intuitive, and rudimentary assessment, and passing this assessment was necessary for any domain method to be considered useful for materials. These results show that $d$ effectively provided privileged information regarding the chemical classes and physics that separate materials. We are confident that any positive outcomes observed for $A^{|y-\hat{y}|/MAD_{y}}$, $A^{RMSE/\sigma_{y}}$, or $A^{area}$ are not solely attributable to numerical artifacts. These readily interpretable results from $A^{chem}$ establish a firm, intuitive, and chemically informed foundation before exploring more general numerical approaches, which we do now.

\subsection*{Relationship Between Absolute Residuals ($E^{|y-\hat{y}|/MAD_{y}}$) and Distance ($d$) from the Residual Assessment ($A^{|y-\hat{y}|/MAD_{y}}$)}
\label{results_res}

\begin{figure}[H]
    \centering
    
    \begin{subfigure}{0.42\textwidth}
        \includegraphics[width=\linewidth]{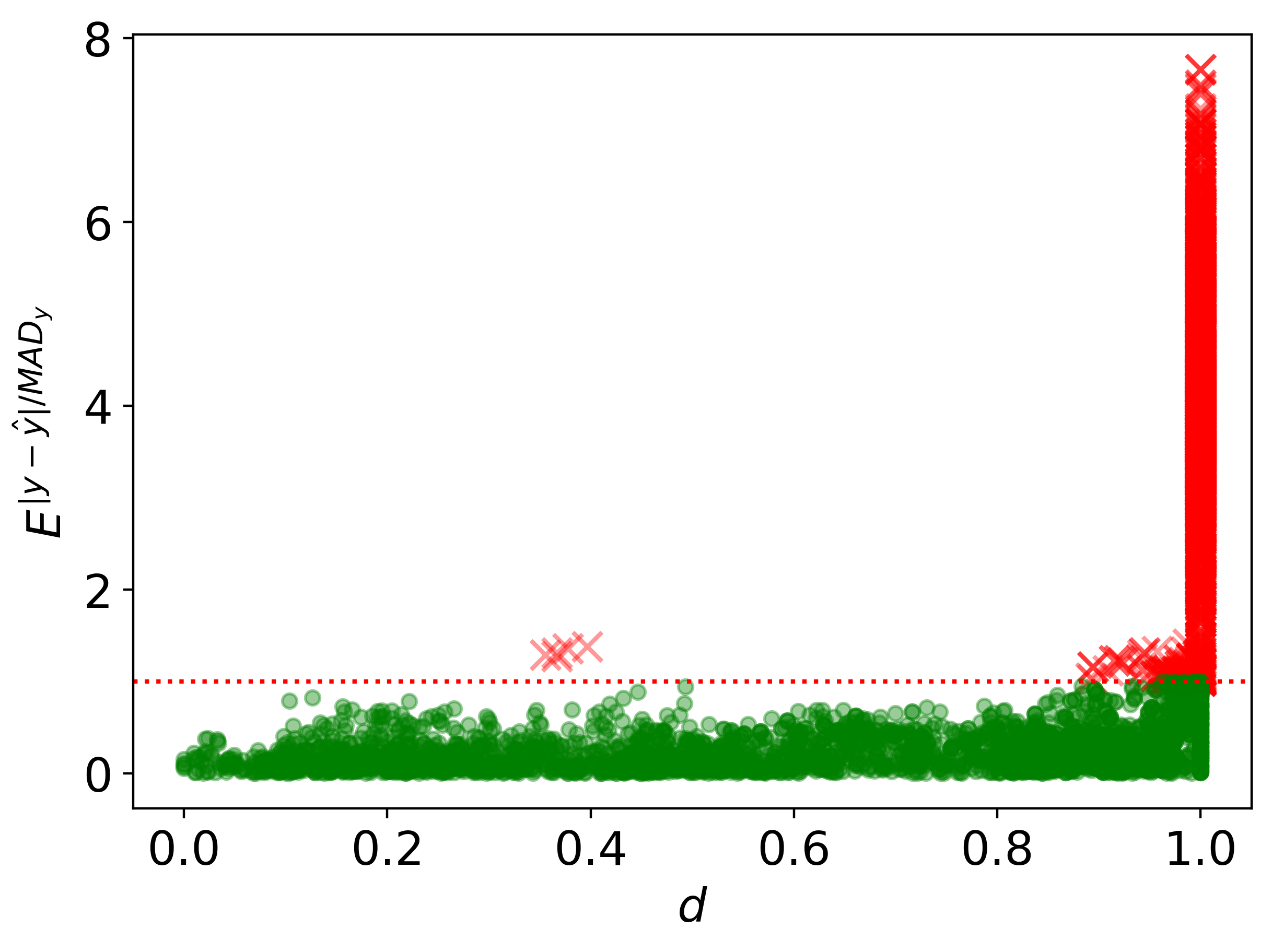}
        \caption{Diffusion}
        \label{diffusion_res}
    \end{subfigure}
    
    \vspace{1cm}
    
    \begin{subfigure}{0.42\textwidth}
        \includegraphics[width=\linewidth]{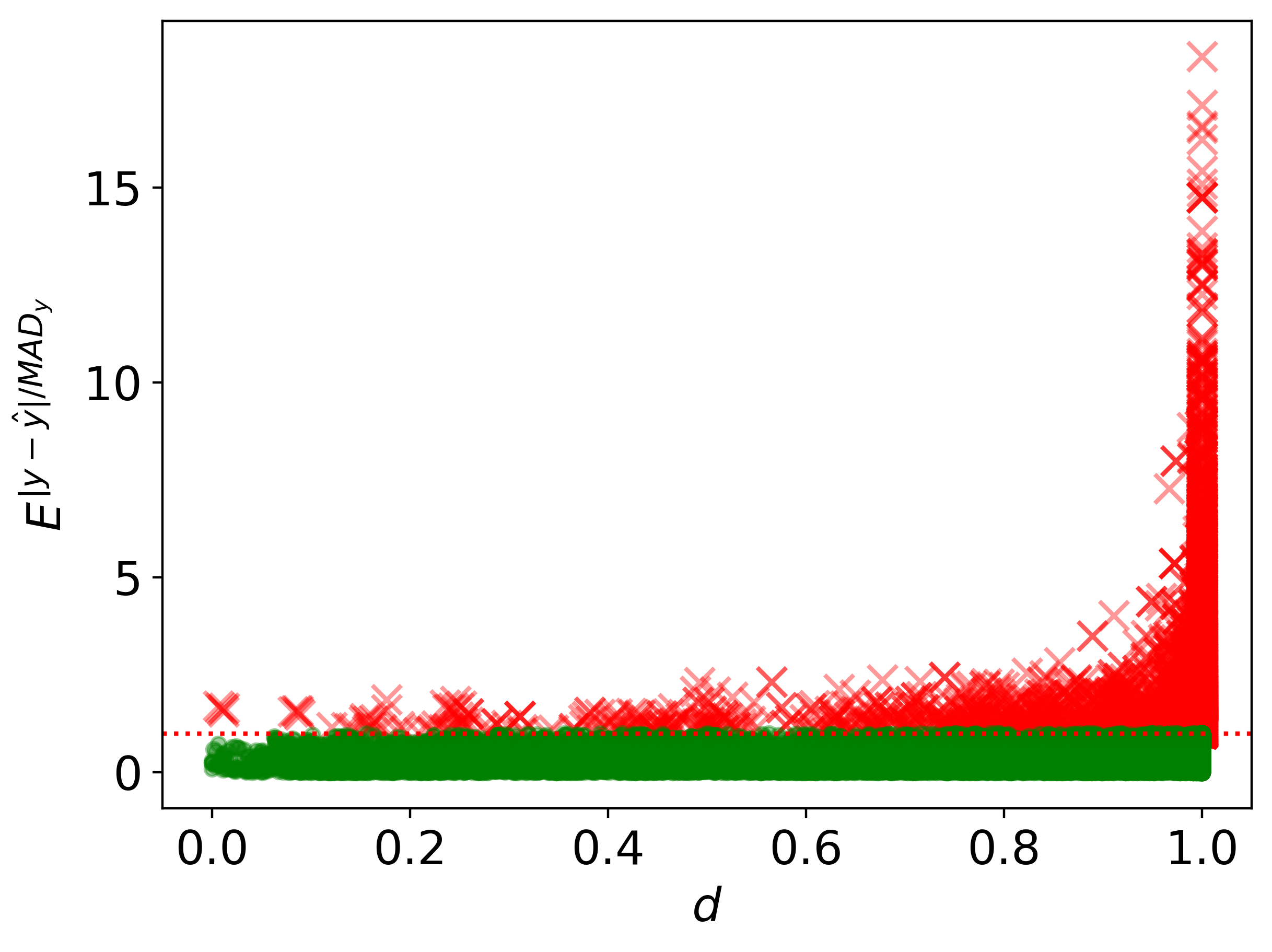}
        \caption{Fluence}
        \label{fluence_res}
    \end{subfigure}
    \hfill
    \begin{subfigure}{0.42\textwidth}
        \includegraphics[width=\linewidth]{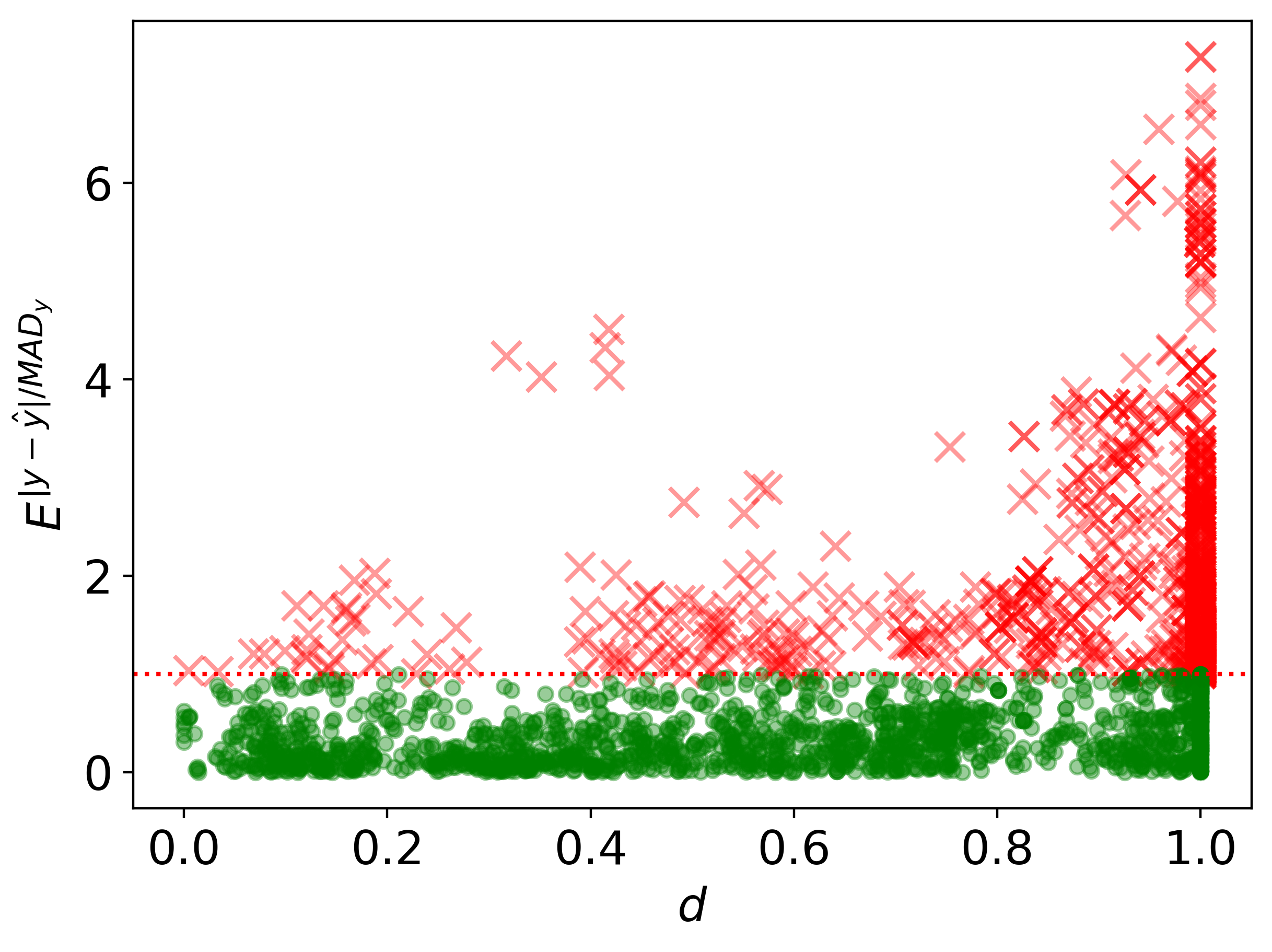}
        \caption{Steel Strength}
        \label{steel_res}
    \end{subfigure}

    \vspace{1cm}

    \begin{subfigure}{0.42\textwidth}
        \includegraphics[width=\linewidth]{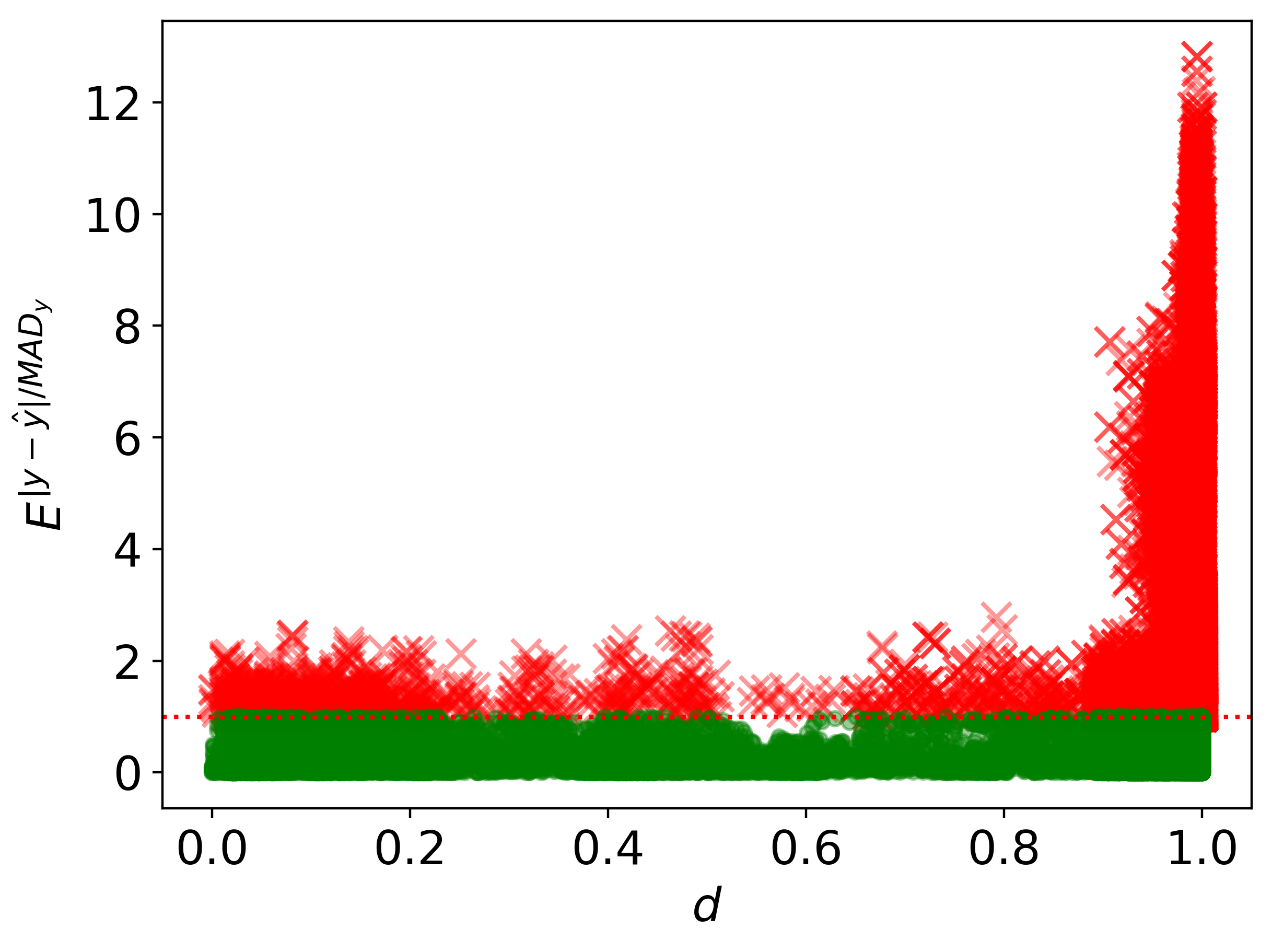}
        \caption{Superconductor}
        \label{cond_res}
    \end{subfigure}
    \hfill
    \begin{subfigure}{0.42\textwidth}
        \includegraphics[width=\linewidth]{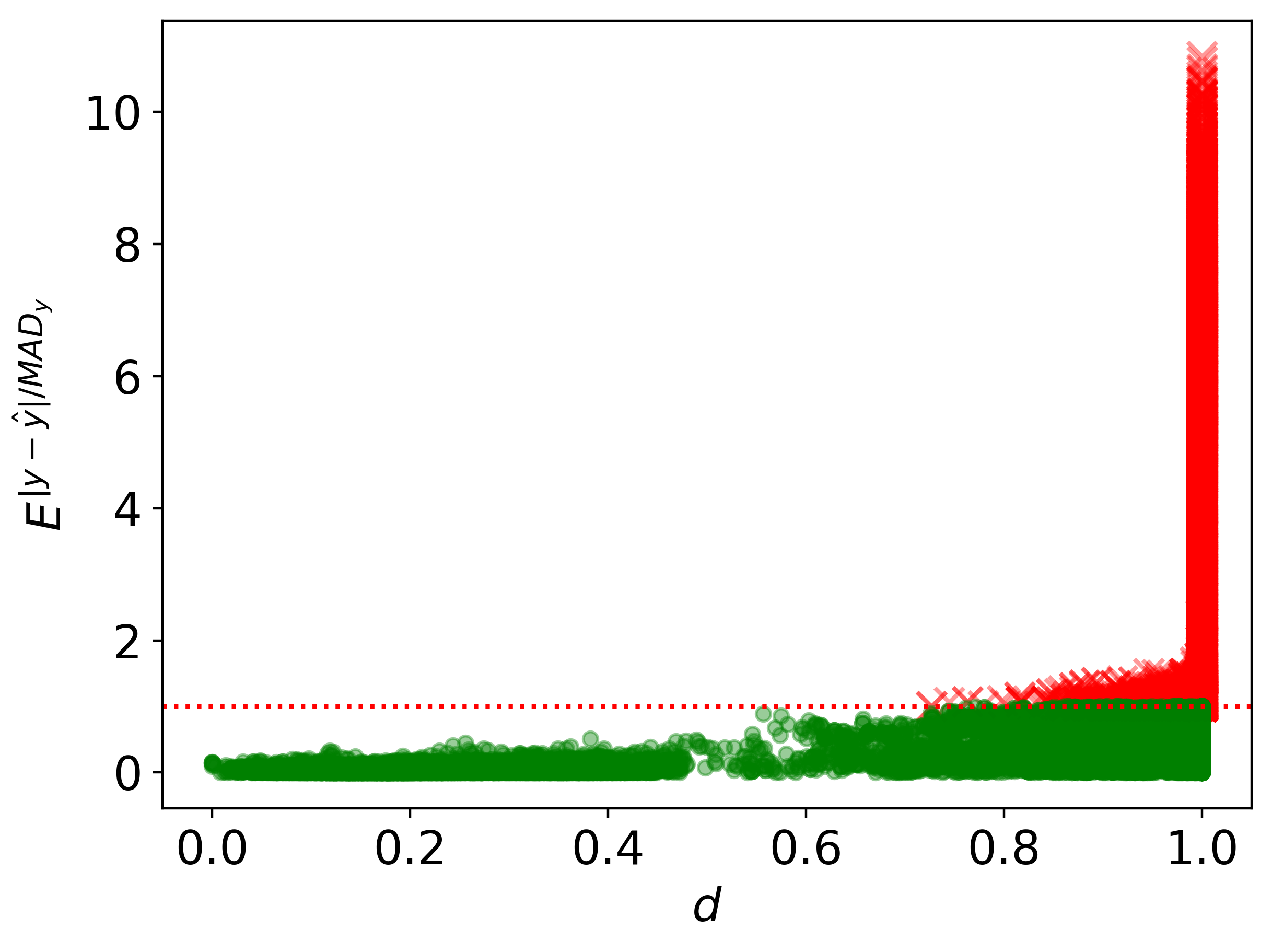}
        \caption{Friedman}
        \label{friedman_res}
    \end{subfigure}

    \caption{\textbf{Absolute residuals grow as OOB data becomes increasingly dissimilar.} The relationship between $E^{|y-\hat{y}|/MAD_{y}}$ and $d$ for the RF model type is shown. Generally, $E^{|y-\hat{y}|/MAD_{y}}$ increases with an increase in $d$. $E^{|y-\hat{y}|/MAD_{y}}_{c}$ is shown by the horizontal red line, which separates our $OD$ (red) and $ID$ (green) cases. The data set is denoted by the captions.}
    \label{res_test_results}
\end{figure}

In Fig.~\ref{res_test_results}, we illustrate how $E^{|y-\hat{y}|/MAD_{y}}$ is related to $d$ for the RF model type. The same set of plots for other model types are offered in the Supplemental Materials, but we summarize their results in Table~\ref{table_results}. We observe that $E^{|y-\hat{y}|/MAD_{y}}$ increases when the likelihood of observing similar points to ITB data decreases (larger $d$). All AUC-Baseline scores in Table~\ref{table_results} are positive for the relevant assessment, meaning that our $d$ measure gives more information than a na{\"i}ve guess provided by our baseline. Nearly all $F1_{max}$ scores are above 0.7, which means that $d$ can be used to significantly separate $ID$ and $OD$ points. Only 2 out of 20 measures of $F1_{max}$ fall below 0.7. As an example of application, we can examine Fig.~\ref{diffusion_res} and select $d^{t}_{c}$=1.00 for which we reject predictions (i.e., label as $\widehat{OD}$) from any $M^{prop}$ according to Eq.~\ref{eq_mdom}. Most points with large $E^{|y-\hat{y}|/MAD_{y}}$ tend to occur at $d$=1.00. By selecting a $d^{t}_{c}$=1.00, a user can filter out the majority of those points from a study, thereby saving time by excluding untrustworthy cases. Other data sets in Figs.~\ref{fluence_res}, \ref{steel_res}, and \ref{cond_res} have more $OD$ cases above $E^{|y-\hat{y}|/MAD_{y}}_{c}$ at lower $d$ compared to the previous example. However, the fraction of $OD$ cases greatly exceeds that of $ID$ cases when $d$=1.00 and the fraction of $ID$ cases is higher than the number of $OD$ cases when $d$<1.00. Therefore, the $F1_{max}$ occurs at $d^{t}_{c}$=1.00, where data have a zero likelihood based on the KDE constructed from ITB data. A similar methodology can be applied for $d^{t}_{c}$ values across other data set and model combinations. Essentially, we can select predictions likely to be better than na{\"i}ve. We note that $E^{|y-\hat{y}|/MAD_{y}}$ is a statistical quantity that will almost certainly be challenging to predict by $d$ or any other domain method as it is quite stochastic. Even if one has a method that was essentially perfect at identifying data that was $ID$/$OD$, some residuals would likely be large for $ID$ data and small for $OD$ data by chance. Given this intrinsic limitation of predicting $E^{|y-\hat{y}|/MAD_{y}}$, we consider the present results quite strong. The statistical quantity $E^{RMSE/\sigma_{y}}$ is essentially the same information as $E^{|y-\hat{y}|/MAD_{y}}$, but averaged over many points and therefore less stochastic, and we will see below that $E^{RMSE/\sigma_{y}}$ is predicted more robustly by $d$.

\pagebreak

\begin{figure}[H]
	\centering

    \begin{subfigure}{0.42\textwidth}
        \includegraphics[width=\linewidth]{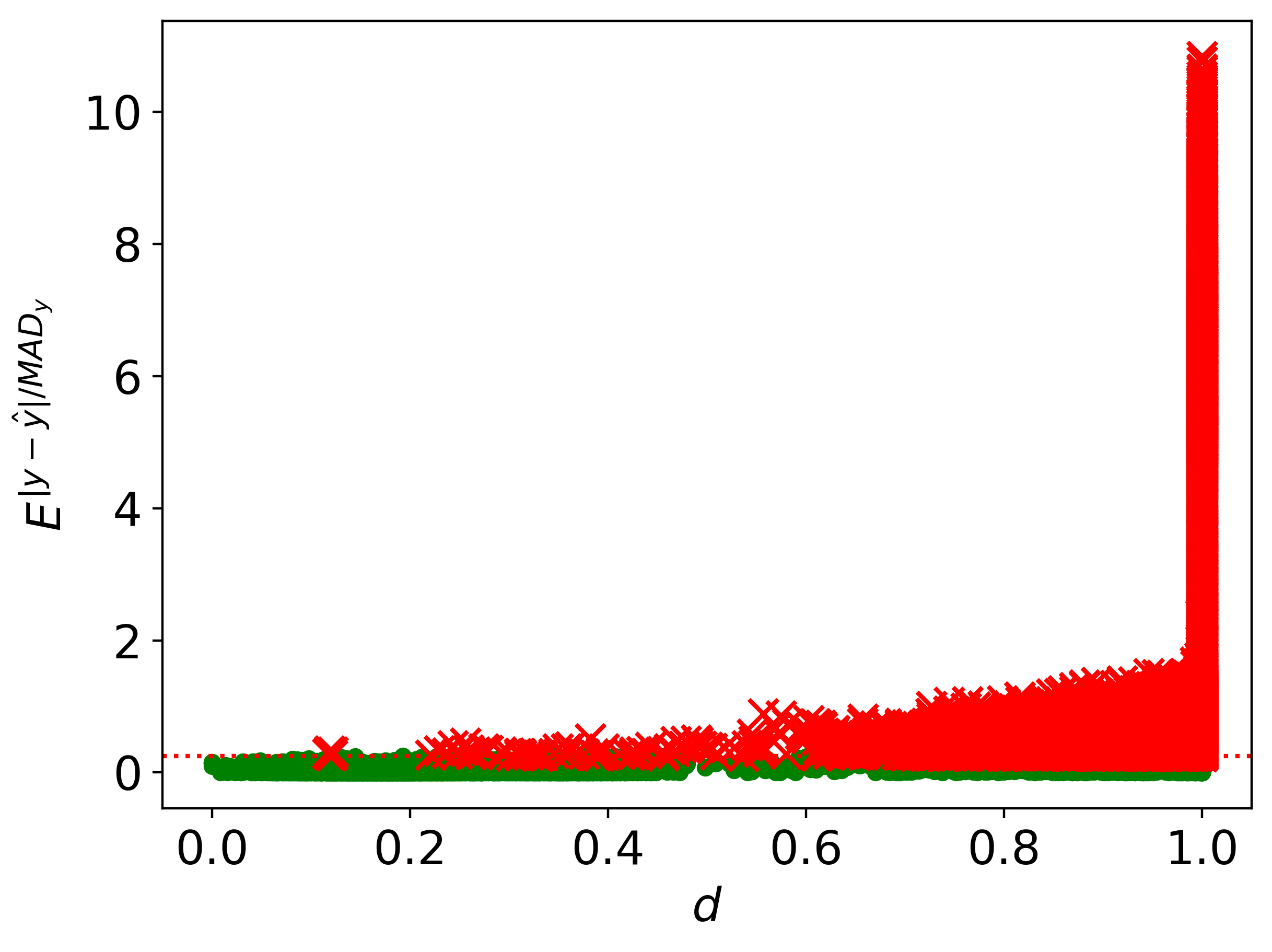}
        \caption{$E^{|y-\hat{y}|/MAD_{y}}_{c}=0.25$}
        \label{lower_gt}
    \end{subfigure}
    \hfill
    \begin{subfigure}{0.42\textwidth}
        \includegraphics[width=\linewidth]{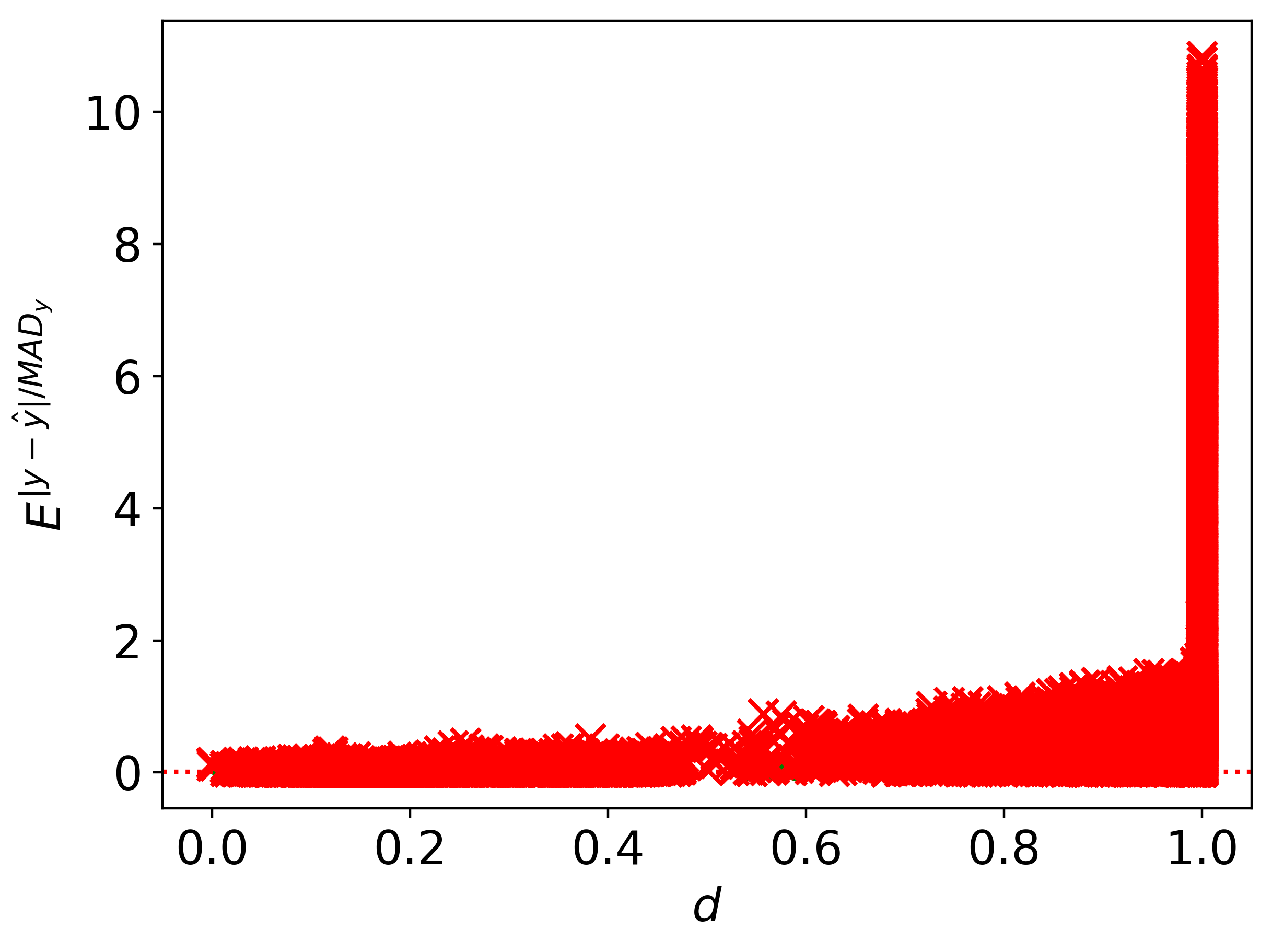}
        \caption{$E^{|y-\hat{y}|/MAD_{y}}_{c}=0.01$}
        \label{even_lower_gt}
    \end{subfigure}

	\caption{\textbf{Lowering Ground Truth Error.} Lowering $E^{|y-\hat{y}|/MAD_{y}}_{c}$ has a large impact on how the threshold for $M^{dom}$ is chosen and its corresponding metrics like precision. These data points, identical to those in Fig.~\ref{friedman_res}, have been reclassified as $ID/OD$ based on lower ground truth thresholds of 0.25 and 0.01 for Figs.~\ref{lower_gt} and \ref{even_lower_gt}, respectively (indicated by the horizontal red line in each figure). $ID$ points are green while $OD$ points are red.}
    \label{lowering_gt}
\end{figure}

Classification between $ID$ and $OD$ cases should become increasingly challenging when all considered data points are more similar. We examine the ability of $d$ at discerning $ID$ from $OD$ data points when differences between the two are less by lowering $E^{|y-\hat{y}|/MAD_{y}}_{c}$ for the Friedman data using the RF model type for $M^{prop}$. All remaining $ID$ and $OD$ points become more similar and can be seen in Fig.~\ref{lower_gt} where many $ID$ and $OD$ points now have the same value of $d$ (i.e., $ID$ and $OD$ points are more on top of each other and are more similar). Note that the labeling of $ID/OD$ is independent of $d$ for $A^{|y-\hat{y}|/MAD_{y}}$. Re-evaluating data from Fig.~\ref{friedman_res} for $E^{|y-\hat{y}|/MAD_{y}}_{c}=0.25$, the recall, precision, and $d^{t}_{c}$ changes from 0.86, 0.93, and 1.00 to 0.73, 0.97, and 0.74, respectively. $M^{dom}$ becomes more conservative with a lower ground truth by selecting a lower $d^{t}_{c}$ but is still a viable model for discerning domain. If $E^{|y-\hat{y}|/MAD_{y}}_{c}$ is further lowered to 0.01, then the overwhelming majority of points are $OD$ and $M^{dom}$'s recall, precision, and $d^{t}_{c}$ become 0.72, 0.19, and 0.27. $M^{dom}$ becomes imprecise at discerning $ID$ from $OD$ points when they are extremely similar (Fig.~\ref{even_lower_gt}).

\subsection*{Relationship Between $RMSE$ ($E^{RMSE/\sigma_{y}}$) and Distance ($d$) from the $RMSE$ Assessment ($A^{RMSE/\sigma_{y}}$)}
\label{results_rmse}

\begin{figure}[H]
    \centering
    
    \begin{subfigure}{0.42\textwidth}
        \includegraphics[width=\linewidth]{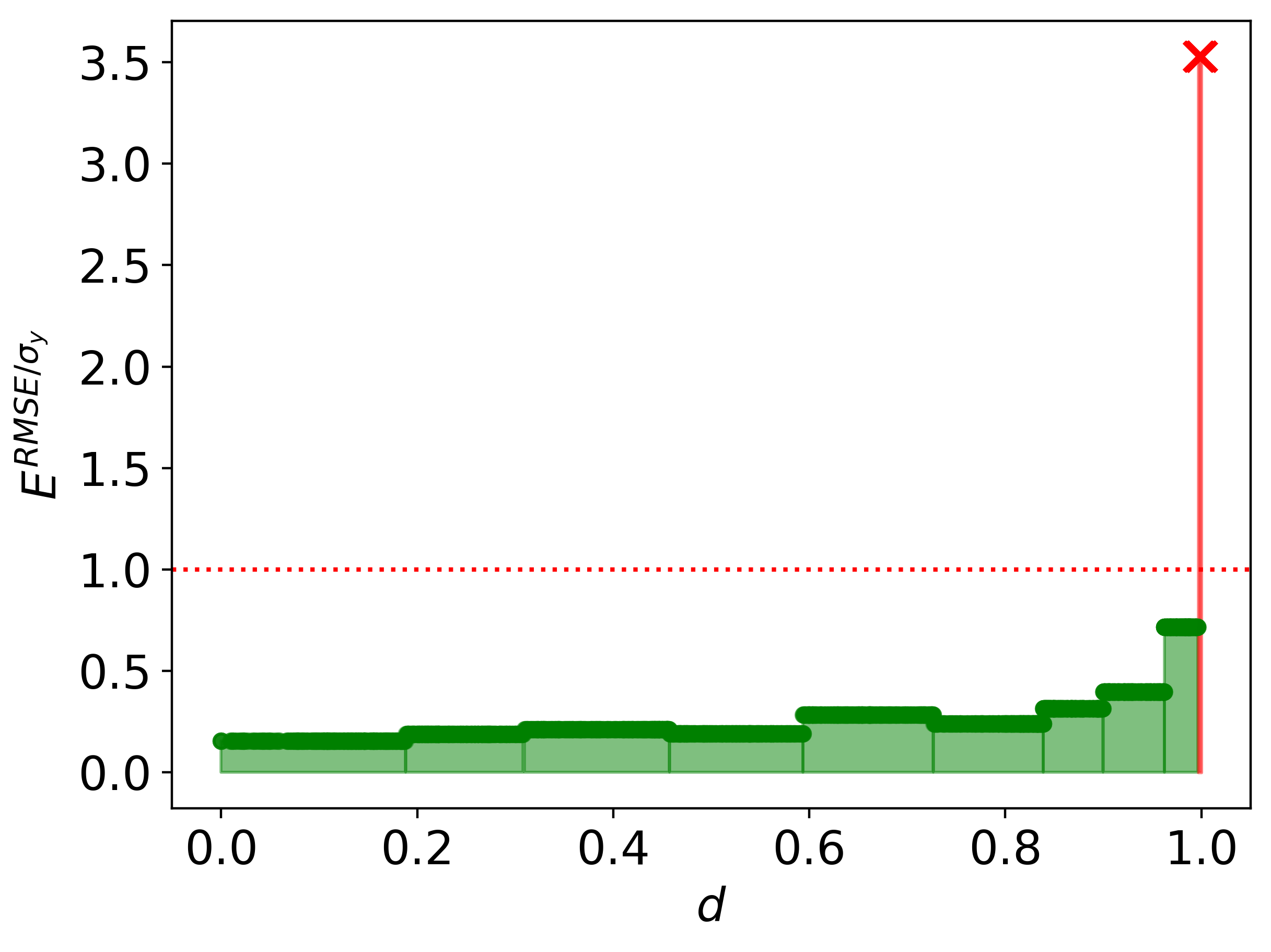}
        \caption{Diffusion}
        \label{diffusion_single}
    \end{subfigure}
    
    \vspace{1cm}
    
    \begin{subfigure}{0.42\textwidth}
        \includegraphics[width=\linewidth]{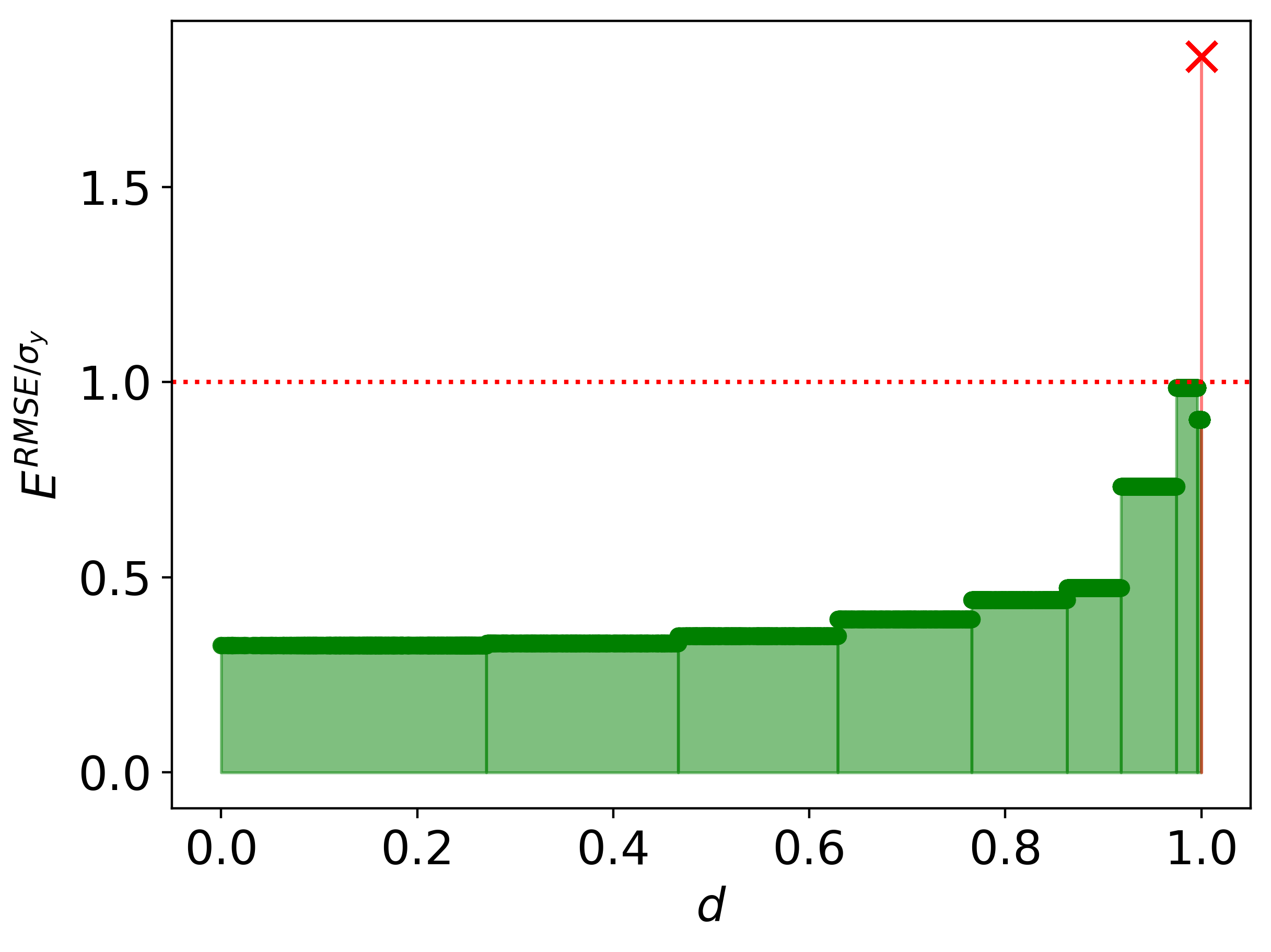}
        \caption{Fluence}
        \label{fluence_single}
    \end{subfigure}
    \hfill
    \begin{subfigure}{0.42\textwidth}
        \includegraphics[width=\linewidth]{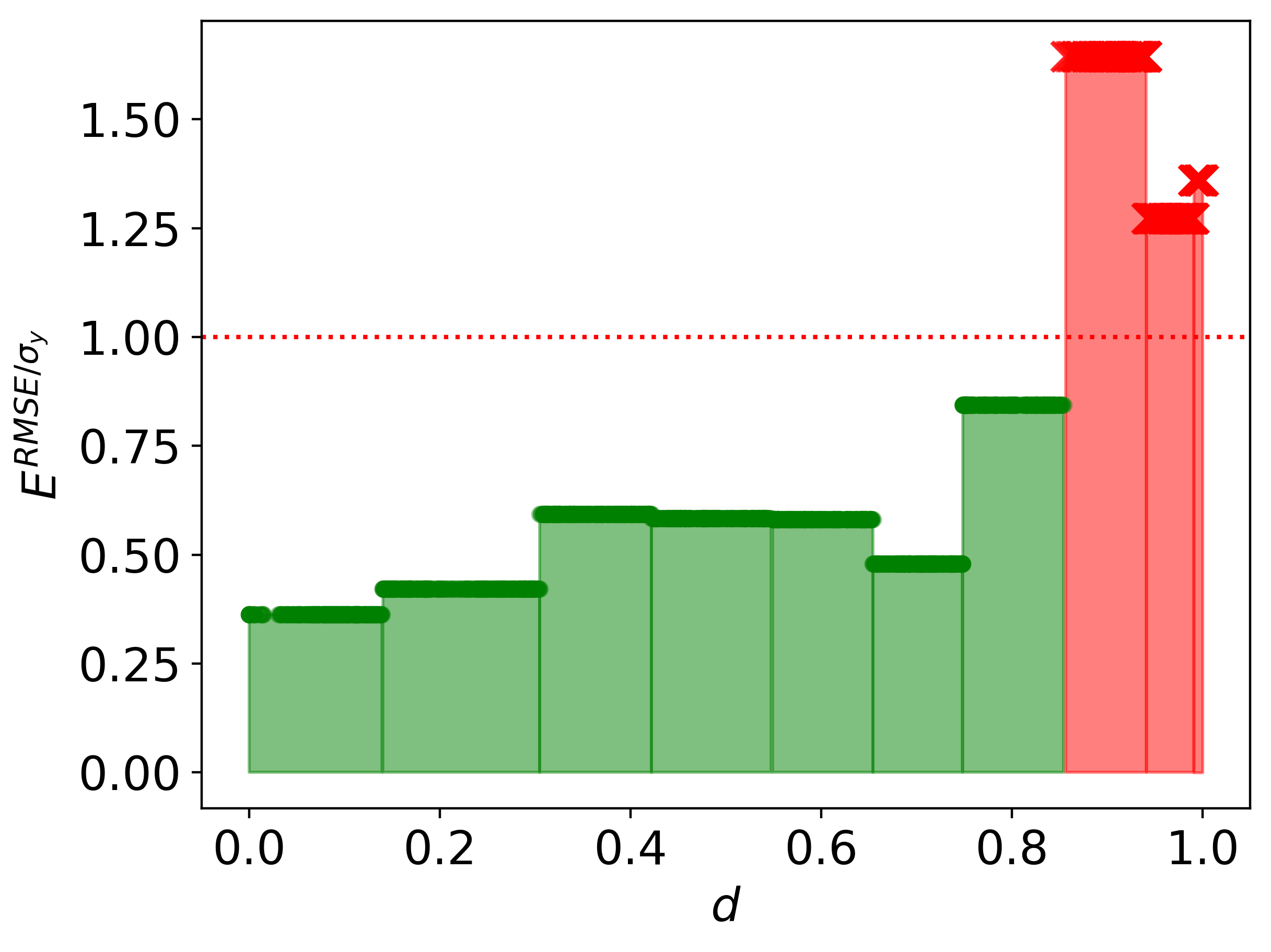}
        \caption{Steel Strength}
        \label{steel_single}
    \end{subfigure}
    
    \vspace{1cm}
    
    \begin{subfigure}{0.42\textwidth}
        \includegraphics[width=\linewidth]{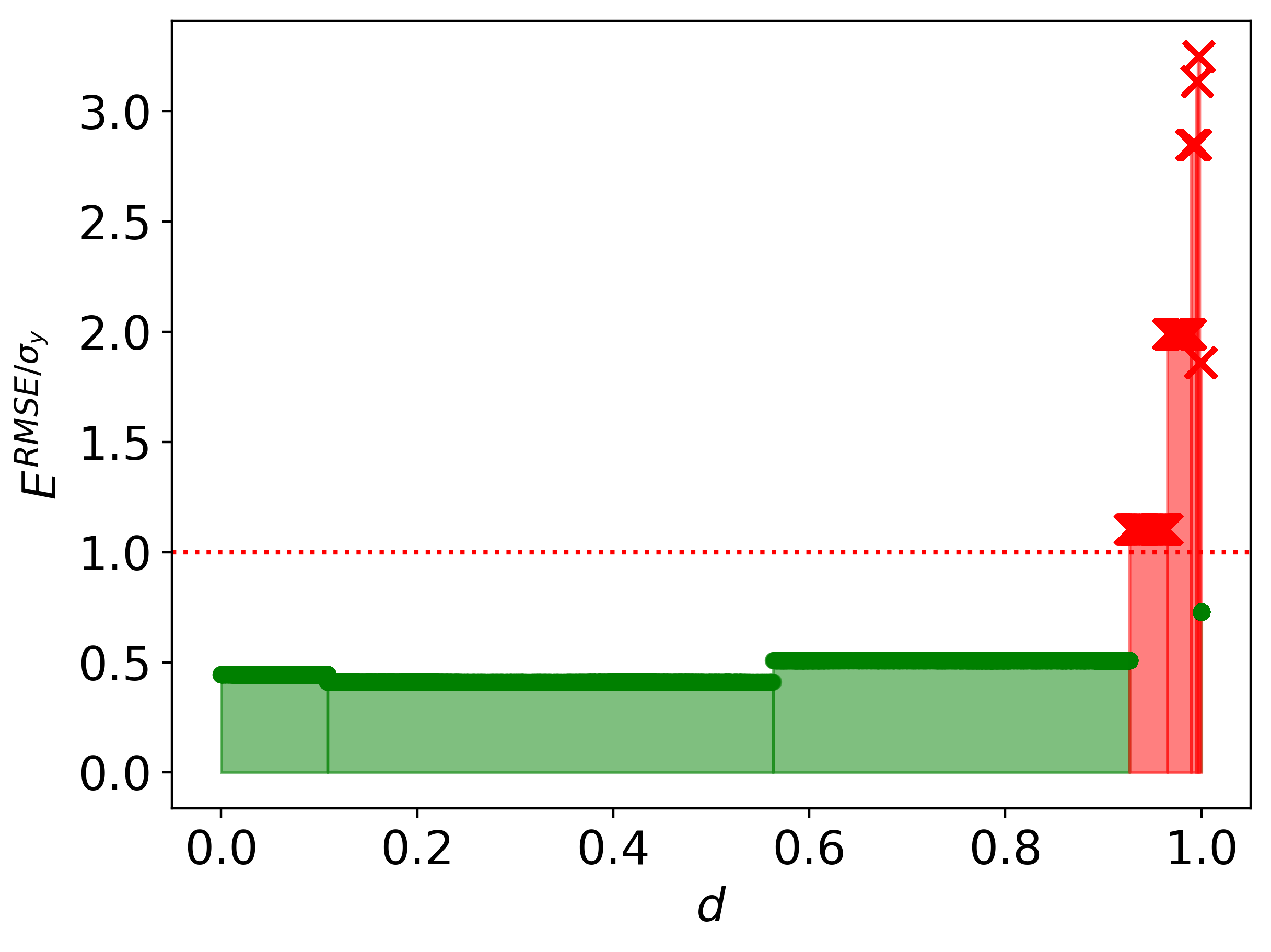}
        \caption{Superconductor}
        \label{cond_single}
    \end{subfigure}
    \hfill
    \begin{subfigure}{0.42\textwidth}
        \includegraphics[width=\linewidth]{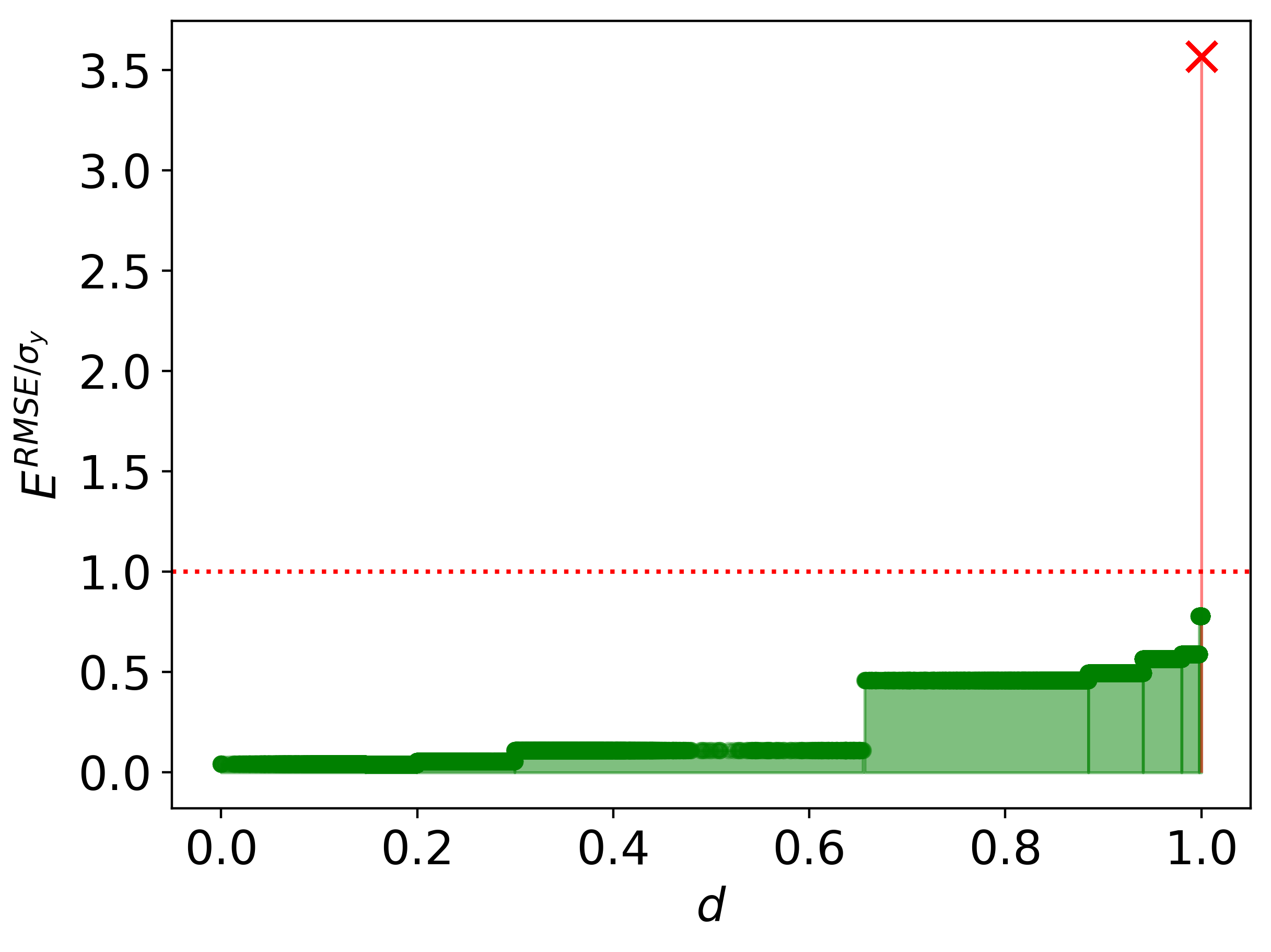}
        \caption{Friedman}
        \label{friedman_single}
    \end{subfigure}
    
    \caption{\textbf{RMSE grows as OOB data becomes increasingly dissimilar.} The relationship between $E^{RMSE/\sigma_{y}}$ and $d$ for the RF model type is shown. Generally, $E^{RMSE/\sigma_{y}}$ increases with an increase in $d$. $E^{RMSE/\sigma_{y}}_{c}$ is shown by the horizontal red line, which separates our $OD$ (red) and $ID$ (green) bins. The data set is denoted by the captions.}
    \label{residual_test_results}
\end{figure}

In Fig.~\ref{residual_test_results}, we illustrate how $E^{RMSE/\sigma_{y}}$ is related to $d$ for the RF model type. The same set of plots for other model types are offered in the Supplemental Materials, but we summarize their results in Table~\ref{table_results}. Similar to the results from the previous section, we observed that $E^{RMSE/\sigma_{y}}$ increased when the likelihood of observing similar points to ITB data decreased (larger $d$). All AUC-Baseline scores in Table~\ref{table_results} are positive for the relevant assessment, meaning that our $d$ measure gives more information than a na{\"i}ve guess provided by our baseline. Nearly all $F1_{max}$ scores are 1.00, which means that $ID$ and $OD$ bins were nearly perfectly separated (i.e., only 4 out of 20 were not perfect but still above 0.7). As an example of application, we can examine Fig.~\ref{steel_single} and select $d^{t}_{c}$=0.85 for Eq.~\ref{eq_mdom}. Three bins at $d$>0.85 have $E^{RMSE/\sigma_{y}}$ above $E^{RMSE/\sigma_{y}}_{c}$ and are $OD$. All other bins have $E^{RMSE/\sigma_{y}}$ below $E^{RMSE/\sigma_{y}}_{c}$ and are $ID$. If $M^{dis}$ yields $d$=0.8 for a data point, the point falls in a bin that is $ID$. If $M^{dis}$ yields $d$=0.95 for a data point, the point falls in a bin that is $OD$. A similar methodology can be applied for $d^{t}_{c}$ values across other data set and model combinations. We can use $d$ to distinguish points where $E^{RMSE/\sigma_{y}}$ is expected to be low from those likely to have high $E^{RMSE/\sigma_{y}}$ (i.e., we can discern predictions likely to be better than a na{\"i}ve). While it is true that some cases exhibit perfect discrimination between $ID$ and $OD$ classifications, this outcome can be attributed to the specific binning strategies used in our analysis (unlike the results in Fig.~\ref{res_test_results}). The threshold we used to separate these bins is selected to maximize the $F1$ score ($F1_{max}$), which can inadvertently lead to overfitting. However, it is important to highlight that when examining OOB predictions (i.e., data that were not utilized during model training) the $E^{RMSE/\sigma_{y}}$ generally increases with respect to the dissimilarity metric $d$.

\pagebreak

\subsection*{Relationship Between Miscalibration Area ($E^{area}$) and Distance ($d$) from the Uncertainty Quality Assessment ($A^{area}$)}
\label{results_area}

\begin{figure}[H]
    \centering

    \begin{subfigure}{0.42\textwidth}
        \includegraphics[width=\linewidth]{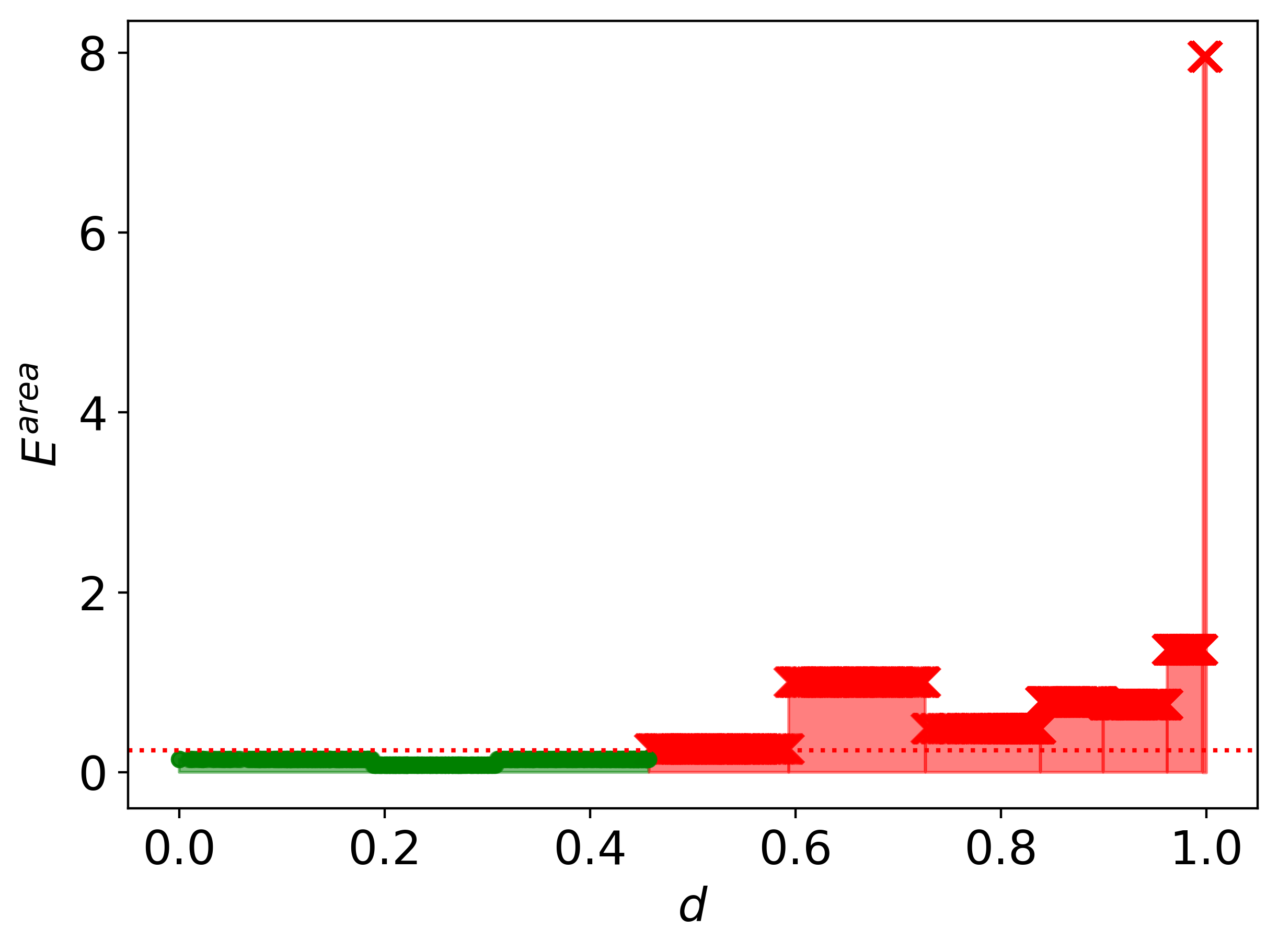}
        \caption{Diffusion}
        \label{diffusion_stat}
    \end{subfigure}

    \vspace{1cm}

    \begin{subfigure}{0.42\textwidth}
        \includegraphics[width=\linewidth]{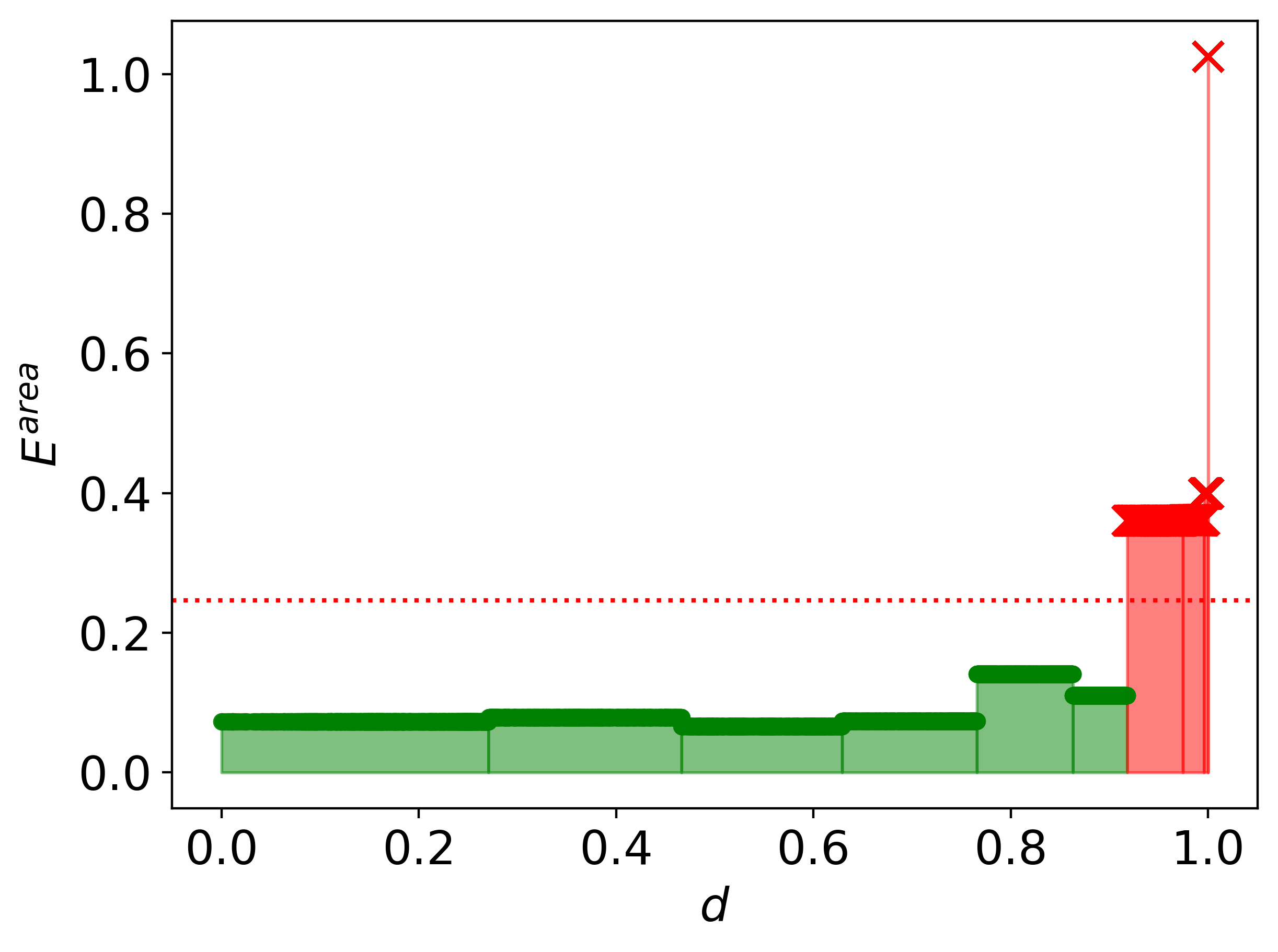}
        \caption{Fluence}
        \label{fluence_stat}
    \end{subfigure}
    \hfill
    \begin{subfigure}{0.42\textwidth}
        \includegraphics[width=\linewidth]{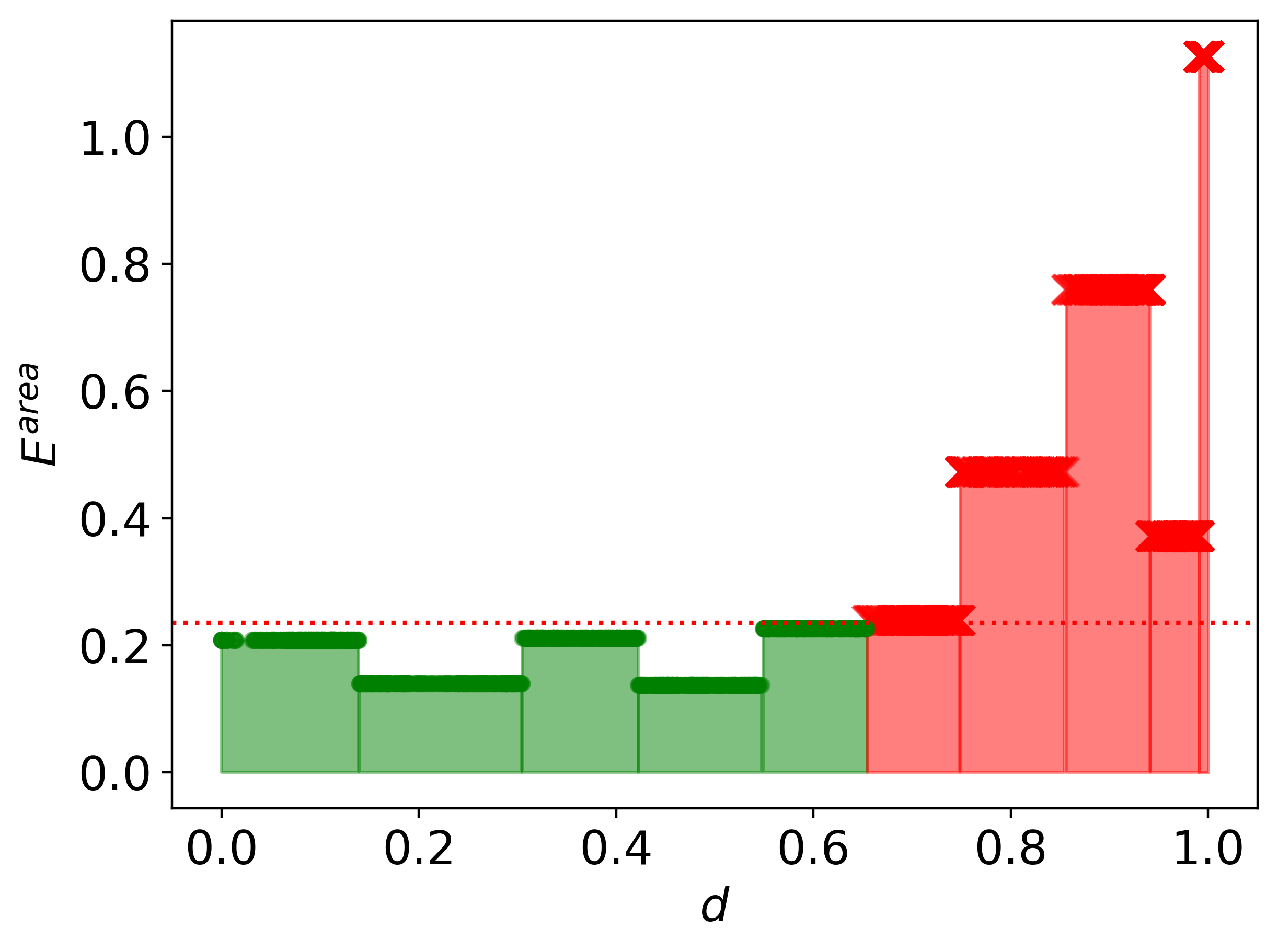}
        \caption{Steel Strength}
        \label{steel_stat}
    \end{subfigure}

    \vspace{1cm}

    \begin{subfigure}{0.42\textwidth}
        \includegraphics[width=\linewidth]{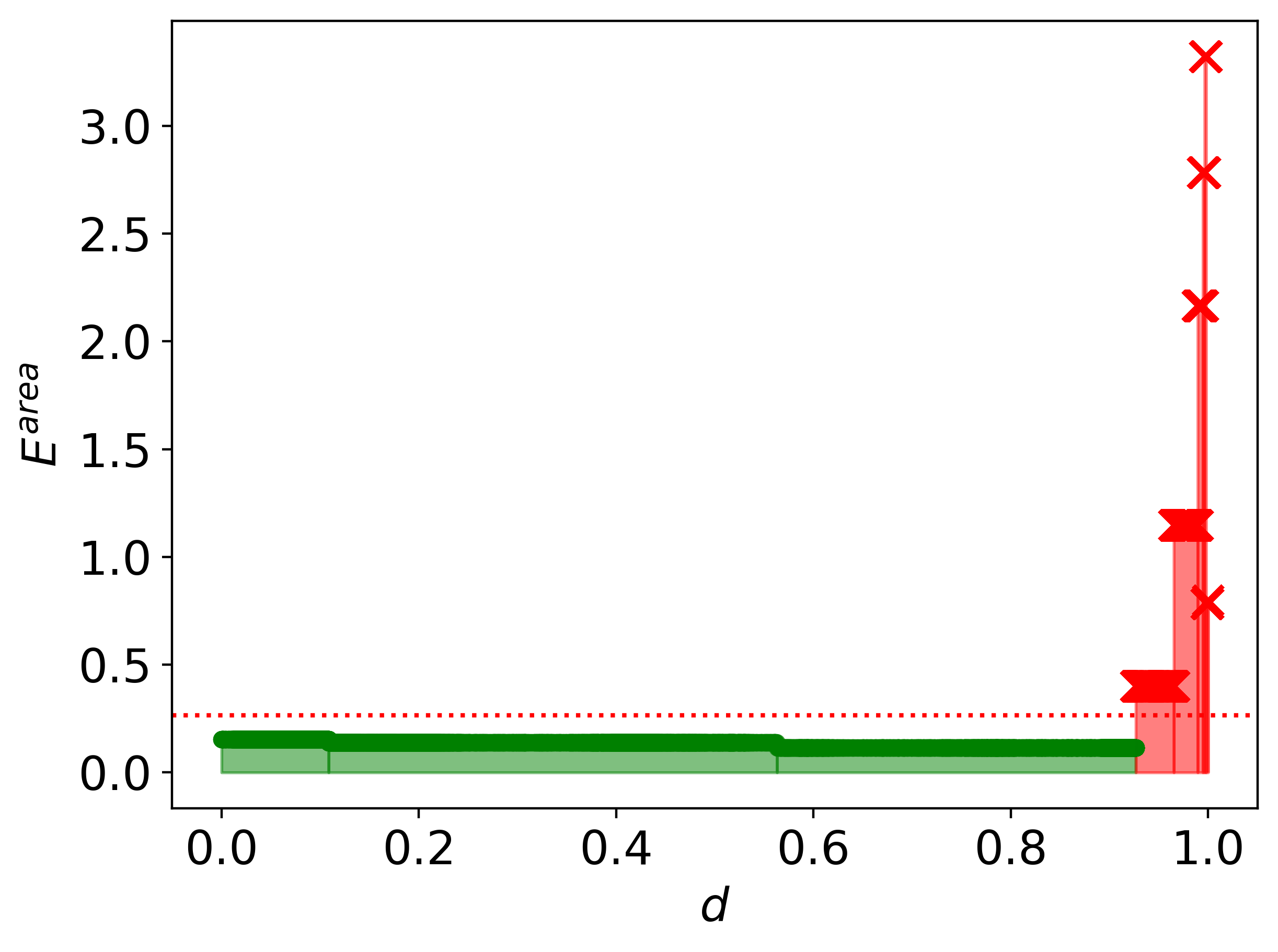}
        \caption{Superconductor}
        \label{cond_stat}
    \end{subfigure}
    \hfill
    \begin{subfigure}{0.42\textwidth}
        \includegraphics[width=\linewidth]{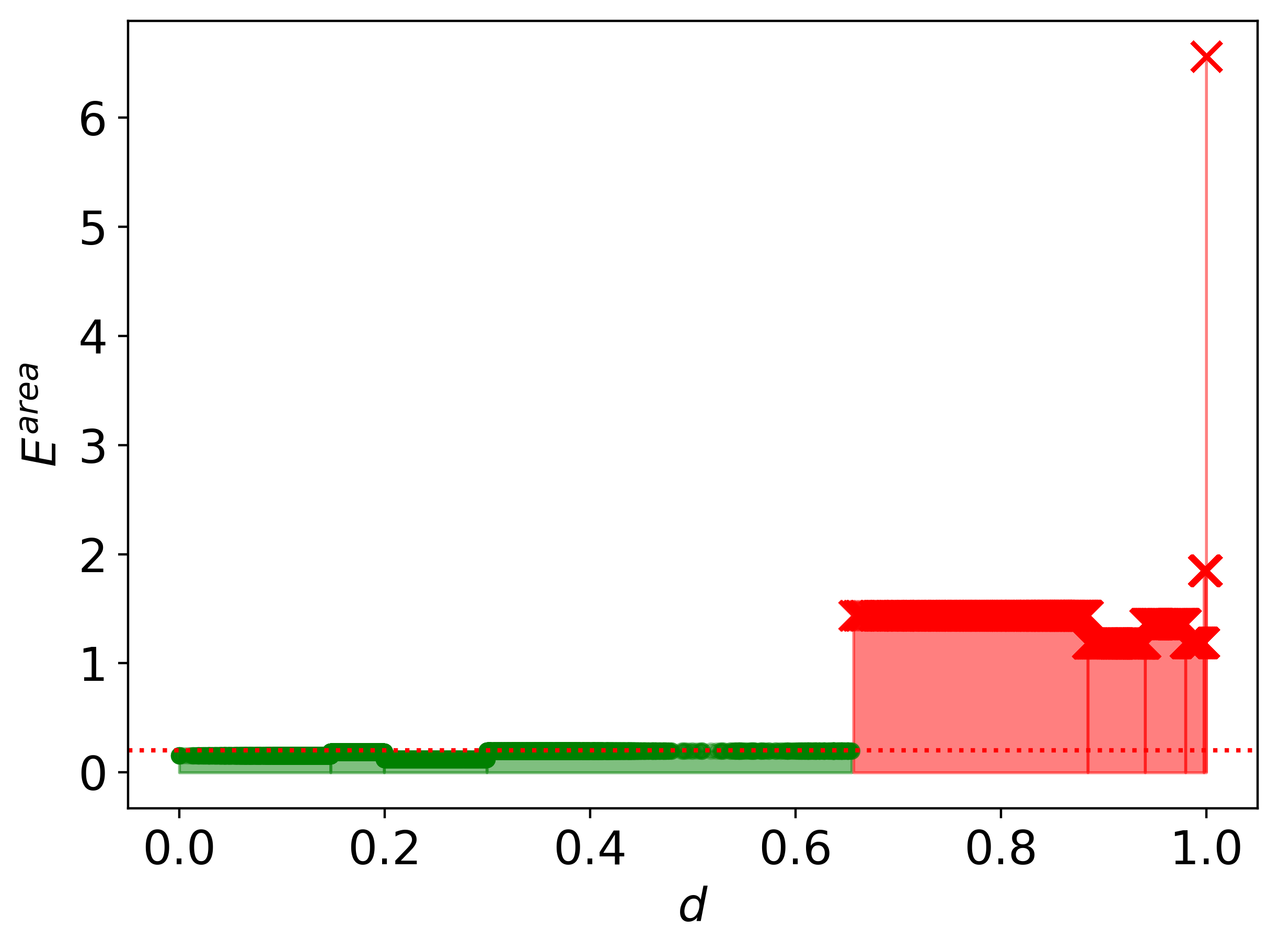}
        \caption{Friedman}
        \label{friedman_stat}
    \end{subfigure}

    \caption{\textbf{Uncertainty estimates deteriorate as OOB data becomes increasingly dissimilar.} The relationship between $E^{area}$ and $d$ for the RF model type is shown. Generally, $E^{area}$ increases with an increase in $d$. $E^{area}_{c}$ is shown by the horizontal red line, which separates our $OD$ (red) and $ID$ (green) bins. The data set is denoted by the captions.}
    \label{uncertainty_test_results}
\end{figure}

We now provide a similar analysis for $E^{area}$ as previously provided for $E^{RMSE/\sigma_{y}}$. In Fig.~\ref{uncertainty_test_results}, we illustrate how $E^{area}$ and $d$ are related for the RF model type. The same set of plots for the other model types are offered in the Supplemental Materials, but we summarize their results in Table~\ref{table_results}. We observed that $E^{area}$ for bins with small $d$ values tend to be relatively small compared to bins with larger $d$ values for all the data presented in Fig.~\ref{uncertainty_test_results}. All AUC-Baseline scores are greater than zero except for two, indicating that two data set and model combinations performed worse than naively predicting the number of $ID$ bins (i.e., only 2 out of 20 results were undesirable). For the most part, our $d$ measure offers substantial insight into the quality of $E^{area}$ for the reported data. Approximately half (9 out of 20) of the model type and data combinations yield precision and recall scores of 1.00, indicating their ability to perfectly discern points in $ID$/$OD$ bins. A total of 11 out of the 20 entries report $F1_{max}$ above 0.7, which is relatively good but not nearly as great as the results from $A^{|y-\hat{y}|/MAD_{y}}$ and $A^{RMSE/\sigma_{y}}$. An additional 4 entries were only slightly below 0.7. The only entries with $F1_{max}$ scores worse than na{\"i}ve were Steel Strength with BNN and Diffusion with BOLS. No data for the entry for Friedman with BOLS was $ID$, which was the reason for an AUC-Baseline of zero. The quality of $M^{unc}$ affects the ability of $M^{dom}$ to discern domain, a problem we discuss in detail as case (ii) in the Notes of Caution for Domain Prediction section and in the Supplemental Materials. As an example of application, we can examine Fig.~\ref{fluence_stat} and select $d^{t}_{c}$=0.92 for as a threshold to discern between $ID$/$OD$ bins. If $M^{dis}$ yields $d$=0.5 for a point, the point falls in a bin that is $ID$. If $M^{dis}$ yields $d$=0.99 for a point, the point falls in a bin that is $OD$. Choices for $d^{t}_{c}$ can be similarly made for other models. These results show that we can usually use $d$ to distinguish points where $E^{area}$ is expected to be low from those likely to have high $E^{area}$ (i.e., we can discern predictions likely to be better than a na{\"i}ve). It is important to note that each bin utilizes the exact same data to calculate $E^{RMSE/\sigma_{y}}$ and $E^{area}$. This implies that bins that were $ID$ based on $E^{RMSE/\sigma_{y}}_{c}$ but $OD$ based on $E^{area}_{c}$ had low absolute residuals but poor uncertainty quantification accompanying the predictions.  

\pagebreak
\pagebreak

\subsection*{Time Complexity of Established Methods}

\begin{figure}[H]
	\centering
    
    \begin{subfigure}{0.49\textwidth}
	    \includegraphics[width=\linewidth]{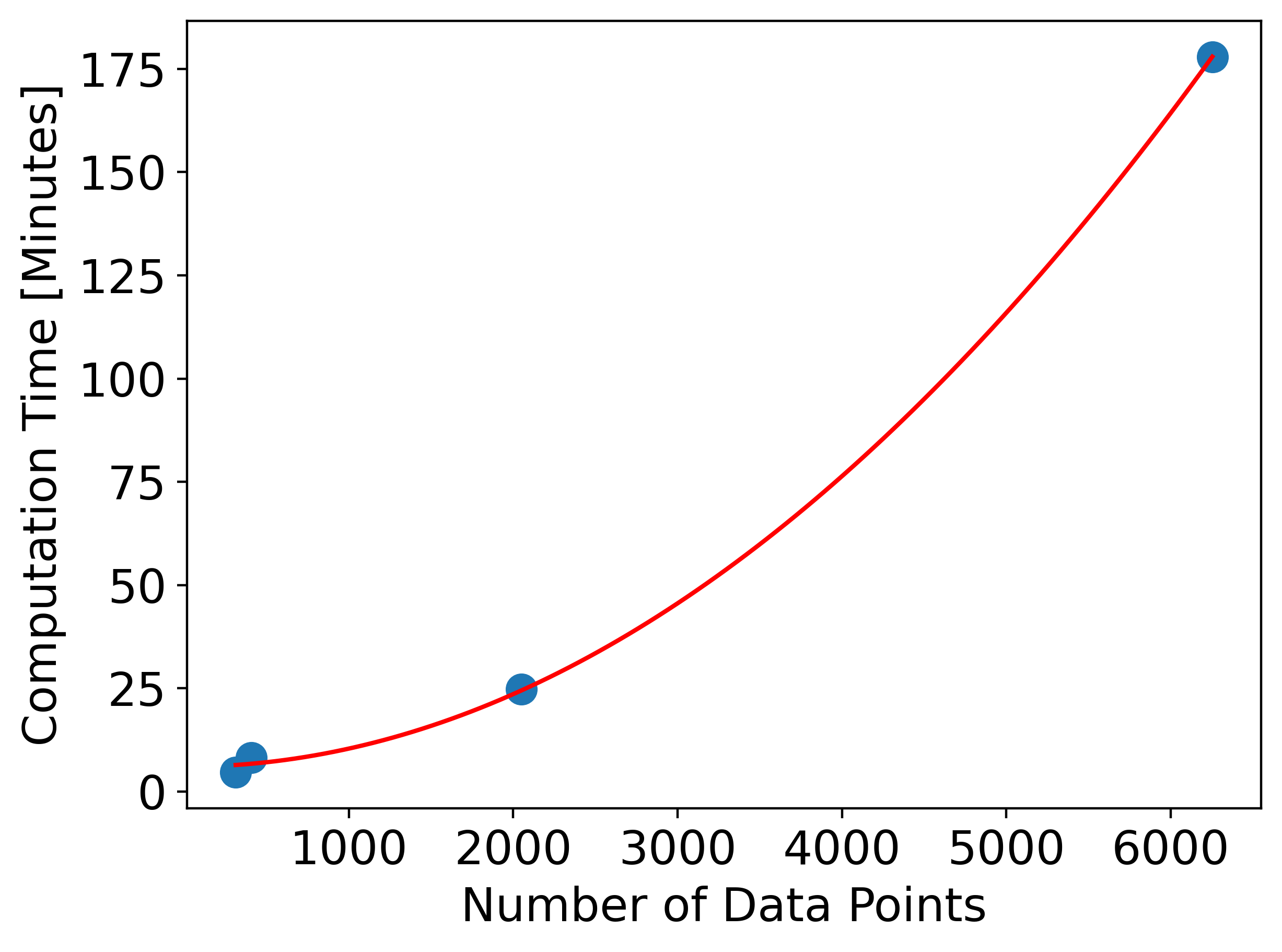}
	    \caption{Data Points}
        \label{scaling_points}
    \end{subfigure}
    \begin{subfigure}{0.49\textwidth}
	    \includegraphics[width=\linewidth]{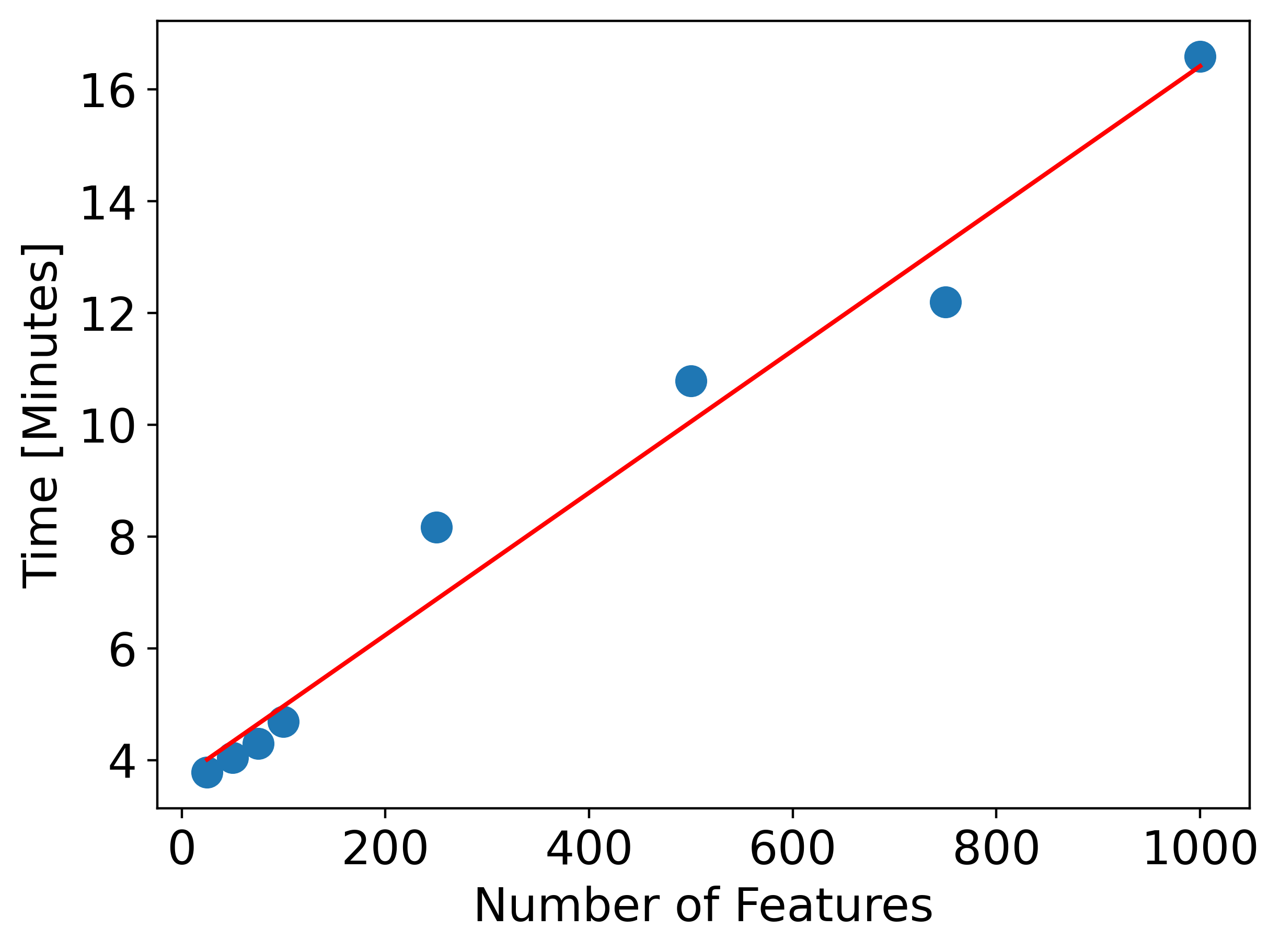}
	    \caption{Features}
        \label{scaling_features}
    \end{subfigure}

    \caption{\textbf{Time complexity with respect to the number of data points and features.} Figs.~\ref{scaling_points} and \ref{scaling_features} show the time scaling of the developed methods as a function of the number of data points and features, respectively. The RF model type was used. Only the Diffusion data was used for Fig.~\ref{scaling_features}. The red line is the best fit to the blue data points gathered from completed calculations (Fig.~\ref{scaling_points} is a quadratic fit and Fig.~\ref{scaling_features} is a linear fit).
    }
\end{figure}

Here we describe the time scaling of our domain method with respect to the number of features and the number of data points. The scikit-learn implementation of KDE can scale as either $\mathcal{O}(n log(n))$ (Ball-Tree algortihm) or $\mathcal{O}(log(n))$ (KD Tree algorithm) depending on the size of training $X$ \supercite{kde_time, de_ball_trees}. $n$ is the number of data points used in training. However, our method includes pre-clustering, cross validation, the underlying model type for $M^{prop}$, and a multitude of other functions which will alter the scaling of the overall codebase. Using the computation times from our assessments (i.e., $A^{|y-\hat{y}|/MAD_{y}}$, $A^{RMSE/\sigma_{y}}$, and $A^{area}$ combined) with $M^{prop}$ being an RF model type, the scaling of our assessment with respect to the number of data points appears to be approximately $\mathcal{O}(n^2)$ (Fig.~\ref{scaling_points}). A number of fuctions (constant, logarithmic, linear, logarithmic linear, and quadratic) were fit to the computation time versus number of data points using the curve\_fit optimizer from SciPy. The quadratic function fit the data the best, so the empirical scaling is estimated to be approximately $\mathcal{O}(n^2)$.

We can also examine how our methods scale with respect to the number of features. Using the Diffusion data, we re-run our analysis with the 25 features as discussed in the Diffusion Data Set section. Additionally, we run the same analysis with 50, 75, 100, 250, 500, 750, and 1000 features to compare the run times. Features were added by feature importance rankings established in the Diffusion Data Set section up to 100. We did not rank the remaining features. Because no feature beyond 25 altered $E^{RMSE/\sigma_{y}}$ significantly, features beyond 100 were added randomly. With an $M^{prop}$ of RF and following the methodology used for scaling with respect to the number of data points, we establish that our method scales approximately linearly (i.e., $\mathcal{O}(n)$). The maximum number of features is about 1,000 and was set by the number of features from our feature generation (covered in the Diffusion Data Set section). Note that the set of computers used for the data point and feature scaling assessments were conducted with separate sets of computers. For example, computer set 0 was used for the data point scaling and computer set 1 was used for feature scaling. Every computer in each corresponding set had the same type of hardware, but detailed results may differ significantly with different hardware.

\subsection*{Notes of Caution for Domain Prediction}
\label{fail}

We have identified four scenarios that could, but not necessarily, lead to the breakdown of developed models (although others likely exist): (i) the trivial case where OOB $y$ cannot be learned from $X$ (i.e., $M^{prop}$ breaks), (ii) the case where uncertainties are bad (i.e., $M^{unc}$ breaks), (iii) the case where Bootstrapped Leave-One-Cluster Out (BLOCO) fails to select regions of $X$ that are sufficiently distinct as OOB sets, and (iv) the case where KDE does not provide adequate information about ITB data because of a very large number of features. We will show the first 3 failure modes with easily controlled data generated with Eq.~\ref{friedman_eq} (Friedman and Friedman WithOut Distinct Clusters (FWODC)). BLOCO and FWODC are defined in the Methods section. To show case (i), we shuffled $y$ with respect to $X$. $M^{prop}$ could not learn $y$ from $X$ in this scenario. $E^{RMSE/\sigma_{y}}$ will be above $E^{RMSE/\sigma_{y}}_{c}$ for all (or nearly all) bins of $d$. To show case (ii), we used uncalibrated ($\sigma_{u}$) instead of calibrated ($\sigma_{c}$) uncertainties. If $M^{unc}$ is not accurate, then $d$ will show no reasonable trend between $E^{area}$ and $d$. To show case (iii), we used Uniform Manifold Approximation and Projection (UMAP) to show how insufficient clustering of distinct spaces provided values of $d$ that could not be used to build $M^{dom}$. This is a failure in producing $ID$/$OD$ labels for OOB data. In cases (i)-(iii), the failure is not really a problem with the fundamental domain method, as we now explain. For case (i), it is reasonable to assume an $M^{prop}$ model has no domain (or, equivalently, all data is $OD$) if its predictions are poor, and therefore no domain prediction method can reasonably be expected to work. Similarly, an $M^{unc}$ incapable of generating accurate uncertainties as outlined in case (ii) would have most, if not all, data as being $OD$. For case (iii), we do not separate data with our BLOCO procedure. This is not a failure of the fundamental approach, but a failure of our specific strategy for splitting leading to poor generation of $ID$/$OD$ samples. However, case (iv) would be a true failure of the underlying KDE approach, although we have not seen it occur in our tests.

\pagebreak
\begin{figure}[H]
    \centering

    \begin{subfigure}{0.42\textwidth}
        \includegraphics[width=\linewidth]{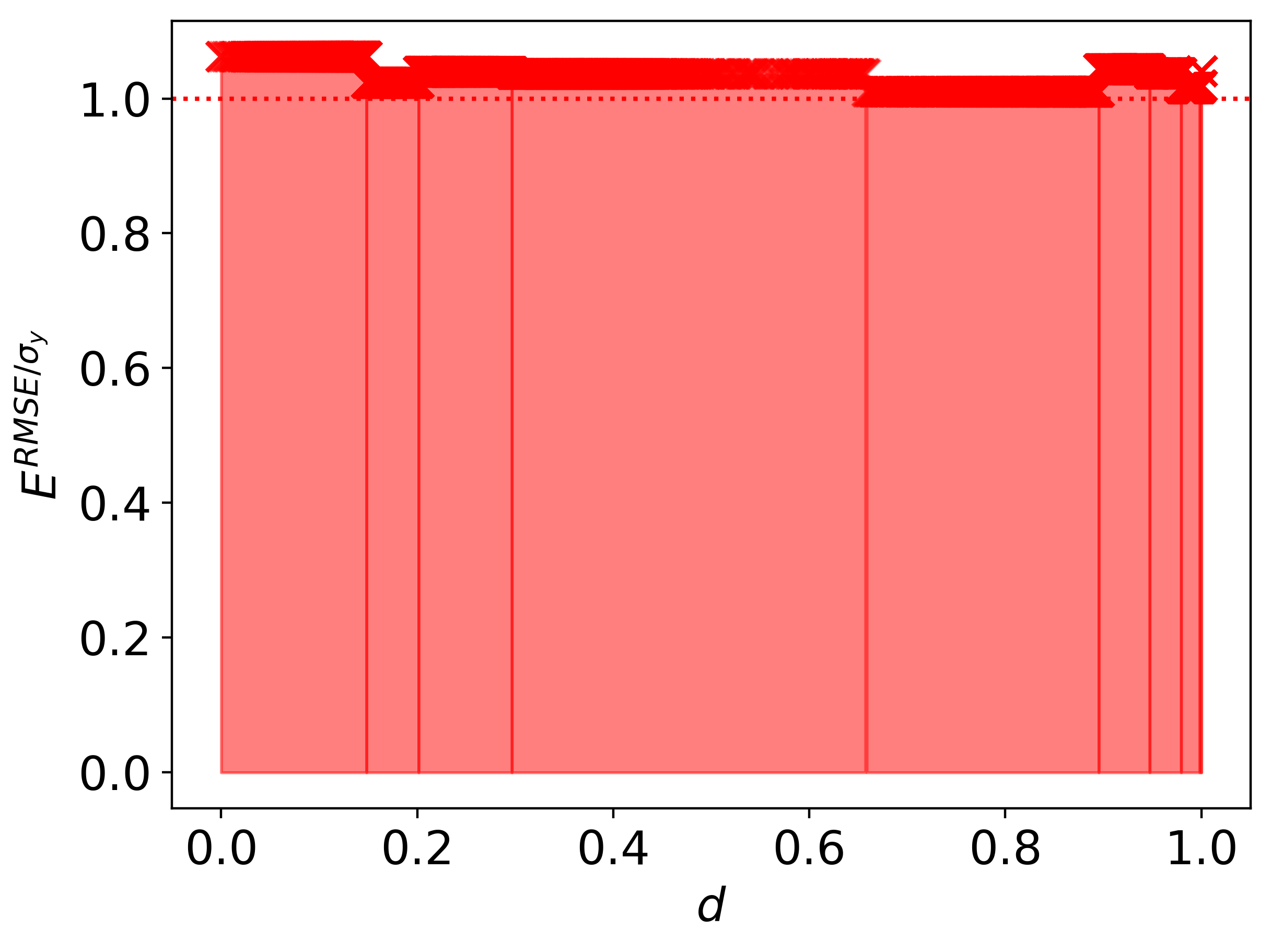}
        \caption{Shuffled $y$}
        \label{failure_target_res}
    \end{subfigure}
    \hfill
    \begin{subfigure}{0.42\textwidth}
        \includegraphics[width=\linewidth]{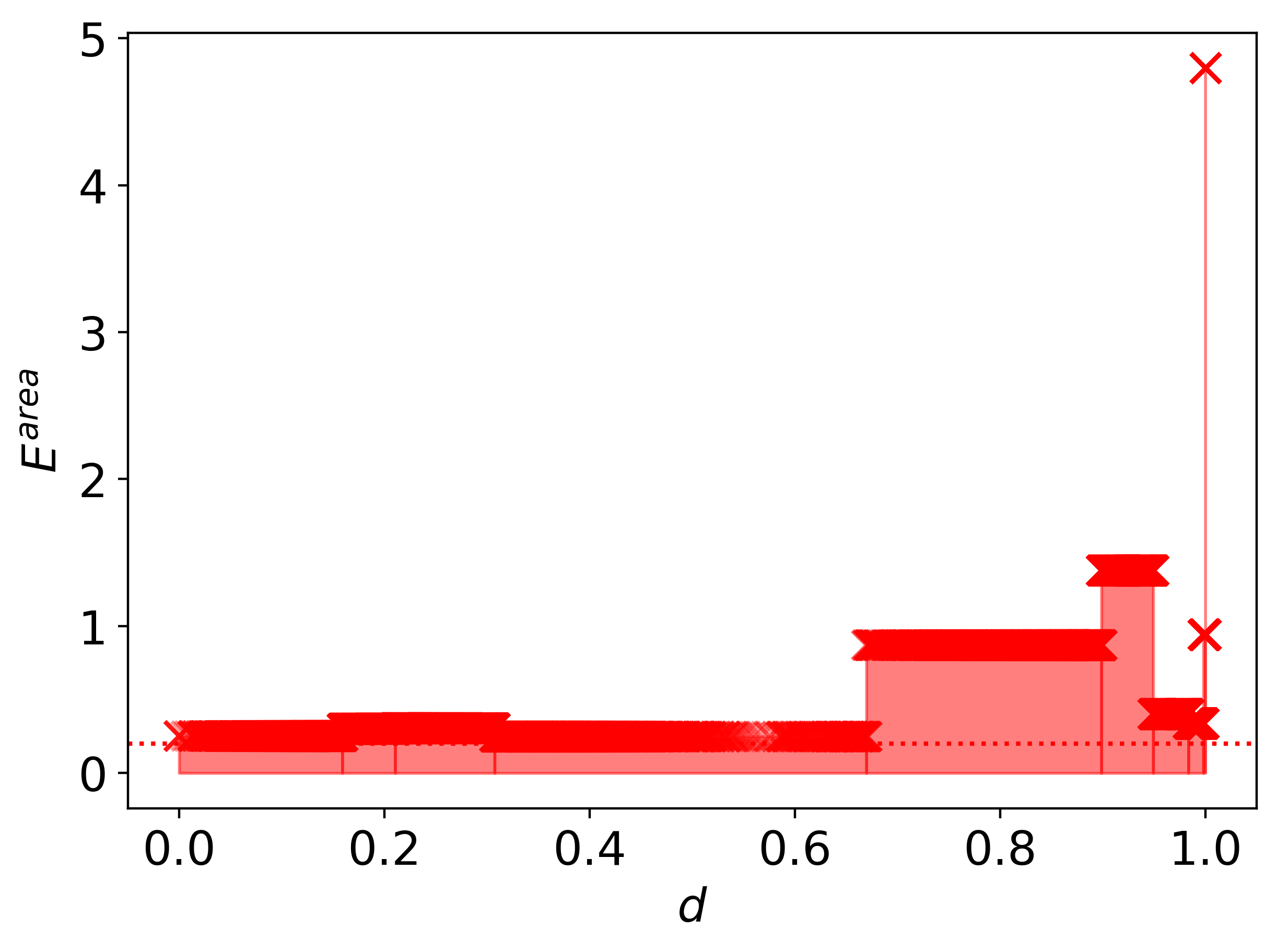}
        \caption{Uncertainties from $\sigma_{u}$ Instead of $\sigma_{c}$}
        \label{failure_uq}
    \end{subfigure}

    \vspace{1cm}

    \begin{subfigure}{0.42\textwidth}
        \includegraphics[width=\linewidth]{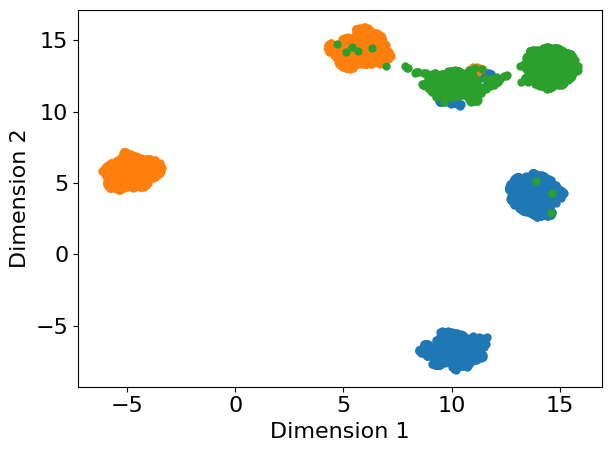}
        \caption{UMAP of Friedman}
        \label{umap_yes_sub}
    \end{subfigure}
    \hfill
    \begin{subfigure}{0.42\textwidth}
        \includegraphics[width=\linewidth]{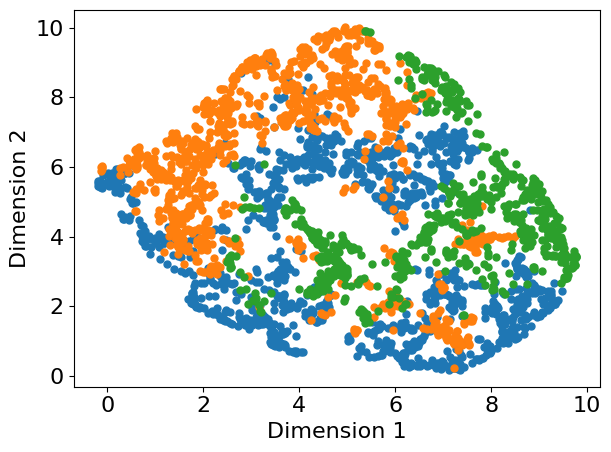}
        \caption{UMAP of FWODC}
        \label{umap_no_sub}
    \end{subfigure}

    \vspace{1cm}

    \begin{subfigure}{0.42\textwidth}
        \includegraphics[width=\linewidth]{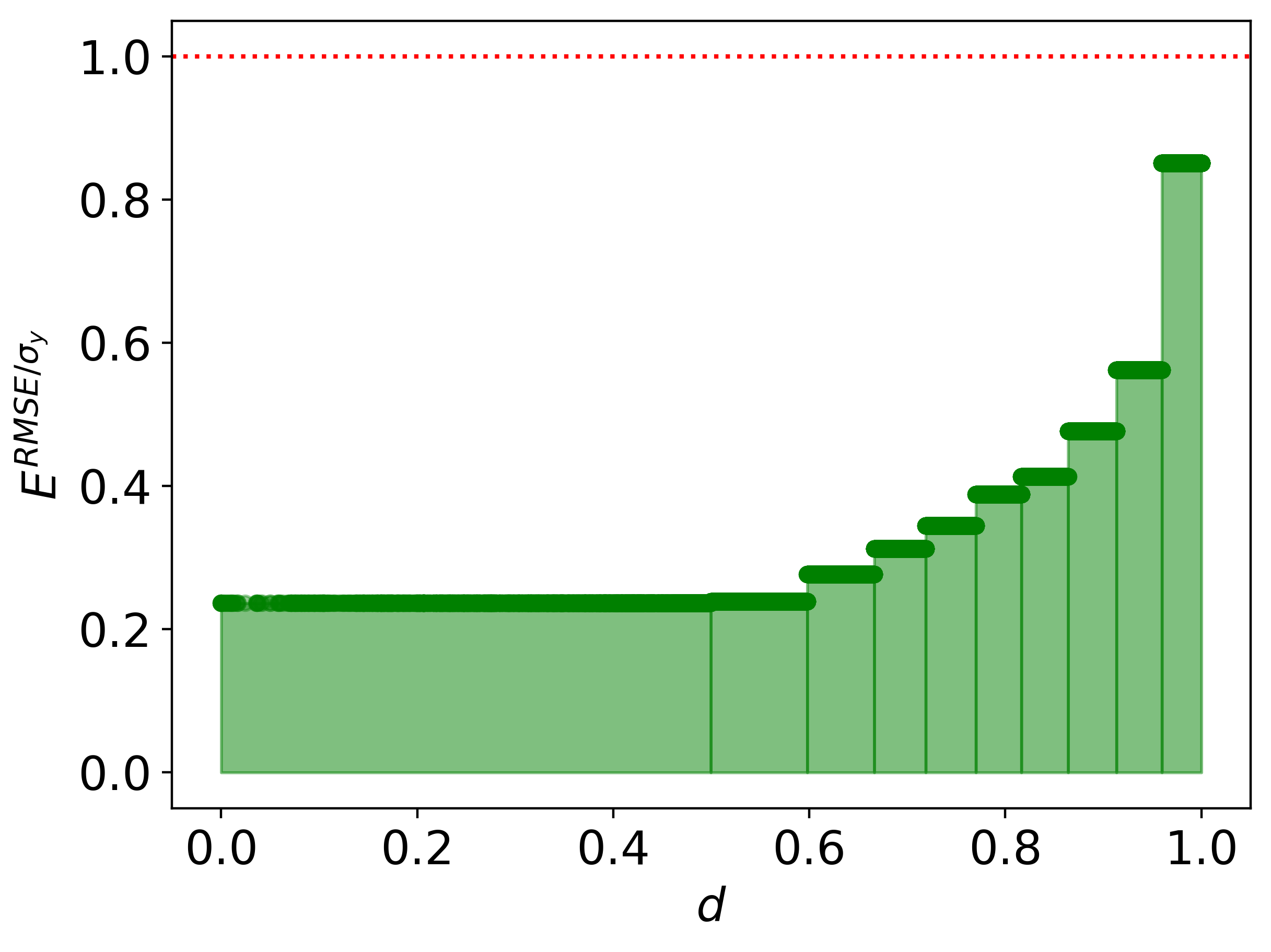}
        \caption{$A^{RMSE/\sigma_{y}}$ on FWODC}
        \label{failure_space_rmse}
    \end{subfigure}
    \hfill
    \begin{subfigure}{0.42\textwidth}
        \includegraphics[width=\linewidth]{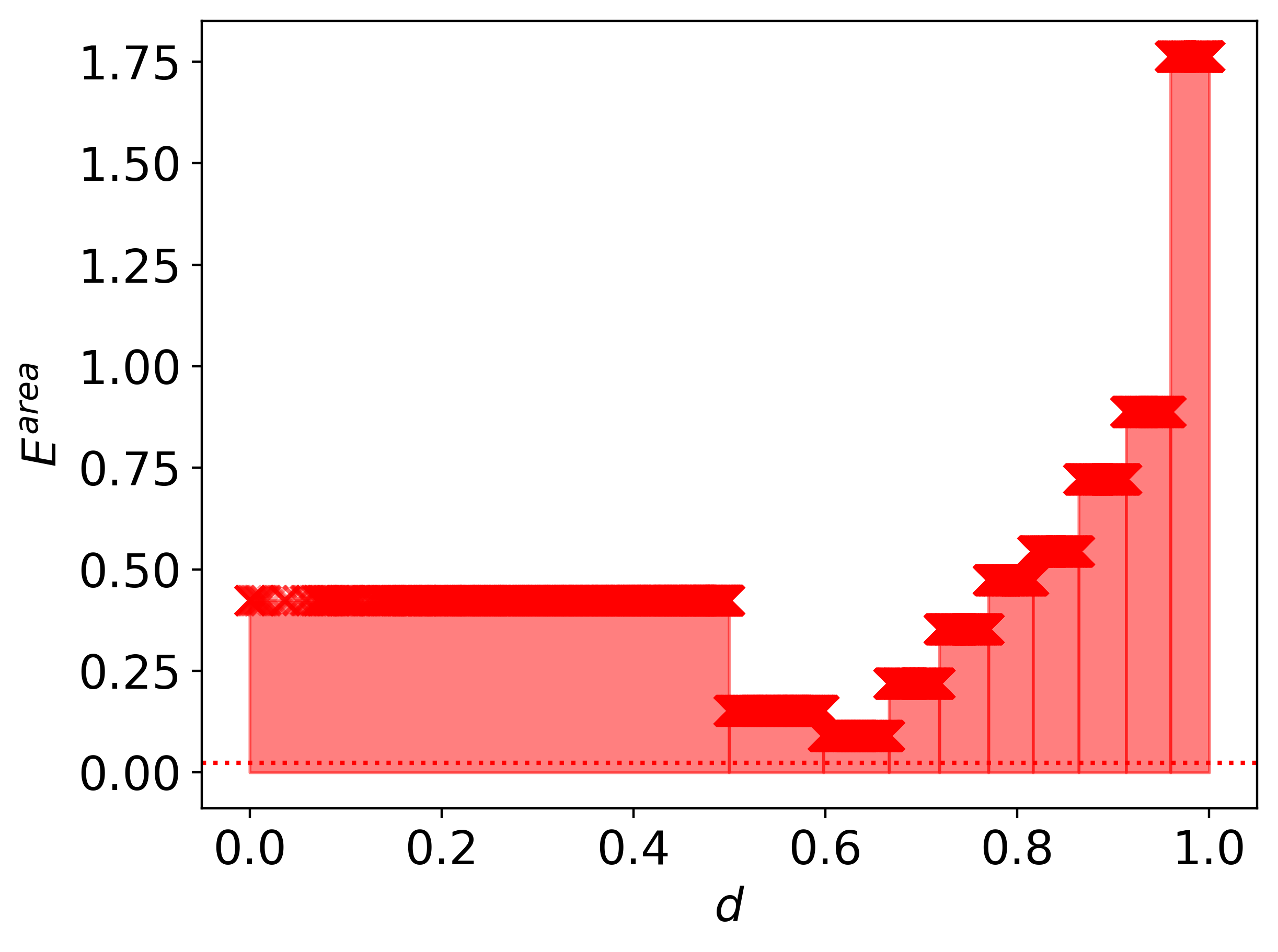}
        \caption{$A^{area}$ on FWODC}
        \label{failure_space_area}
    \end{subfigure}

    \caption{\textbf{Explanation of the limits of KDE to determine domain.} Fig.~\ref{failure_target_res} shows the assessment of the Friedman data fit with an RF model type where we purposefully shuffled $y$ to acquire an $M^{prop}$ with no predictive ability. Because $X$ no longer has a strong relationship with $y$, all data are $OD$. Fig.~\ref{failure_uq} shows the relationship between $d$ and $E^{area}$ for using $\sigma_{u}$ instead of $\sigma_{c}$ for $M^{unc}$. Because $M^{unc}$ is poor at estimating uncertainties, all data are $OD$. The UMAP projections of Friedman and FWODC onto two-dimensions are shown in Figs.~\ref{umap_yes_sub} and \ref{umap_no_sub}, respectively. One sampling of $X$ yields distinct regions in features (left) and the other does not (right). The colors represent the labels for three clusters acquired through agglomerative clustering. Figs.~\ref{failure_space_rmse} and \ref{failure_space_area} show the relationship between $d$, $E^{RMSE/\sigma_{y}}$, and $E^{area}$ for the poorly clustered FWODC data. Note that at least the bin with the highest $d$ for $A^{RMSE/\sigma_{y}}$ should be $OD$. We observe that data closest to our $X_{ITB}$ for $A^{area}$ are marked as $OD$ domain by $E^{area}_{c}$, but should ideally be $ID$.}
\end{figure}
\pagebreak

The RF model type was used for $M^{prop}$ for cases (i)-(iii) with either Friedman or FWODC data. We start with case (i) using the Friedman data. By shuffling $y$ with respect to $X$, the results from our assessments of $A^{RMSE/\sigma_{y}}$ breakdown (Fig.~\ref{failure_target_res}). Note that all predictions from $M^{prop}$ are worse than naively predicting $\bar{y}$ for all points. A $d^{t}_{c}$ cannot be established for separating $ID$/$OD$ cases from $E^{RMSE/\sigma_{y}}$ since all data are $OD$. Fig.~\ref{failure_uq} shows how $M^{dom}$ fails if $M^{unc}$ fails via case (ii). We use $\sigma_{u}$ instead of $\sigma_{c}$ as the estimates provided by $M^{unc}$. Errors in our uncertainties are high, and no value of $d^{t}_{c}$ separates $ID$/$OD$ points well ($M^{dom}$ failure) because no bins exist that are $ID$. Note that the portion of $M^{dom}$ trained with $E^{RMSE/\sigma_{y}}$ data should remain unaffected as long as $M^{prop}$ functions well.

Failure case (iii) for our domain methodology stems from a specific failure case of the BLOCO procedure. First, we use the UMAP approach to project $X$ into a two-dimensional space and visualize the cluster labels assigned by agglomerative clustering for Friedman (Fig.~\ref{umap_yes_sub}) and FWODC (Fig.~\ref{umap_no_sub}) data. Each color from the figures represents the cluster labels for three clusters. Because there are intervals of sampling for $X$ that exclude other intervals for Friedman in Fig.~\ref{umap_yes_sub}, we can cluster subspaces of $X$ that are distinct. If we consider FWODC as shown in Fig.~\ref{umap_no_sub}, all data that are OOB by BLOCO are close to each other, which results in an insufficient distinctiveness of clusters for pseudo-label generation ($ID$/$OD$). The $X$ space from Fig.~\ref{umap_yes_sub} led to the previously seen results on $A^{RMSE/\sigma_{y}}$ and $A^{area}$ (Figs.~\ref{friedman_single} and \ref{friedman_stat}). Data used to make Fig.~\ref{umap_no_sub} led to the results shown in Figs.~\ref{failure_space_rmse} and \ref{failure_space_area}. For Fig.~\ref{failure_space_rmse}, we know that data at $d$=1.00 can produce $OD$ points as seen in Fig.~\ref{friedman_single}. For Fig.~\ref{failure_space_area}, $E^{area}$ is much larger for our first bin than that from Fig.~\ref{friedman_stat}. Conversely, $E^{area}$ is much smaller for our last bin in Fig.~\ref{failure_space_area} compared to Fig.~\ref{friedman_stat}. Unusual trends of $d$ with respect to $E^{area}$ emerge when the data set lacks diversity (i.e., data that cannot be clustered well).

\begin{figure}[H]
	\centering
	\includegraphics[width=0.75\textwidth]{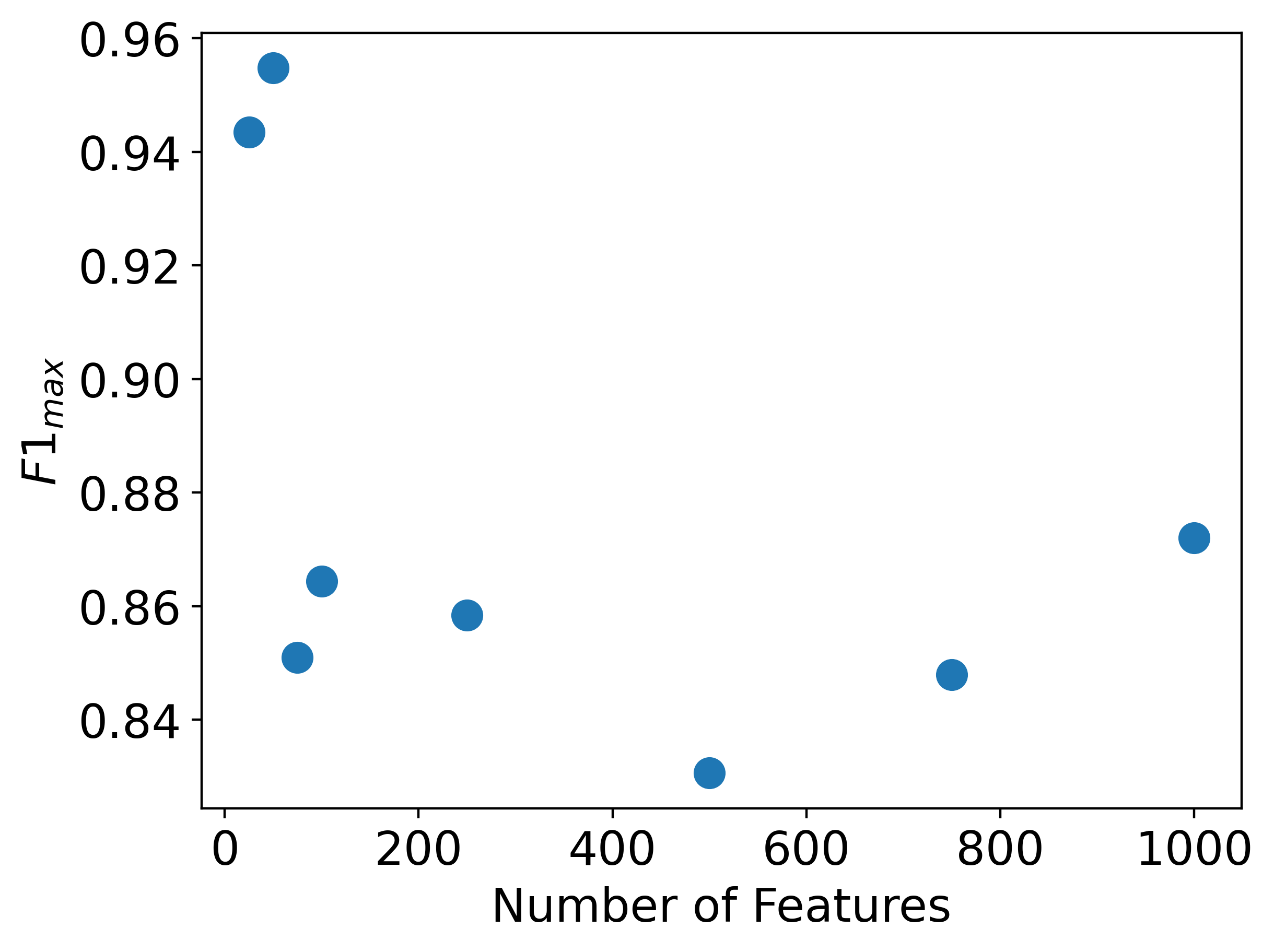}
	\caption{\textbf{Number of features and classification performance.} The $F1_{max}$ score from classifying $ID$ points from $OD$ points decreases for larger numbers of features.}
    \label{features_f1}
\end{figure}

Finally, we discuss failure mode (iv). KDE is known to have challenges in getting accurate representations for a large number of dimensions \supercite{NAGLER201669}. For models with many features, it is possible that KDE will give a poor representation of the distance to the $X_{ITB}$ data and the approach taken in this work will break down (i.e., yield inaccurate $\widehat{ID}$/$\widehat{OD}$ labels). Here we explore how the $F1_{max}$ scoring metric for $A^{|y-\hat{y}|/MAD_{y}}$ is affected by the number of features included in the assessment. We use the exact same tests as used in the feature scaling results described in the Time Complexity of Established Methods section. RF was the model type for $M^{prop}$ and the data used was the Diffusion data set. As shown in Fig.~\ref{features_f1}, the smaller features sets of 25 and 50 have an approximate 0.09 increase in $F1_{max}$ scores compared to larger feature sets. Based on these results, it is recommended to perform feature selection to reduce the number of features to a modest level. We do not know how many features would cause issues, but these results suggest that 50 and below is optimal and it is reasonable to assume the inclusion of thousands of irrelevant features will further degrade scores. Although this result shows that the established methods work best with a smaller number of relevant features for $M^{prop}$, the degradation with even 20 times a larger feature set is not very large. However, the degree of performance degradation with respect to the number of features is likely to depend on the data analyzed and $M^{prop}$ used.

In summary, there are three known conditions and a fourth one hypothesized that will lead to a non-functional $M^{dom}$. First, $M^{prop}$ cannot properly predict $y$ from $X$. Second, $M^{unc}$ fails to provide accurate measures of uncertainty. Third, the data cannot be clustered to produce $ID$/$OD$ labels. Fourth, the KDE fails to provide a good density due to a large number of features. So long as the aforementioned conditions are not met, $M^{dom}$ has been found to effectively predict domain in our tests.

\section*{Discussion}

Our work addresses a significant concern in the deployment of machine learning models: the potential for these models to produce inaccurate or imprecise predictions without warning. We have shown that kernel density estimates provide valuable insights into where in the feature space model performance degrades significantly or where predictions fall outside the model's domain of applicability. The central idea of our approach is to use kernel density estimation to define a dissimilarity measure $d$ and then classify data as in-domain or out-of-domain based on $d$. We assessed the approach with a range of machine learning models (random forest, bagged neural network, bagged support vector regressor, and bagged ordinary least squares) on data sets with diverse physical properties and chemistries (dilute solute diffusion in metals, ductile-to-brittle transition temperature shifts in irradiated steel alloys, yield strengths of steel alloys, and superconducting critical temperatures) as well as the synthetic Friedman data set. We demonstrated qualitatively that as $d$ increased, data became more chemically distinct, residual magnitudes increased, and uncertainty estimates deteriorated, validating that $d$ was a powerful descriptor for these critical ways of thinking about domain. Our quantitative assessment compared our predicted domain categorization to ground truth values based on chemistry, residuals, and uncertainty estimates and generally found good improvement over na{\"i}ve models and high $F1$ scores. This approach can be easily applied to many problems and allows categorization of in-domain or out-of-domain for any test data point during inference. The approach can be applied easily through its stand-alone implementation or its implementation in the MAterials Simulation Toolkit - Machine Learning (MAST-ML) package (see Data and Code Availability). Researchers can easily use this method to provide automated guardrails for their machine learning models, greatly enhancing their reliable application.

\section*{Methods}

\subsection*{Software Tools Used}

In our study, we utilized several software tools for analysis. We employed the MAST-ML to generate and select features, $X$, across multiple data sets \supercite{Jacobs2020}. Furthermore, we utilized various models and subroutines from scikit-learn \supercite{scikit-learn}. For neural network (NN) implementations, we relied on Keras \supercite{chollet2015keras}. Visualizations from projecting $X$ onto lower dimensions used the UMAP package \supercite{umap}.

\subsection*{Model Assessments}
\label{domain_tests}

We can assess how well $M^{dom}$, or equivalently, categorizing the data based on $d$ (Eq.~\ref{eq_mdom}), can predict $ID$/$OD$. Specifically, we evaluate the ability of $d$ obtained from $M^{dis}$ to predict domain labels by building precision-recall curves on OOB data sets. For each precision-recall curve, a na{\"i}ve baseline area under the curve (AUC) was constructed by considering a baseline model that predicts $ID$ for every case, which yields a baseline AUC of the ratio of $ID$ cases over the total number of cases \supercite{Saito2015}. The difference between the true and baseline AUC was used as an assessment metric, which we call AUC-Baseline. AUC was calculated using the sklearn.metrics.average\_precision\_score from scikit-learn \supercite{scikit-learn}. AUC-Baseline shows how much additional information $d$ provided for domain classification above the na{\"i}ve baseline model. Any AUC-Baseline above zero indicates domain information retained by $d$. The values of precision, recall, and $F1$ were reported for the acquired metric of $F1_{max}$, which evaluates how effectively the $ID$ points were separated from the $OD$ points by the dissimilarity measure $d$.

\subsubsection*{Assessing Our Domain Prediction Based on A Ground Truth Determined by Chemical Intuition}

In this section, we describe our method to evaluate how $d$ provides information with respect to $E^{chem}$ and call this assessment $A^{chem}$. We started with a set of ITB data that were used to construct $M^{prop}$ and additional data with chemically distinct groups. These data are called the Original and non-Original sets, respectively. The non-Original set can be further subdivided into other chemistries and was always treated as OOB. Each case from all data was labeled as $ID$/$OD$ based on chemical intuition. We wanted to evaluate the ability of $d$ to discern $ID$/$OD$ from groups of chemistries that are OOB. To do this, we built sets of OOB data that contained $ID$/$OD$ labels. For every observation in the Original set, $i$, the following sets of steps were performed: $i$ was taken out of the Original set (i.e., placed in the OOB data), an $M^{dis}$ model was trained on the remaining Original ITB data, then $M^{dis}$ was used to calculate $d$ on all OOB data (i.e., $i$ and all cases from the non-Original set). This formed one prediction set. We repeated the procedure for all data points in the Original set and aggregated that data. In other words, we performed Leave-One-Out (LOO) CV on the Original set while treating the non-Original set as an OOB set for every iteration. If there are $n$ cases from the Original set, then there should be $n$ different models and prediction sets by the time the procedure finished. See Fig.~\ref{chem_cv} for a diagram of the splitting procedure. Violin plots were generated to show the distribution of $d$ values for each chemical group.

This methodology often produced a large class imbalance between the number of $ID$ and $OD$ cases, which can make interpreting the assessment of the domain classifier difficult. Therefore, we used a resampling procedure to obtain a balanced number of $ID$ and $OD$ data points in the OOB aggregated set. More specifically, if the number of $ID$ cases exceeded the number of $OD$ cases, we randomly sampled a subset of the $ID$ data such that the number of $ID$ cases was equal to the number of $OD$ cases. Conversely, if the number of $OD$ cases surpassed the number of $ID$ cases, we randomly sampled a subset of the $OD$ data to match the number of $ID$ cases. However, if the $ID$ and $OD$ data sets contained an equal number of cases, no subsampling was performed. These samplings of data were then used to assess the ability of $d$ to predict $ID$/$OD$ with precision and recall.

\subsubsection*{Assessing Our Domain Prediction Based on A Ground Truth Determined by Normalized Residuals and Errors in Predicted Uncertainties}
\label{residual_test}

\begin{figure}[H]
    \centering
    \begin{subfigure}{0.4\textwidth}
        \centering
        \includegraphics[width=0.9\textwidth]{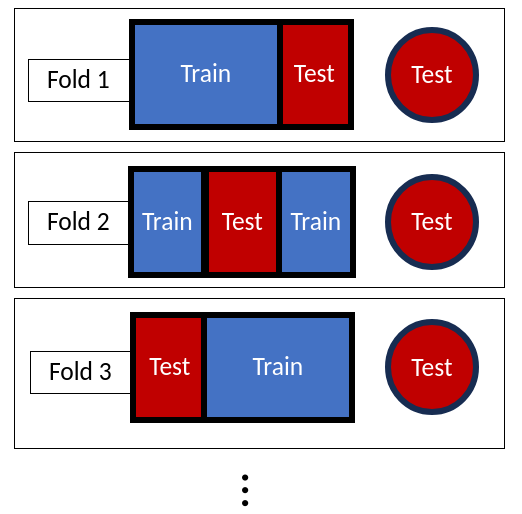}
        \caption{Generating OOB for $A^{chem}$}
        \label{chem_cv}
    \end{subfigure}
    \hfill
    \begin{subfigure}{0.45\textwidth}
        \centering
        \includegraphics[width=0.9\textwidth]{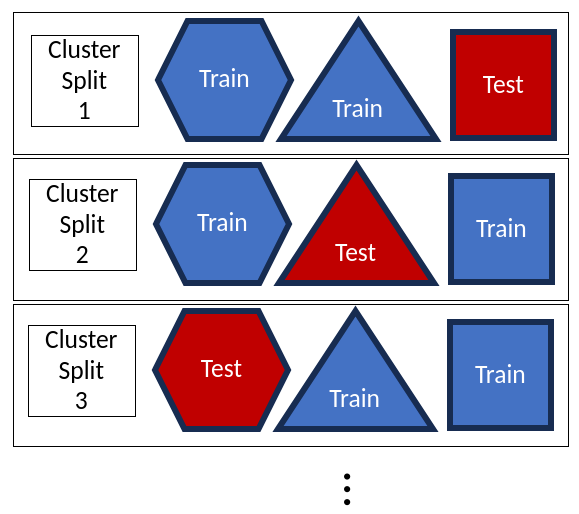}
        \caption{Generating OOB from BLOCO}
        \label{bloco_cv}
    \end{subfigure}

    \vspace{1cm}
    
    \begin{subfigure}{\textwidth}
        \centering
        \includegraphics[width=0.9\textwidth]{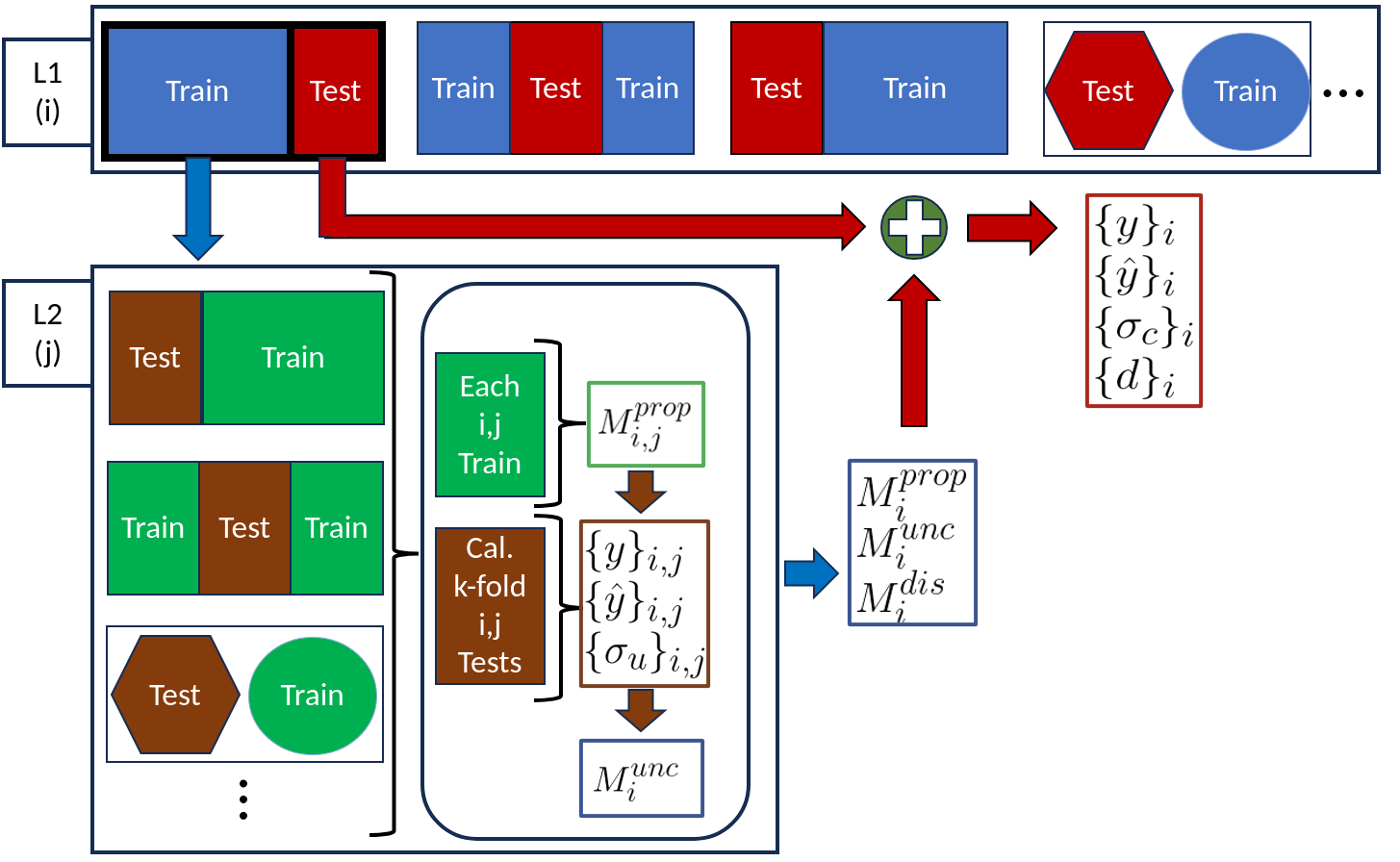}
        \caption{Nested CV}
        \label{cv}
    \end{subfigure}
    \caption{\textbf{Methods to generate OOB data.} Shown in Fig.~\ref{chem_cv} is the splitting methodology for $A^{chem}$. The square and circle sets are from LOO CV and a fixed OOB set, respectively. Shown in Fig.~\ref{bloco_cv} is the BLOCO splitting methodology. Each shape represents a specific cluster. Clusters were iteratively swapped between ITB and OOB sets. Shown in Fig.~\ref{cv} is the depiction of nested CV. The upper level (L1) was used to produce ITB (blue Train) and OOB (red Test) data from k-fold and BLOCO splits. From each L1 Train, we further divided data in a nested manner (L2). From L2, we calibrated model uncertainties on a set of k-fold splits and produced $M^{unc}_{i}$. $M^{prop}_{i}$ and $M^{dis}_{i}$ were fit to all Train for each split in L1. The predictions of $M^{prop}_{i}$, $M^{unc}_{i}$ and $M^{dis}_{i}$ on each respective L1 Test set produced the OOB data used to measure $E^{|y-\hat{y}|/MAD_{y}}$, $E^{RMSE/\sigma_{y}}$, and $E^{area}$.}
\end{figure}

In this section we describe our method to evaluate how $d$ provides information with respect to $E^{|y-\hat{y}|/MAD_{y}}$, $E^{RMSE/\sigma_{y}}$, and $E^{area}$ and call these assessments $A^{|y-\hat{y}|/MAD_{y}}$, $A^{RMSE/\sigma_{y}}$ and $A^{area}$, respectively. We must generate a set of OOB data that contain $ID$/$OD$ labels to assess the ability of $d$ to separate $ID$/$OD$. To do this, data containing both $X$ and $y$ were split using several methods for CV. First, data were split by 5-fold CV where models were iteratively fit on 4 folds and then predicted $\hat{y}$, $\sigma_{c}$, and $d$ on OOB data. Second, data were pre-clustered using agglomerative clustering from scikit-learn \supercite{scikit-learn}. One cluster was left out for OOB data, and all other ITB clusters were used to train models. Then, models predicted $\hat{y}$, $\sigma_{c}$, and $d$ on the OOB data. Like the 5-fold CV methodology, we sequentially left out each cluster, which was similar to the approaches applied in Refs.~\cite{loco} and \cite{Meredig2018}. However, applying this clustering to the original data gave only a limited number of clusters. To generate many more OOB data, we generated a series of bootstrap data sets from clusters and then applied the same leave out cluster approach. To clarify, we first clustered the data, and then performed bootstrapping separately on each cluster, rather than bootstrapping on all the data and then clustering. We call this a BLOCO approach. We performed BLOCO with 3 and 2 total clusters with the aim of providing a large amount of OOB data that are increasingly dissimilar to ITB data and hopefully $OD$. See Fig.~\ref{bloco_cv} for an illustration of BLOCO.

Both 5-fold CV and the BLOCO splitting strategies were repeated 5 times for nested CV on most data and model combinations. Only BNN models had the number of repeats decreased to 3, 3, and 2 for the Fluence, Friedman, and Superconductor data sets, respectively. The amount of RAM consumed by BNNs for these larger data sets lead to a practical limitation on acquirable statistics. The workflow for producing the OOB data is shown in Fig.~\ref{cv}. OOB data were aggregated and binned with respect to $d$ into $N$ bins with equal (or close to equal) number of points. If data could not be divided evenly (e.g., many repeated points with $d$=1), then the extra points went to the bin with the highest average $d$. Our choice of $N=10$ gave robust results. The number of bins should strike a balance between being small enough for each bin to encompass a substantial number of data points for reliable statistics, yet large enough to effectively discern shifts in trends across different bins of OOB data. $E^{RMSE/\sigma_{y}}$ was calculated for all OOB data in each bin. We subsequently compared $O(z)$ to $\Phi(0,1)$ for each bin using Eq.~\ref{misscalibration_area}, which provided $E^{area}$ for each bin. The measure of $E^{|y-\hat{y}|/MAD_{y}}$ did not require binning and was measured for each individual data point. Note that normalization of $E^{|y-\hat{y}|/MAD_{y}}$ and $E^{RMSE/\sigma_{y}}$ used the $MAD_y$ and $\sigma_{y}$, respectively, from ITB data, not the OOB data. To avoid data leakage, we prevented models from learning on OOB data by incorporating information solely from ITB data during the model development phase, which may yield slightly different values for $MAD_y$ and $\sigma_{y}$ according to the splits considered. Class labels were acquired, $E^{|y-\hat{y}|/MAD_{y}}$ versus $d$ plots were generated, $E^{RMSE/\sigma_{y}}$ versus $d$ plots were generated, $E^{area}$ versus $d$ plots were generated, and precision and recall scores were recorded. Each $M^{dom}$ for $E^{|y-\hat{y}|/MAD_{y}}$, $E^{RMSE/\sigma_{y}}$, and $E^{area}$ were built similarly. Any $M^{dom}$ model checks the value of $M^{dis}$ and returns a corresponding $\widehat{ID}$/$\widehat{OD}$ value.

Unfortunately, the above steps meant that the choice of $N$ impacted which data points were averaged together and therefore impacted the values of $E^{RMSE/\sigma_{y}}$ and $E^{area}$ and the grid of points on which $d$ can be evaluated. Thus, $N$ impacts both this assessment of $M^{dom}$ and the $M^{dom}$ that would be developed for a model to be used at inference. While we have found that using 10 bins can be effective, it is important to note that this choice may not necessarily be optimal for these or general models. Future work should establish an approach for this binning that automatically selects $N$ to obtain optimal results, or avoids binning altogether in the determination of $d^{t}_{c}$ for $M^{dom}$. Note that $E^{|y-\hat{y}|/MAD_{y}}$ is unaffected by $N$.

\pagebreak
\pagebreak

\subsection*{Data Curation}
\label{data_curation}

We applied all previous definitions of domain and assessments with five data sets consisting of four physical data sets from the field of materials science and one synthetic data set (all discussed in detail below). Among these, the four physical property data sets exhibit well-defined chemical domains, with samples that are readily categorized as $ID$/$OD$. The assigned categories are shown in Table~\ref{chemical_gt}. For three of the physical data sets, the data curation process involved generation of features, followed by a careful down-selection of a relevant $X$ that held significance for $y$ (see Supplemental Materials for the feature learning curves). It is important to remember that certain groups were denoted as Original and Perturbed Original across various data sets. The Original and Perturbed Original sets have the same number of data points. It is essential to interpret these labels within the specific context of the material data set under investigation.

We assert that the diverse range of target properties (experimental, simulated, and synthetic target values), feature characteristics, and dataset sizes (both in terms of sample counts and feature dimensions) contributes to a rigorous assessment of the developed methods. Our study focuses exclusively on tabular data with well-defined input feature sets, rather than unstructured data types such as text, images, or graphs commonly used in deep learning, where feature extraction is an integral part of the model fitting process. This limitation raises a valid point about the applicability of our method to such data types. While it's not immediately apparent how to extend our approach to these domains, one potential strategy could involve leveraging the latent space representations from deep learning models as a feature space for KDE in our model. Although this approach, or similar variants, may prove effective, we have not yet investigated the efficacy of our method in these scenarios. Exploring such applications would extend beyond the scope of the current paper and warrants future research.

\begin{table}
    \centering
    \caption{Tabulated are the ground truth labels for $A^{chem}$. The first and second values of entries containing a comma specify values for the Steel Strength and Fluence data sets, respectively.}
    \label{chemical_gt}
    \begin{tabular}{|l|lllr|}

    \hline
    Data Set                            & Chemical Group         & Number of Points & Material & Label  \\
    \hline
    Diffusion                           &                        &                                            &        \\

                                        & Original               & 408      & Many                            & $ID$   \\
                                        & Alkaline earth metals  & 36       & Many                            & $ID$   \\
                                        & Transition metals      & 1156     & Many                            & $ID$   \\
                                        & Post-transition metals & 81       & Many                            & $ID$   \\
                                        & Metalloids             & 36       & Many                            & $OD$   \\
                                        & Alkali metals          & 36       & Many                            & $OD$   \\
                                        & Reactive nonmetals     & 121      & Many                            & $OD$   \\
                                        & Noble gases            & 36       & Many                            & $OD$   \\
                                        & Lanthanides            & 225      & Many                            & $OD$   \\
                                        & Actinides              & 225      & Many                            & $OD$   \\
                                        & Manual                 & 6        & Many                            & $OD$   \\

    \hline
    Steel Strength                      &                        &          &                                 &        \\
    and Fluence                         &                        &          &                                 &        \\

                                        & Original               & 312, 2049      & Many                            & $ID$   \\
                                        & Perturbed Original     & 312, 2049      & Many                            & $ID$   \\
                                        & Copper-Based           & 1        & $Cu98.05Be1.7Co0.25$            & $OD$   \\
                                        & Copper-Based           & 1        & $Cu85Zn15$                      & $OD$   \\
                                        & Copper-Based           & 1        & $Cu89.8Sn10P0.2$                & $OD$   \\
                                        & Copper-Based           & 1        & $Cu92Ni4Sn4$                    & $OD$   \\
                                        & Copper-Based           & 1        & $Cu85Zn5Pb5Sn5$                 & $OD$   \\
                                        & Aluminum-Based         & 1        & $Al95Si5$                       & $OD$   \\
                                        & Aluminum-Based         & 1        & $Al90Mg10$                      & $OD$   \\
                                        & Aluminum-Based         & 1        & $Al92.5Mg1.5Ni2Cu4$             & $OD$   \\
                                        & Aluminum-Based         & 1        & $Al83.5Si12Mg1Ni2.5Cu1$         & $OD$   \\
                                        & Aluminum-Based         & 1        & $Al88Cu3.5Si8.5$                & $OD$   \\
                                        & Iron-Based-Far         & 1        & $Fe99.1C0.2Mn0.45Si0.25$        & $OD$   \\
                                        & Iron-Based-Far         & 1        & $Fe96.49C0.21Mn0.75Si0.25Ni2.3$ & $OD$   \\
                                        & Iron-Based-Far         & 1        & $Fe60.6Ni35Si2C2.4$             & $OD$   \\
                                        & Iron-Based-Far         & 1        & $Fe42.45Cr18Ni39C0.55$          & $OD$   \\
                                        & Iron-Based-Far         & 1        & $Fe63Cr28Ni9$                   & $OD$   \\

    \hline
    Cuprates                            &                        &          &                                &        \\

                                        & Cuprates               & 3809     & Many                           & $ID$   \\
                                        & Iron-Based             & 1182     & Many                           & $OD$   \\
                                        & Low-${T_{c}}$          & 1261     & Many                           & $OD$   \\

    \hline
    Iron-Based                          &                        &          &                                &        \\
                                        & Iron-Based             & 1182     & Many                           & $ID$   \\
                                        & Cuprates               & 3809     & Many                           & $OD$   \\
                                        & Low-${T_{c}}$          & 1261     & Many                           & $OD$   \\

    \hline
    Low-${T_{c}}$                       &                        &          &                                &        \\

                                        & Low-${T_{c}}$          & 1261     & Many                           & $ID$   \\
                                        & Iron-Based             & 1182     & Many                           & $OD$   \\
                                        & Cuprates               & 3809     & Many                           & $OD$   \\

    \hline
    \end{tabular}
\end{table}

\subsubsection*{Diffusion Data Set}
\label{diffusion_data}

Diffusion activation energies ($y$) of single atom (dilute solute) impurities in metal hosts were acquired from Refs.~\cite{Lu2019} and \cite{Wu2017}. The activation energies were calculated using Density Functional Theory (DFT) methods. The total number of host-impurity pairs are 408. We refer to these compositions as the Original set. We generated host and impurity combinations across many chemical groups in the periodic table, but made sure to only include compositions outside the Original data set (no $y$ are available). Chemical groups were assigned $ID$/$OD$ based on intuition (see Table~\ref{chemical_gt}). The Original group was labeled $ID$ because they were data used to build $M^{prop}$ models. Host metals for the Original group include elements in the following groups: alkaline earth, transition, and post-transition metals. Any data from the aforementioned groups that were not included in the Original data are considered to be $ID$ due to their chemical proximity and shared physical properties. None of the remaining chemical groups in Table~\ref{chemical_gt} were included as the host element in our data, and are therefore considered to be $OD$. Furthermore, we included a Manual group which mixed elements across chemical groups designated as [host][impurity] and are as follows: [$Fe$][$O$], [$Fe$][$Cl$], [$Na$][$Cl$], [$Al$][$Br$], [$Ca$][$P$], and [$Sr$][$I$]. This provided another $OD$ group to test our $d$ measure against. Mixing elements across the periodic table groups created a set of data with distinct physics compared to our $ID$ data.

Features were generated using the MAST-ML elemental property feature generator (named ElementalFeatureGenerator in the package) \supercite{Jacobs2020}. The resulting features consist of the composition average, arithmetic average, maximum, and minimum values of elemental features for the host and impurities. While the DFT calculations used to create this database investigated dilute solutes in a metallic host, the host and impurity elements were weighted equally, following previous work in Refs.~\cite{Lu2019} and \cite{Wu2017}. The EnsembleModelFeatureSelector from MAST-ML was used to reduce the number of features to 25 ($X$) for the data containing $y$. We select the same $X$ for data without $y$. The $X$ we study has a strong relation with the $y$ of interest. We do not concern ourselves with possible overfitting from feature selection on all the data as our aim is not to make a maximally robust model but to study how well we can predict the domain of the model. We refer to data in this section as ``Diffusion''.

\subsubsection*{Fluence Data Set}
\label{fluence_data}

Data of Reactor Pressure Vessel (RPV) steel embrittlement were acquired from Ref.~\cite{ASTM2024} and described in detail from the work in Ref.~\cite{jacobs2023predictions}, where the ductile-to-brittle transition temperature shift, DT41J $[C]$, is $y$. The total number of cases is 2,049. These steels are over 96.5 wt\% $Fe$ with small amounts of alloying additions less than around 3.5 wt\%. No alloy has more than 98.6 wt\% $Fe$. For materials that should be $ID$, each element fraction (not percent) from each material of the Original set was incremented by a random number $\pm 0.01$ (a change of $1\%$) and then normalized such that the sum of element fractions is 1. In other words, minor adjustments were made to the elemental fractions of each composition by adding or subtracting a small number, resulting in materials that closely resemble the Original set. These data have $X$ that were slightly perturbed from the Original set and are called Perturbed Original, but do not have target variable values ($y$).

For $OD$ data, a subset of materials without $y$ from Ref.~\cite{nickel_institute} were added (Table~\ref{chemical_gt}). These are randomly chosen metal alloys that have been previously manufactured, but contain a majority of either $Al$, $Cu$, or $Fe$. These data should be $OD$ since much of the data contain non-trace amounts of elements not present in the Original set and contain less $Fe$ in all cases except one. The exception  contains a higher percentage of $Fe$ than any other alloy in the Original set. Furthermore, all $OD$ $Fe$-based steels are characterized by the absence of $Cu$, which contrasts with the Original set where $Cu$ is present in all alloys, albeit at significantly lower weight percentages than in the $OD$ $Cu$-based alloys. Small additions of elements often have a minor or negligible effect on the physical properties of a material. However, it is important to distinguish between the physical behavior of materials and the knowledge limitations of a machine learning model. A machine learning model, constrained by its training on a specific compositional range, cannot inherently predict the impact of introducing new elements or encountering concentrations beyond its training boundaries. Therefore, such a model should adopt a conservative approach by automatically classifying these unexplored compositional scenarios as $OD$ cases to ensure reliable predictions. A similar reasoning was made for $OD$ designation in the Steel Strength Data Set section.

For these data, elemental feature generation was not needed and thus not conducted. Instead, the features for the Fluence data include the weight percentages of elements in each alloy ($Fe$, $Cr$, $Al$, $Be$, $Co$, $Si$, $Mn$, $Zn$, $Sn$, $Pb$, $Ni$, $P$, $Cu$, $Mg$, and $C$). Additional features include irradiation fluence (in $log$ values), flux (in $log$ values), and temperature for a total of 18 features used for $A^{chem}$. Note that this feature set is a natural extension of the successfully used features in Ref.~\cite{jacobs2023predictions}. For cases that do not provide irradiation fluence or flux (i.e., all data sets other than the Original set), we performed mean imputation. The mean value for each feature was taken from the 2049 data points in the Original set and then used as the value for each data point for the corresponding feature. For $A^{|y-\hat{y}|/MAD_{y}}$, $A^{RMSE/\sigma_{y}}$, and $A^{area}$, only the features that are chemically relevant for the Original set were used. Features of elemental weights for elements not seen in the Original set were dropped since the feature columns would be equal to zero for all data points. We also omitted the weight percent of $Fe$ because it trivially represented the residual weight percent in the alloy composition. In other words, the weight percentages for $Cu$, $Ni$, $Mn$, $P$, $Si$, and $C$ were kept with fluence, flux, and temperature, which constitute 9 features. We refer to data in this section as ``Fluence''.

\subsubsection*{Steel Strength Data Set}
\label{strength_data}

Data of steel strengths were acquired from Ref.~\cite{citrination_steel} and our $y$ is yield strength. The total number of cases is 312. These alloys are majority $Fe$ with the minimum (maximum) percent of $Fe$ being 62\% (86\%). The minimum (maximum) number of alloying components is 10 (13). Similar to our Fluence data, we create an $ID$ set of cases by following the same elemental perturbation method. The same cases of $OD$ materials from Table~\ref{chemical_gt} were used as $OD$. The $OD$ alloys either have majority $Cu$, $Fe$, or $Al$, or, if they have majority $Fe$, then the $Fe$ percent is generally outside the range of our Original data (i.e., either $Fe$ alloying percent is less or greater than the $Fe$ percent in Ref.~\cite{citrination_steel}). Furthermore, the minimum number of alloying components of 10 for $ID$ alloys is much greater than the maximum of 5 for $OD$ alloys. Our designation of $OD$ materials follows a similar convention as in the previous Fluence Data Set section. Similar to our study of the Diffusion data set, we used MAST-ML to generate elemental features \supercite{Jacobs2020}. However, we have the percent composition of constituent elements, so we included the weighted averages of properties based on elemental fractions (i.e., taking the properties for each pure element and combining them with weights equal to element fractions). Because these features are based on elemental fractions, the features or material properties generated for each material in the Perturbed Original set will be different than the features from the Original set (i.e., the fractions are different so their features will differ). Application of the same feature selection method as used on the Diffusion data yielded 15 final features ($X$). We refer to data in this section as ``Steel Strength''.

\subsubsection*{Superconductor Data Set}
\label{supercond_data}

Material compositions and superconducting critical temperatures were acquired from Ref.~\cite{Stanev2018} and totaled 6,252 cases. Our $y$ was the maximum temperature at which a material is capable of superconduction, $T_{c}$. We split our data into three sets for $A^{chem}$ according to Ref.~\cite{Stanev2018}. These sets are cuprates (called Cuprates), materials containing $Fe$ (called Iron-Based), and materials left out by the previous two classifications (called Low-${T_{c}}$). Stanev et al. have demonstrated that $M^{prop}$ models trained exclusively on one of the aforementioned material classes cannot accurately predict $T_{c}$ for the others \supercite{Stanev2018}. We compared $d$ from $M^{dis}$ on models trained only on subsets of each aforementioned material class and how it compared to the other two. As an example, consider Cuprates. We perform $A^{chem}$ treating Cuprates as $ID$, Iron-Based as $OD$ and Low-${T_{c}}$ as $OD$ (see Table~\ref{chemical_gt}). Features were generated and selected similar to our Steel Strength data which yielded 25 features ($X$). We refer to data in this section as ``Superconductor''. 

\subsubsection*{Friedman Data Set}
\label{friedman_data}

Most data considered in our study are from the physical sciences. To test the generalizability of our developed methods and assessments, we also study a synthetic data set constructed from Eq.~\ref{friedman_eq}.

\begin{equation}\label{friedman_eq}
    y = 10sin(\pi x_{1}x_{2})+20(x_{3}-0.5)^{2}+10x_{4}+5x_{5}
\end{equation}

Eq.~\ref{friedman_eq} is a Friedman function from Ref.~\cite{friedman}. Each $x_{i}$ is the feature for the dimension $i$. Like the original Friedman data set, each $x_{i}$ was generated from a uniform distribution on the $[0.0, 1.0)$ (or $0.0 \leq x_{i} < 1.0$) interval to generate $y$. We chose 500 samples. Because we wanted distinct clusters to exist in our data, we also uniformly generated 500 points for each $x_{i}$ from the following intervals: $[0.0, 0.2)$, $[0.2, 0.4)$, $[0.4, 0.6)$, $[0.6, 0.8)$, and $[0.8, 1.0)$. Data generated from these intervals do not overlap, effectively creating distinct subspaces. For example, a sample where $x_{1}=0.1$ and $x_{2}=0.4$ cannot exist for the exclusive intervals. We call these data ``Friedman''. Additionally, we generated 3,000 points on the $[0.0, 1.0)$ interval to explain one of our failure modes discussed in the Notes of Caution for Domain Prediction section. We call these data FWODC. Note that FWODC is only used in the context of explaining when our method fails. No feature selection was performed because the 5 features completely explain $y$.

\section*{Data Availability}

The raw and processed data required to reproduce these findings are available to download from figshare at doi: https://doi.org/10.6084/m9.figshare.25898017.v3.

\section*{Code Availability}

The developed code is available in four places: a static code version in GitHub at \url{https://github.com/leschultz/materials_application_domain_machine_learning.git}, a continuously developed code in GitHub at \url{https://github.com/uw-cmg/materials_application_domain_machine_learning.git}, an implementation in MAST-ML at \url{https://github.com/uw-cmg/MAST-ML.git}, and a PyPI package at \url{https://pypi.org/project/madml/}.

\section*{Acknowledgements}
\label{acknowledgements}

Lane E. Schultz is grateful for the Bridge to the Doctorate: Wisconsin Louis Stokes Alliance for Minority Participation National Science Foundation (NSF) award number HRD-1612530, the University of Wisconsin–Madison Graduate Engineering Research Scholars (GERS) fellowship program, and the PPG Coating Innovation Center for financial support for the initial part of this work. The other authors gratefully acknowledge support from the NSF Collaborative Research: Framework: Machine Learning Materials Innovation Infrastructure award number 1931306. Lane E. Schultz also acknowledges this award for support for the latter part of this work. Machine learning was performed with the computational resources provided by XSEDE 2.0: Integrating, Enabling and Enhancing National Cyberinfrastructure with Expanding Community Involvement Grant ACI-1548562. We thank former and current members of the Informatics Skunkworks at the University of Wisconsin-Madison for their contributions to early aspects of this work: Angelo Cortez, Evelin Yin, Jodie Felice Ritchie, Stanley Tzeng, Avi Sharma, Linxiu Zeng, and Vidit Agrawal.

\section*{Author Contributions}

L.E.S. developed the methodology, performed the computational analyses, created visualizations, and wrote the initial manuscript draft. Y.W. contributed to the conceptualization of the work. R.J. helped with data curation, statistical analyses, and content review. D.M. conceived the project, supervised the research, acquired funding, and edited the manuscript. All authors discussed the results and contributed to the final manuscript.

\section*{Competing Interests}

The authors declare no competing financial or non-financial interests.

\pagebreak
\sloppy
\printbibliography[heading=bibintoc]

\glsaddall
\begin{tcolorbox}[
    title=Box 1 | Glossary of terms,
    breakable,
    floatplacement=ht,
    float
]

\printunsrtglossary[title={}]

\end{tcolorbox}

\end{document}